\documentclass[dvipdfmx]{pasj01}

\usepackage{graphicx}
\usepackage{natbib}
\usepackage{multirow}
\usepackage{url}
\usepackage{bm}
\usepackage{epsf}
\newenvironment{trueauthors}{\section*{Author contributions}\fontsize{8}{11}\selectfont}{\par}

\begin{document} 
\Received{}
\Accepted{}

\title{Measurements of resonant scattering in the Perseus cluster core with Hitomi SXS
\thanks{The corresponding authors are
Kosuke \textsc{Sato}, 
Irina  \textsc{Zhuravleva}, 
Frits  \textsc{Paerels}, 
Maki   \textsc{Furukawa}, 
Masanori \textsc{Ohno}, 
Megan  \textsc{Eckart}, 
Akihiro \textsc{Furuzawa}, 
Caroline \textsc{Kilbourne},
and
Maurice \textsc{Leutenegger}
}
}
\author{Hitomi Collaboration,
Felix \textsc{Aharonian}\altaffilmark{1,2,3},
Hiroki \textsc{Akamatsu}\altaffilmark{4},
Fumie \textsc{Akimoto}\altaffilmark{5},
Steven W. \textsc{Allen}\altaffilmark{6,7,8},
Lorella \textsc{Angelini}\altaffilmark{9},
Marc \textsc{Audard}\altaffilmark{10},
Hisamitsu \textsc{Awaki}\altaffilmark{11},
Magnus \textsc{Axelsson}\altaffilmark{12},
Aya \textsc{Bamba}\altaffilmark{13,14},
Marshall W. \textsc{Bautz}\altaffilmark{15}
% ...
}

\altaffiltext{1}{Dublin Institute for Advanced Studies, 31 Fitzwilliam Place, Dublin 2, Ireland}
\altaffiltext{2}{Max-Planck-Institut f{\"u}r Kernphysik, P.O. Box 103980, 69029 Heidelberg, Germany}
\altaffiltext{3}{Gran Sasso Science Institute, viale Francesco Crispi, 7 67100 L'Aquila (AQ), Italy}
\altaffiltext{4}{SRON Netherlands Institute for Space Research, Sorbonnelaan 2, 3584 CA Utrecht, The Netherlands}
\altaffiltext{5}{Institute for Space-Earth Environmental Research, Nagoya University, Furo-cho, Chikusa-ku, Nagoya, Aichi 464-8601}
\altaffiltext{6}{Kavli Institute for Particle Astrophysics and Cosmology, Stanford University, 452 Lomita Mall, Stanford, CA 94305, USA}
\altaffiltext{7}{Department of Physics, Stanford University, 382 Via Pueblo Mall, Stanford, CA 94305, USA}
\altaffiltext{8}{SLAC National Accelerator Laboratory, 2575 Sand Hill Road, Menlo Park, CA 94025, USA}
\altaffiltext{9}{NASA, Goddard Space Flight Center, 8800 Greenbelt Road, Greenbelt, MD 20771, USA}
\altaffiltext{10}{Department of Astronomy, University of Geneva, ch. d'\'Ecogia 16, CH-1290 Versoix, Switzerland}
\altaffiltext{11}{Department of Physics, Ehime University, Bunkyo-cho, Matsuyama, Ehime 790-8577}
\altaffiltext{12}{Department of Physics and Oskar Klein Center, Stockholm University, 106 91 Stockholm, Sweden}
\altaffiltext{13}{Department of Physics, The University of Tokyo, 7-3-1 Hongo, Bunkyo-ku, Tokyo 113-0033}
\altaffiltext{14}{Research Center for the Early Universe, School of Science, The University of Tokyo, 7-3-1 Hongo, Bunkyo-ku, Tokyo 113-0033}
\altaffiltext{15}{Kavli Institute for Astrophysics and Space Research, Massachusetts Institute of Technology, 77 Massachusetts Avenue, Cambridge, MA 02139, USA}
% ...

\email{
ksato@rs.tus.ac.jp, ksksato@phy.saitama-u.ac.jp, zhur@stanford.edu, frits.paerels@gmail.com
}


\maketitle

\begin{abstract}

Thanks to its high spectral resolution ($\sim5$ eV at 6 keV), 
the Soft X-ray Spectrometer (SXS) on board Hitomi enables us to measure
the detailed structure of spatially resolved emission lines from highly
ionized ions in galaxy clusters for the first time. In this series of 
papers, using the SXS we have measured the velocities of gas
motions, metallicities and the multi-temperature structure of the gas in
the core of the Perseus cluster. Here, we show that when inferring 
physical properties from line emissivities in systems like Perseus, 
the resonant scattering effect should be taken into account. 
In the Hitomi waveband, resonant scattering mostly affects 
the Fe\emissiontype{XXV} He$\alpha$ line ($w$) - the strongest line 
in the spectrum. The flux measured by Hitomi in this line is suppressed 
by a factor $\sim$1.3 in the inner $\sim$30 kpc, compared to predictions 
for an optically thin plasma; the suppression decreases with the distance 
from the center.
The $w$ line also appears slightly broader than other lines from the same
ion. The observed distortions of the $w$ line flux, shape and distance 
dependence are all consistent with the expected effect of the resonant 
scattering in the Perseus core.
By measuring the ratio of fluxes in optically thick ($w$) and thin
(Fe\emissiontype{XXV} forbidden, He$\beta$, Ly$\alpha$) lines, and
comparing these ratios with predictions from Monte Carlo radiative
transfer simulations, the velocities of gas motions have been obtained.
The results are consistent with the direct measurements of gas velocities 
from line broadening described elsewhere in this series, although 
the systematic and statistical uncertainties remain significant. 
Further improvements in the predictions of line emissivities in plasma 
models, and deeper observations with future X-ray missions offering 
similar or better capabilities to the Hitomi SXS will enable resonant 
scattering measurements to provide powerful constraints on the amplitude 
and anisotropy of clusters gas motions.
\end{abstract}

\section{Introduction}
\label{sec:introduction}

The hot ($10^7 - 10^8$ K) gas in the intra-cluster medium (ICM) 
is optically thin to the continuum X-ray radiation, meaning that 
galaxy clusters are transparent to their own X-ray continuum photons.
However, \citet{Gilfanov87} showed that the strongest X-ray resonance
lines can have significant optical depths, of order unity or larger.
Line photons can therefore undergo resonant scattering (hereafter RS),
that is, they can be absorbed by ions and be almost instantaneously 
re-emitted in a different direction. 
As a result of this scattering, the emission line intensity is reduced 
in the direction of the center of the cluster (generally the region 
of largest optical depth along our line of sight), and enhanced 
towards the outskirts \citep[e.g. see review by][]{Churazov10}. 
Even if the RS effect is not strong, it will affect the spatially 
resolved measurement of elemental abundances in the ICM 
\citep[e.g.][]{Boehringer01,Sanders04}, distort
the profiles of X-ray surface brightness 
\citep[e.g.][]{Gilfanov87,Shigeyama98},
and can lead to up to tens of percent polarization
of the line radiation \citep{Sazonov02,Zhu10}.

There have been numerous attempts to detect the RS effect in X-ray
spectra of the Perseus Cluster 
\citep[e.g.][]{Molendi98,Ezawa01,Churazov04}
and other clusters 
\citep[e.g.][]{Kaastra99,Akimoto00,Mathews01,Sakelliou02,Sanders06}.
However, the results remain somewhat controversial.
More recently it was shown for the Perseus Cluster that the energy 
resolutions of the CCD-type spectrometers on XMM-Newton and Chandra 
are not sufficient to uniquely and robustly distinguish between 
spectral distortions due to RS, different metal abundance profiles, 
and/or levels of gas turbulence \citep{Zhu13}.

Here we present Hitomi observations of the RS effect in the core of
the Perseus Cluster. Due to the superb energy resolution 
(FWHM $\sim 5$ eV at 6 keV) of the non-dispersive Soft X-ray 
Spectrometer (SXS) on-board Hitomi, individual spectral lines are 
resolved \citep{Hitomi16}, allowing us to measure the suppression of 
the flux in the He-like Fe $n=1-2$ resonance line at 6.7 keV 
for the first time. As we discuss below, this suppression is likely 
due to photons having been scattered out of the line of sight.

Given that the optical depth at the center of a line depends on the 
turbulent Doppler broadening, the comparison of fluxes for optically 
thin and thick lines can be used to measure the characteristic amplitude 
of gas velocities in the ICM, complementing direct velocity measurements
via Doppler broadening and centroid shifts. The RS technique has
previously been successfully applied to high-resolution spectra from 
the cool ($kT\sim 1$ keV), dense cores of massive elliptical galaxies 
and galaxy groups, using deep XMM-Newton observations with the
Reflection Grating Spectrometer (RGS). Detailed study of those data
showed that the Fe\emissiontype{XVII} resonance line at 15.01 \AA\,
is suppressed in the dense galaxy cores, but not in the surrounding regions,
while the line at 17.05 \AA\ from the same ion is optically thin and
is not suppressed 
\citep[e.g.][]{Xu02,Kahn03,Hayashi09,Pinto16,Ahoranta16}.
Performing modeling of the RS effect, accounting for different
levels of turbulence, revealed random gas velocities of order 
$\sim 100$ km s$^{-1}$ in many elliptical galaxies and groups 
\citep[e.g.][]{Werner09,dePlaa12,Ogo17}.

Doppler spectroscopy and the RS technique provide complementary,
non-redundant constraints on the velocity field. A measurement of
the Doppler broadening along a given line of sight depends on the
line-of-sight integral of the velocity field weighted by the square
of the density. In contrast, the RS effect probes the integral of 
the velocity field along photon trajectories, weighted by the density 
itself. Even more striking, if turbulence is isotropic, the measurements 
of the Doppler effect and RS should provide the same results.
If the measured velocities differ, this may indicate that the velocity
field is anisotropic. Namely, if motions are radial (tangential) the
scattering efficiency is reduced (increased) compared to the isotropic
case \citep{Zhu11}. It is also important to note that the RS technique
is mostly sensitive to small-scale motions \citep{Zhu11}.
The comparison between the two measurements of the velocity field
can also reveal large scale deviations from spherical symmetry,
and density inhomogeneities.

\citet{Hitomi16} mentioned the presence of the RS effect in the Perseus
core. The measured ratio of the Fe\emissiontype{XXV} He$\alpha$
resonance to forbidden line fluxes, $2.48 \pm 0.16$ with 90\%
statistical uncertainties, is smaller
than the predicted ratio in optically thin plasma with mean gas 
temperature of 3.8 keV\@.
\citet{Hitomi16} also reports velocity dispersions of
$187 \pm 13$ and $164 \pm 10$ km s$^{-1}$ in the core and outer
regions, respectively. Theoretical studies of the RS effect predict 
that the resonance line flux should be still suppressed if gas is moving 
with such velocities in the Perseus Cluster.
In this paper, we measure spatial variations of line ratios and widths 
using the improved calibration data and taking systematic uncertainties 
into account. We confirm the presence of the RS effect and, using 
numerical simulations of radiative transfer in the Perseus Cluster,
infer the velocities of gas motions.
We refer the reader to \citet{atomic, velocity, tz, abundance} papers 
for the most complete analysis of spectroscopic velocity measurements, 
details of the plasma modeling and detailed measurements of the 
temperature structure\footnote{We will refer these papers as the ``Atomic'' 
or ``A'', the ``Velocity'' or ``V'', the ``Temperature'' or ``T'', and 
the ``Abundance'' or ``Z'' papers, respectively.}. 
The ICM parameters present in this paper are consistent 
with the measurements in these papers; small variations of specific 
parameters do not affect our conclusions.

The structure of our paper is as follows. In section \ref{sec:obs},
we describe the observations and data reduction.
In section \ref{sec:indications}, we demonstrate that the complex
coupled spectral and spatial behavior of the emission line intensities
in the Fe\emissiontype{XXV} He${\alpha}$ spectrum are qualitatively
consistent with the presence of RS. In section \ref{sec:spectralfits}, 
we measure line intensity ratios that are sensitive to RS, 
as a function of position in the cluster. 
In section \ref{sec:rs-sim}, we describe the radiative transfer simulations 
performed. We used two independent simulation codes: one based on the 
software packages of {\tt Geant4} tool kit\footnote{http://geant4.cern.ch} 
and
HEAsim\footnote{https://heasarc.gsfc.nasa.gov/ftools/caldb/help/heasim.html},
and one custom-written by one of us (IZ) based on \citet{Sazonov02};
we will refer to this latter code as the ICM Monte Carlo code or
'ICMMC code'. In section \ref{sec:comparison}, we compare the results of
simulations with the measured line ratios, and derive constraints
on the turbulent velocity field. In section \ref{sec:uncertainties},
we discuss the uncertainties associated with the atomic excitation rates,
and possible presence of additional excitation processes
such as charge exchange. The results are summarized and discussed 
in section \ref{sec:conclusions}.

Throughout this paper we adopt AtomDB version
3.0.8 \footnote{http://www.atomdb.org}, and the plasma emission models
in APEC
\footnote{Astrophysical Plasma Emission Code; http://www.atomdb.org}.
All data analysis software tasks refer to the HEAsoft package
\footnote{https://heasarc.nasa.gov/lheasoft}.
We adopt a Galactic hydrogen column density of
$N_{\rm H}=1.38 \times 10^{21}$ cm$^{-2}$ \citep{Kalberla05}
in the direction of the Perseus Cluster, and use the solar abundance
table provided by \citet{Lodders09}. Unless noted otherwise,
the errors are the 68\% (1 $\sigma$) confidence limits for a single
parameter of interest.

\section{Observation \& data reduction}
\label{sec:obs}


\begin{table*}[t]
\caption{Hitomi Observations of the Perseus Cluster}
\label{tab:obs}
\begin{tabular}{lcccr} \hline
Region ID & Seq. No. & Obs. date & \multicolumn{1}{c}{(RA, Dec)$^\ast$} &Exp.$^\dagger$ \\
&&&J2000& ks  \\
\hline
obs. 1 & 100040010 & 2016-02-24T02:19:41 & (\timeform{3h19m29.8s},
 \timeform{+41D29'1.9''})& 48.7 \\
obs. 2 & 100040020 & 2016-02-25T02:14:13 & (\timeform{3h19m43.6s},
 \timeform{+41D31'9.8''})& 97.4 \\
obs. 3 & 100040030/100040040/100040050 & 2016-03-04T02:17:32 & (\timeform{3h19m43.8s}, \timeform{+41D31'12.5''})& 146.1 \\
obs. 4 & 100040060 & 2016-03-06T22:56:20 & (\timeform{3h19m48.2s}, \timeform{+41D30'44.1''})& 45.8 \\
\hline\\[-1ex]
\multicolumn{5}{l}{\parbox{0.9\textwidth}{\footnotesize
\footnotemark[$\ast$]
Average pointing direction of the Hitomi SXS, as recorded in the
RA\_NOM and DEC\_NOM keywords of the event FITS files.}}\\
\multicolumn{5}{l}{\parbox{0.9\textwidth}{\footnotesize
\footnotemark[$\dagger$]
After screening on rise time cut and for events that occur near
in time to other events
}}\\
\end{tabular}
\end{table*}

Hitomi carried out a series of 4 overlapping pointed
observations of the Perseus Cluster core during 
its commissioning phase 
in 2016 February and March, with a total of 340 ks exposure 
time (Table \ref{tab:obs}). 
The Hitomi SXS is a system that combines 
an X-ray micro-calorimeter spectrometer with a Soft X-ray Telescope (SXT) 
to cover a 3 $\times$ 3 arcmin field of view (FOV) with an angular 
resolution of 1.2 arcmin (half power diameter). The micro-calorimeter 
spectrometer provided a spectral energy resolution 
of $\Delta E \sim 5$~eV 
at 6 keV \citep{Kelley16}. 
It is operated inside a dewar, in which a multi-stage cooling system 
maintains a stable environment at 50 mK; 
temperature stability is important to 
give such a high energy resolution. The SXS was originally expected 
to cover the energy range between 0.3 and 12 keV, but
only data in the $E>\sim2$ keV band were available 
during the Perseus observations because the gate valve 
on the SXS dewar, which consists of a Be window and its support 
structure, was still closed at the time of observation. 
The other instruments on Hitomi are described in \citet{hitomiall}.
In this paper, we use only the SXS data for investigating RS in the 
Perseus cluster core.

\begin{figure}
 \begin{center}
  \includegraphics[width=8cm]{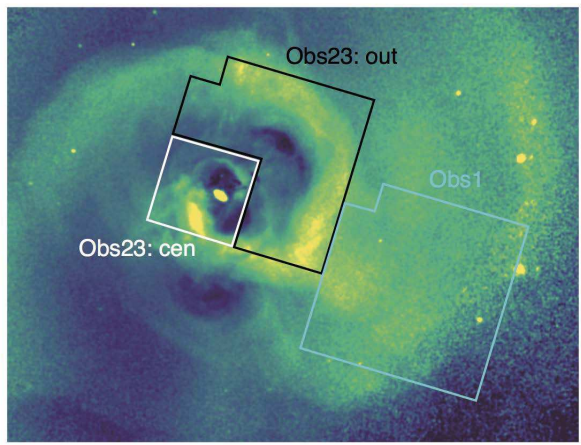} 
 \end{center}
\caption{The Hitomi SXS observation regions overlaid on the 
Chandra X-ray image of the Perseus Cluster in 
the 1.8--9.0 keV band
divided by the spherically-symmetric model for the surface brightness\@. 
In this paper we will consider obs23\_cen as the central region, 
which includes 
the central AGN (white), outer region obs23\_out (black) and obs1\_whole (cyan).
}
\label{fig:image}
\end{figure}

The Perseus Cluster was observed four times with Hitomi over a ten day 
period, but the SXS dewar had not yet achieved thermal equilibrium during 
the first two observations. A drift in temperature of the detector 
implies a drift in the photon energy to signal conversion 
(the so-called 'gain'). For the observations during which the gain drift 
was significant, the photon energy scale was determined using 
data processing routine “sxsperseus” in HEAsoft, which corrects 
the gain scale via an extrapolation of the relationship 
between the relative gain changes on the array and on the continuously 
illuminated calibration pixel from the Perseus observation 
to the later full-array calibration in the official data pipeline processing.
Observations 1 and 2 in table \ref{tab:obs} were affected by this gain drift; 
observations 3 and 4  were obtained under thermal equilibrium in the SXS dewar.
A difference in gain between obs. 2, and the sum of obs. 3 and 4 (full FOV)
of $\sim2$ eV can still be seen \citep{sxscal}. 
It is most clearly seen around the Fe\emissiontype{XXV} 
He$\alpha$ line complex in the official data pipeline processing 
\citep{Angelini16}.
Not surprisingly, obs. 1 has a much larger energy 
scale uncertainty \citep{Porter16}. 
All pixels in the micro-calorimeter array are independent, 
and in principle each has its own energy scale, and energy scale variations.

In our spectral analysis of the central region in section 
\ref{sec:spectralfits} we have to take the contribution
of non-thermal emission from the central AGN in 
the Perseus Cluster, NGC~1275,
explicitly into account. 
Following the T paper, we applied the ``sxsextend'' task 
to register event energies above 16~keV, so that we can construct 
the spectrum up to $\sim$20 keV. This is crucial to discriminate 
the AGN and cluster gas components spectrally, as the former dominates 
the spectrum in the extended energy band.
This method is same as that in the ``T'' paper \citep{tz}.
After having added the high energy events, and having applied the 
extra screenings, we adopted two extra gain corrections, similar to 
the procedures described in the ``A'' paper \citep{atomic}, 
but used the different reference redshift of 0.017284 according to 
the ``V'' paper \citep{velocity}. 
The detailed correction parameters 
were shown in the appendix in the ``T'' paper \citep{tz}.
The first of these corrections is referred to as the ``z-correction'', 
which adjusts the absolute energy scale of each pixel in each data set 
such that the redshift of the Fe\emissiontype{XXV} He$\alpha$ resonance 
line is aligned to the redshift of NGC~1275 at $z=0.017284$. 
The second is referred to as the ``quadratic-curve-correction'', 
which applies a second-order correction, centered on Fe\emissiontype{XXV} 
He$\alpha$, to take out a small apparent offsets in the observed energies 
of the strongest emission lines across the 1.8--9 keV band. 
The intent of the ``z-correction'' is to allow the spectra from different 
pixels and different pointings to be added with minimal broadening of the 
lines from variation in the bulk velocity across the Perseus cluster 
within the SXS FOV.  For the RS analysis we need to measure the ratios 
of line fluxes, thus we use the full available data set to reduce 
statistical uncertainties on measured line ratios, presuming variations 
across the data set are sufficiently small to warrant this approach. 
The uncertainties in the RS analysis associated with the energy scale 
corrections are discussed in section \ref{sec:spectralfits}.

\begin{figure*}[!th]

\begin{minipage}{0.33\textwidth}
\FigureFile(\textwidth,\textwidth){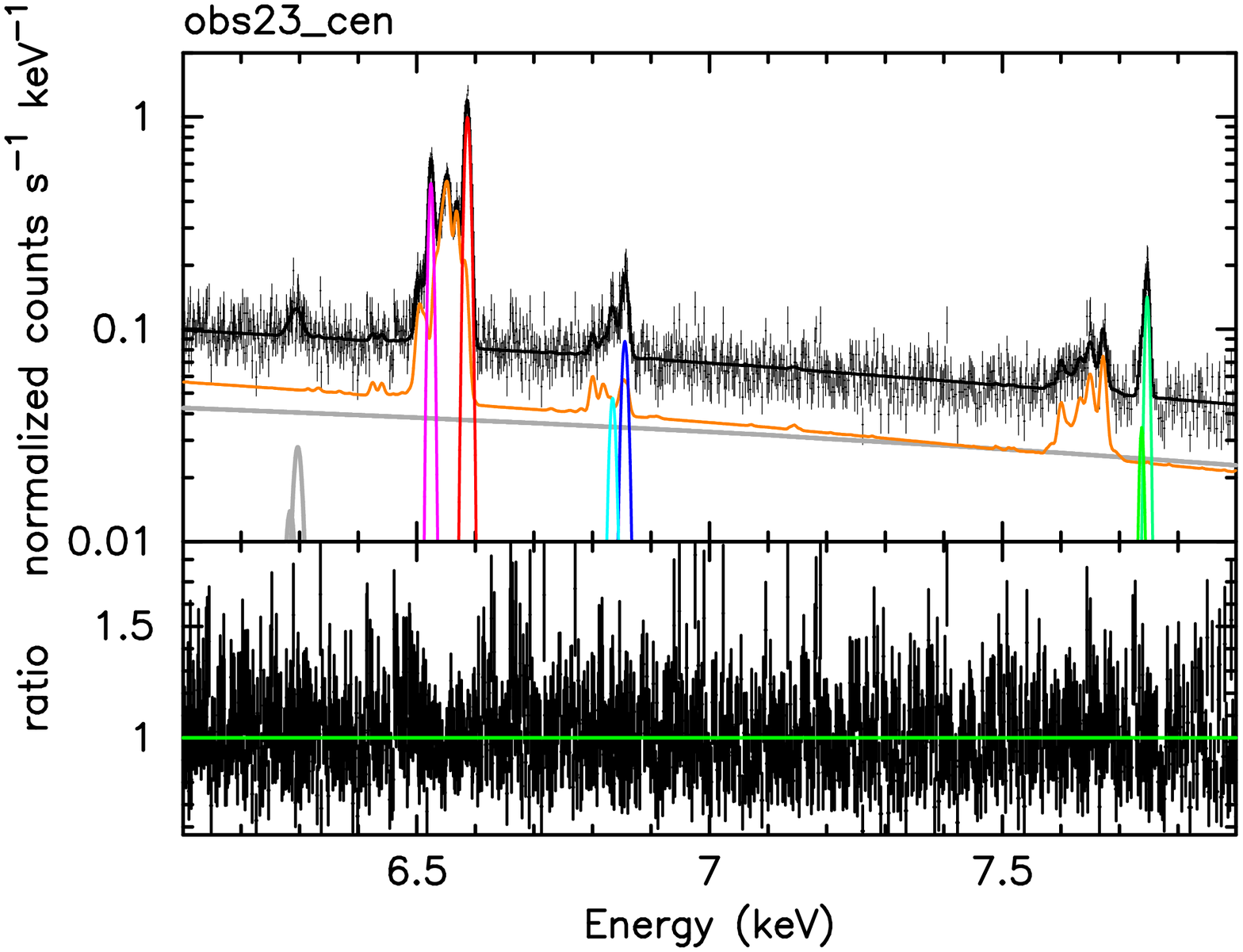}
\end{minipage}\hfill
\begin{minipage}{0.33\textwidth}
\FigureFile(\textwidth,\textwidth){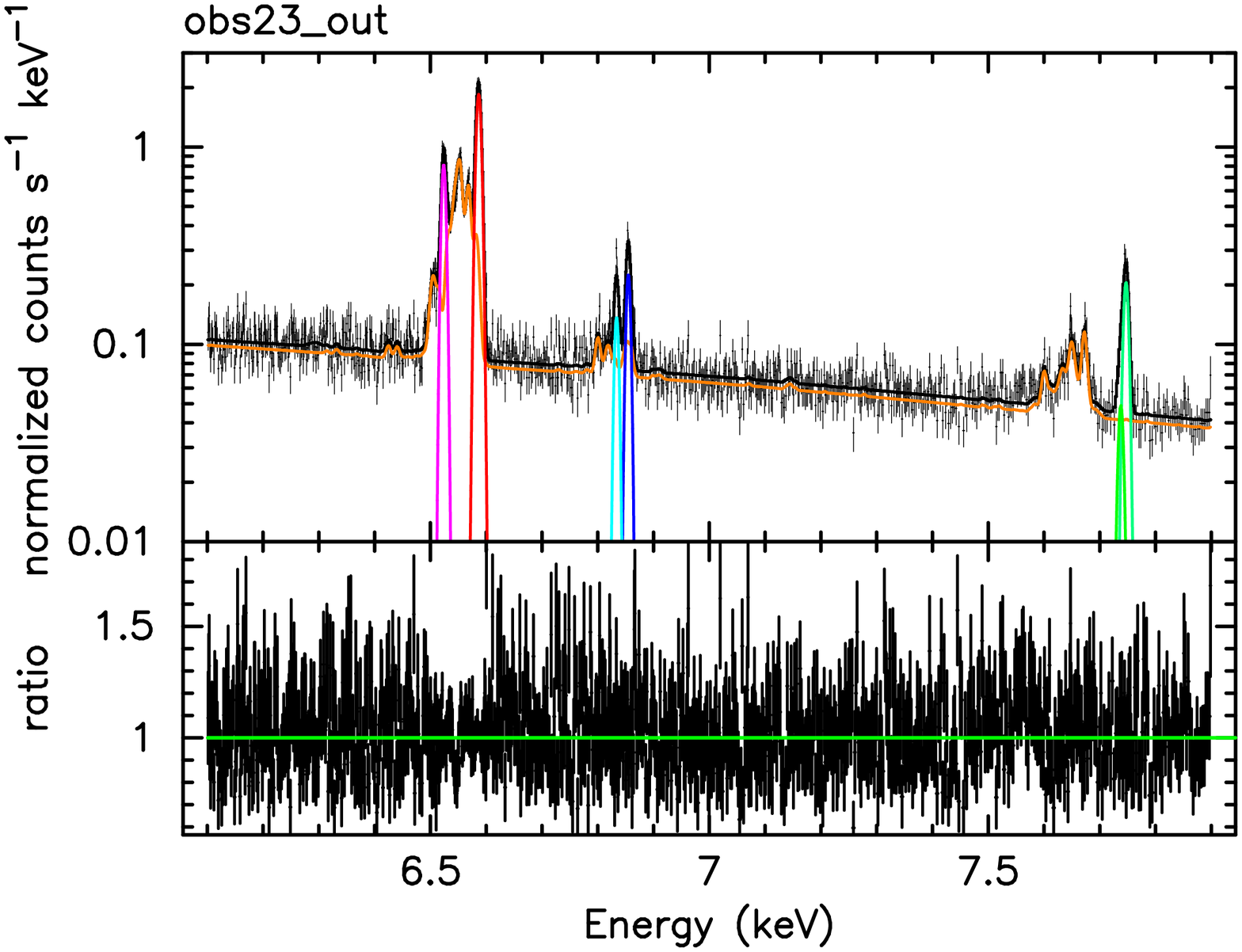}
\end{minipage}\hfill
\begin{minipage}{0.33\textwidth}
\FigureFile(\textwidth,\textwidth){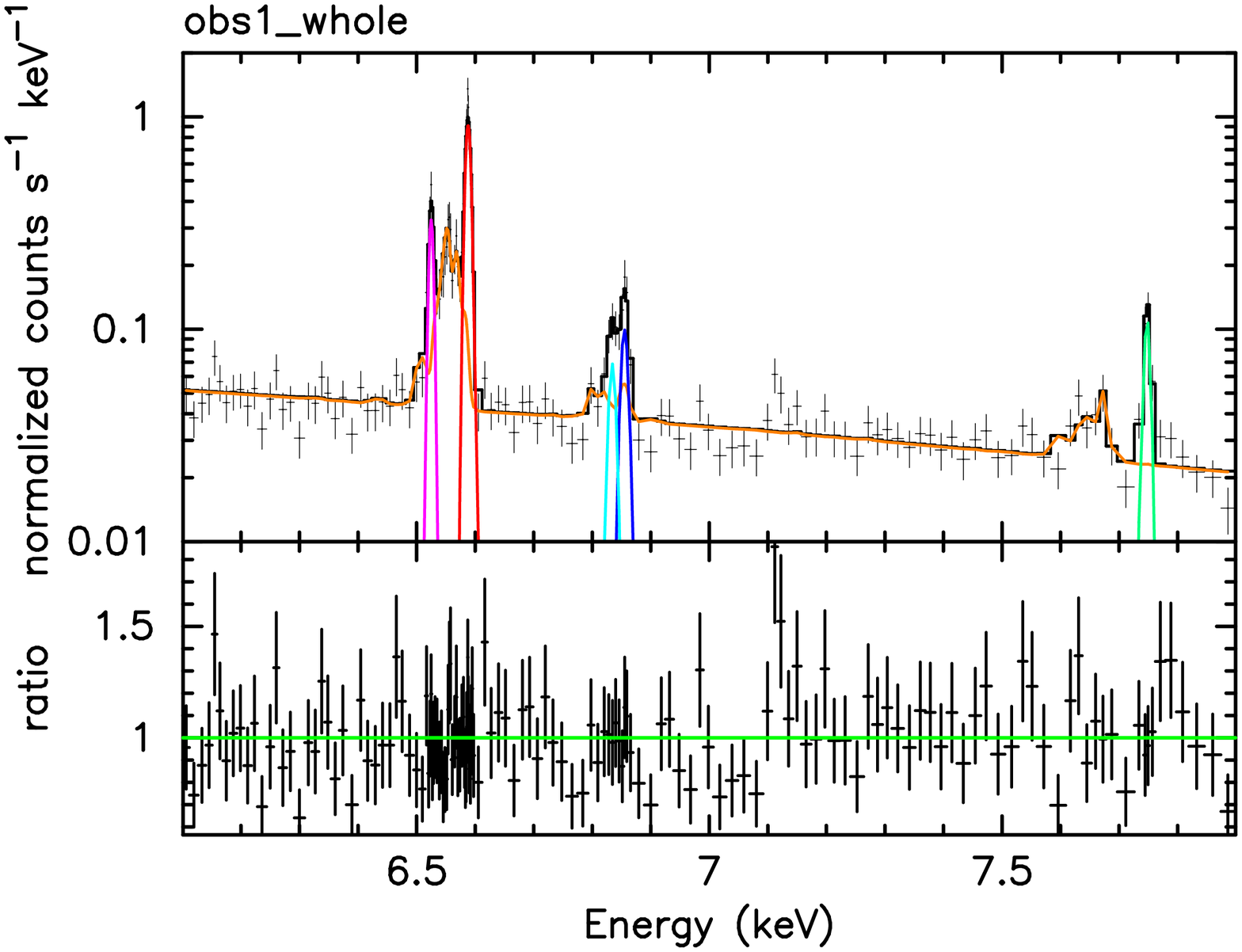}
\end{minipage}

\begin{minipage}{0.33\textwidth}
\FigureFile(\textwidth,\textwidth){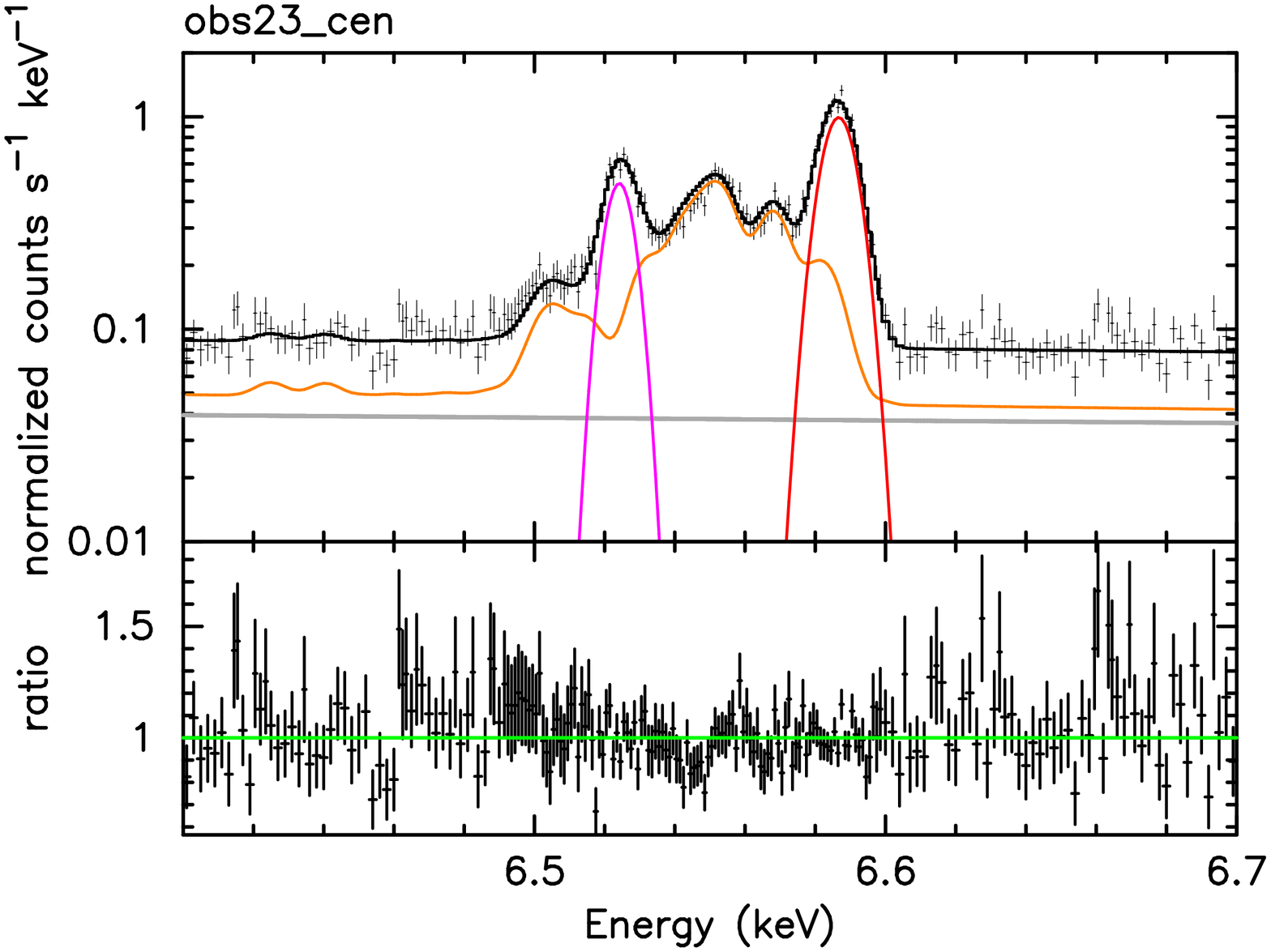}
\end{minipage}\hfill
\begin{minipage}{0.33\textwidth}
\FigureFile(\textwidth,\textwidth){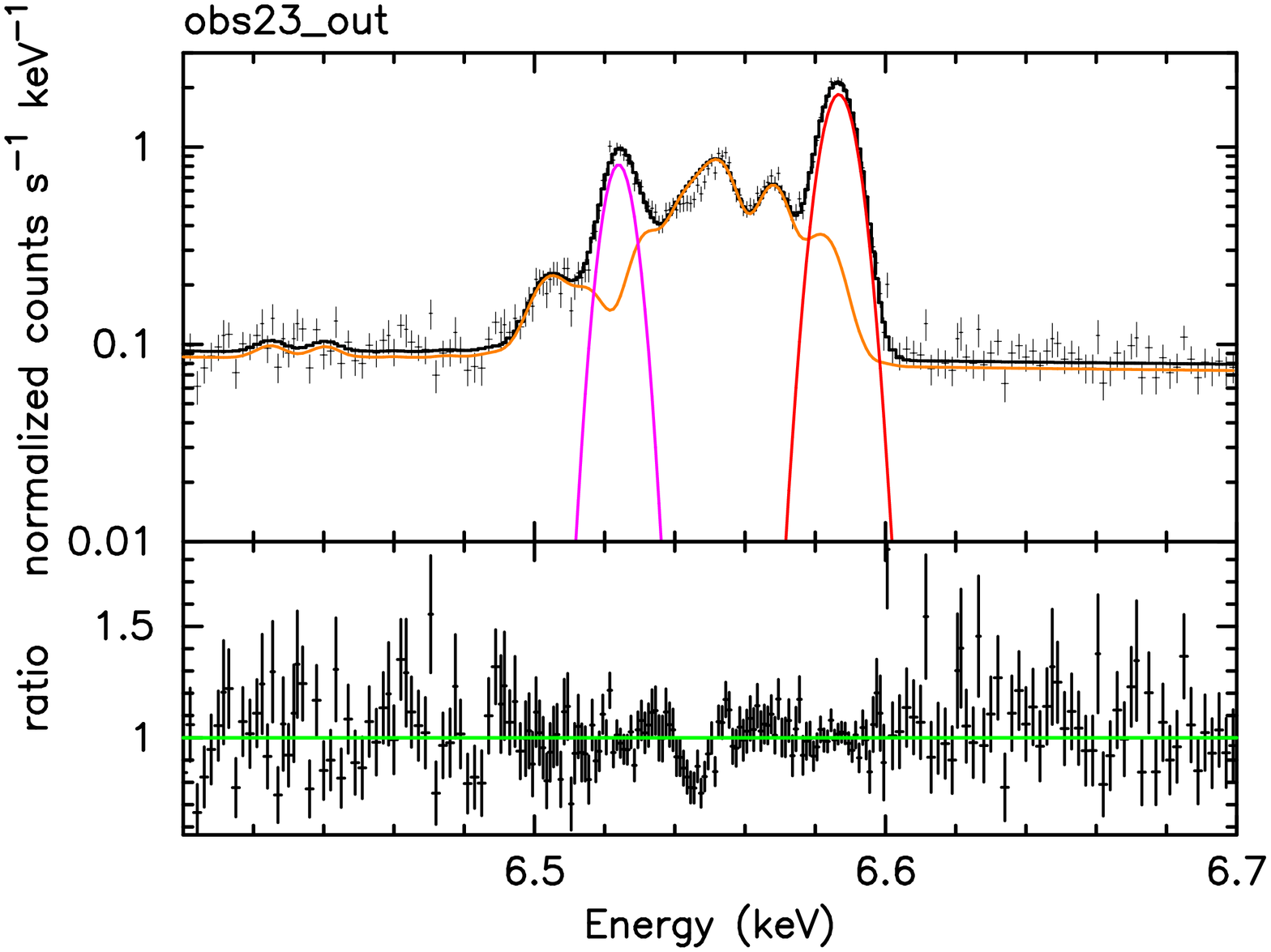}
\end{minipage}\hfill
\begin{minipage}{0.33\textwidth}
\FigureFile(\textwidth,\textwidth){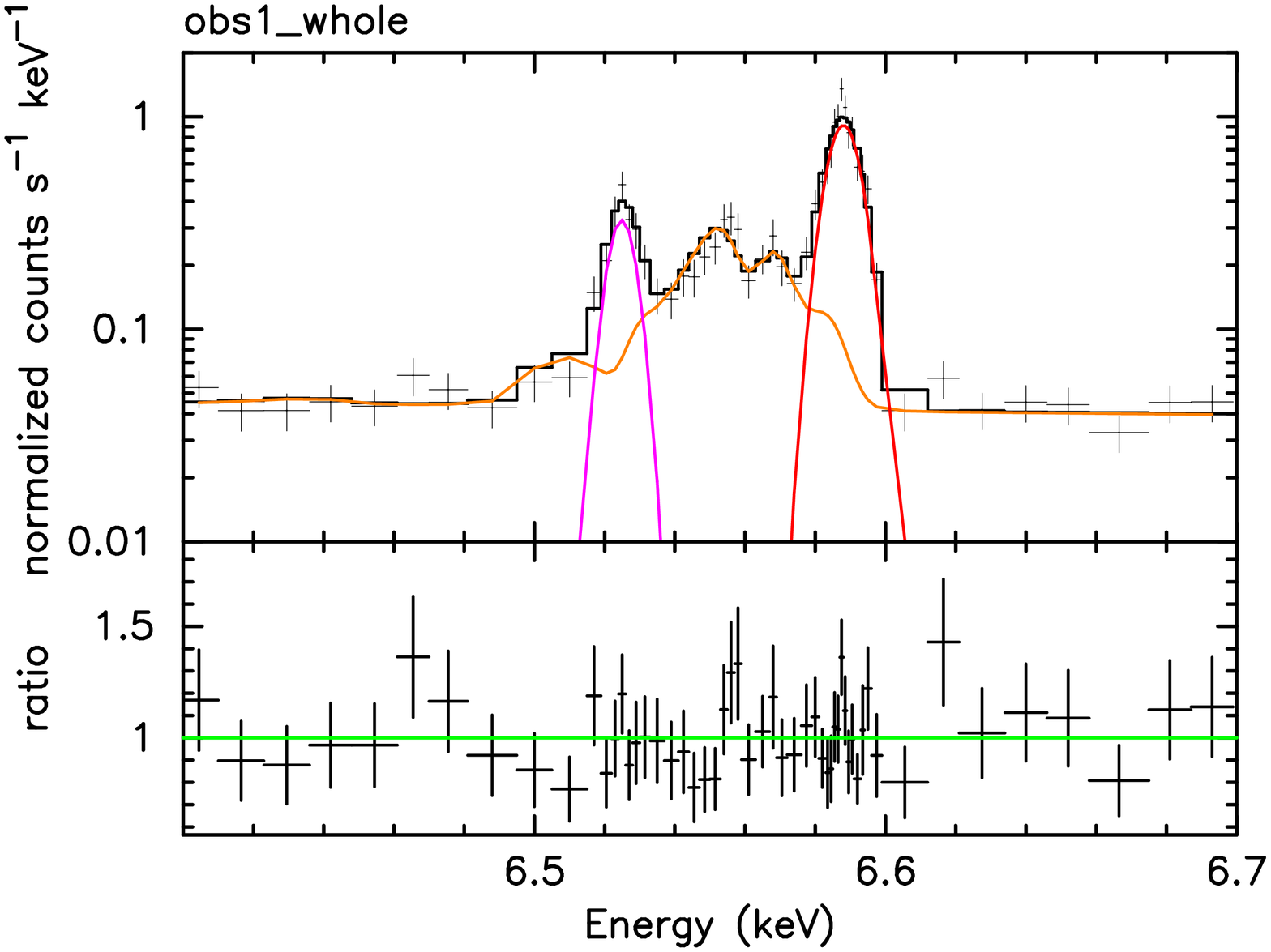}
\end{minipage}

\begin{minipage}{0.33\textwidth}
\FigureFile(\textwidth,\textwidth){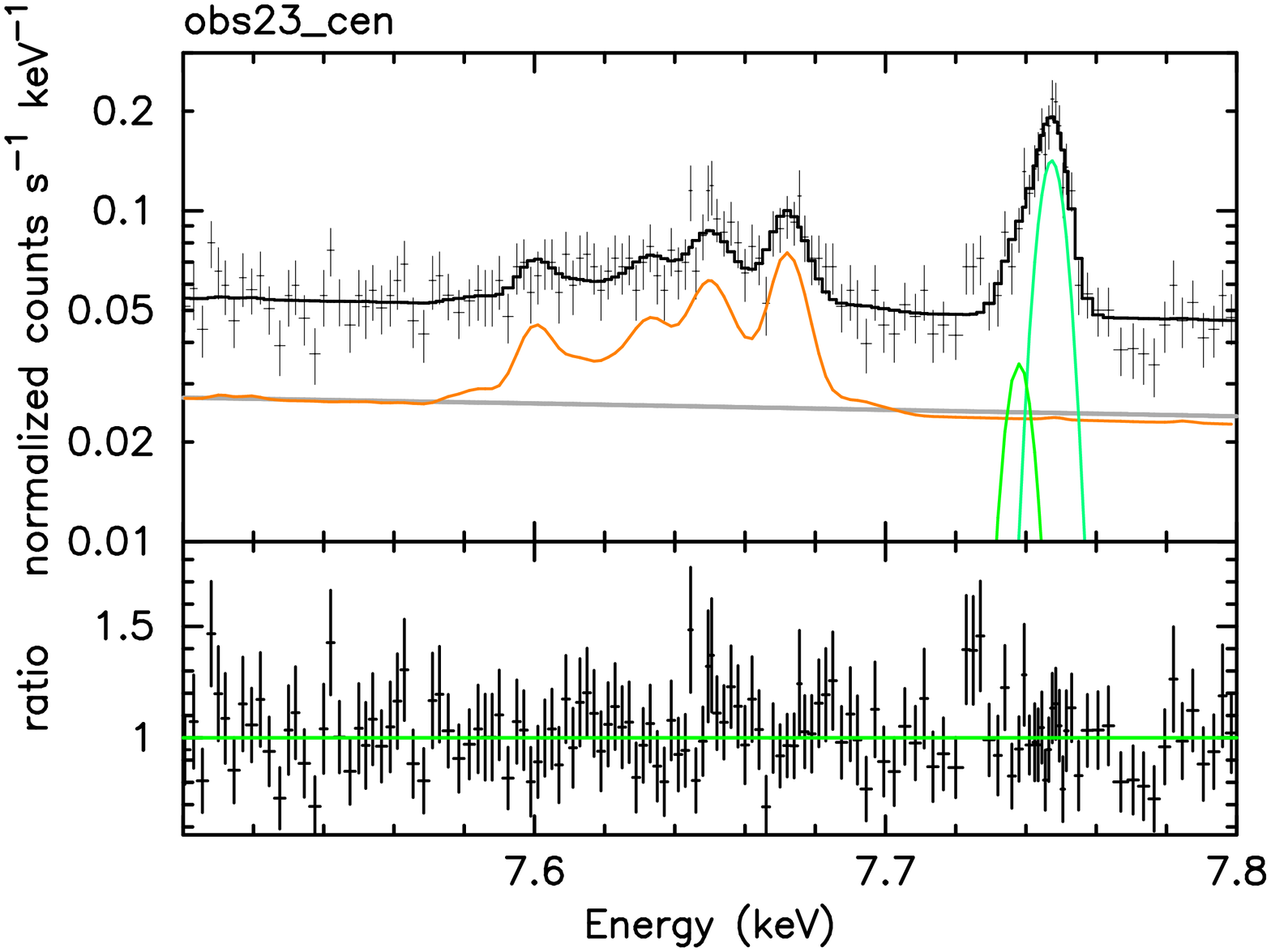}
\end{minipage}\hfill
\begin{minipage}{0.33\textwidth}
\FigureFile(\textwidth,\textwidth){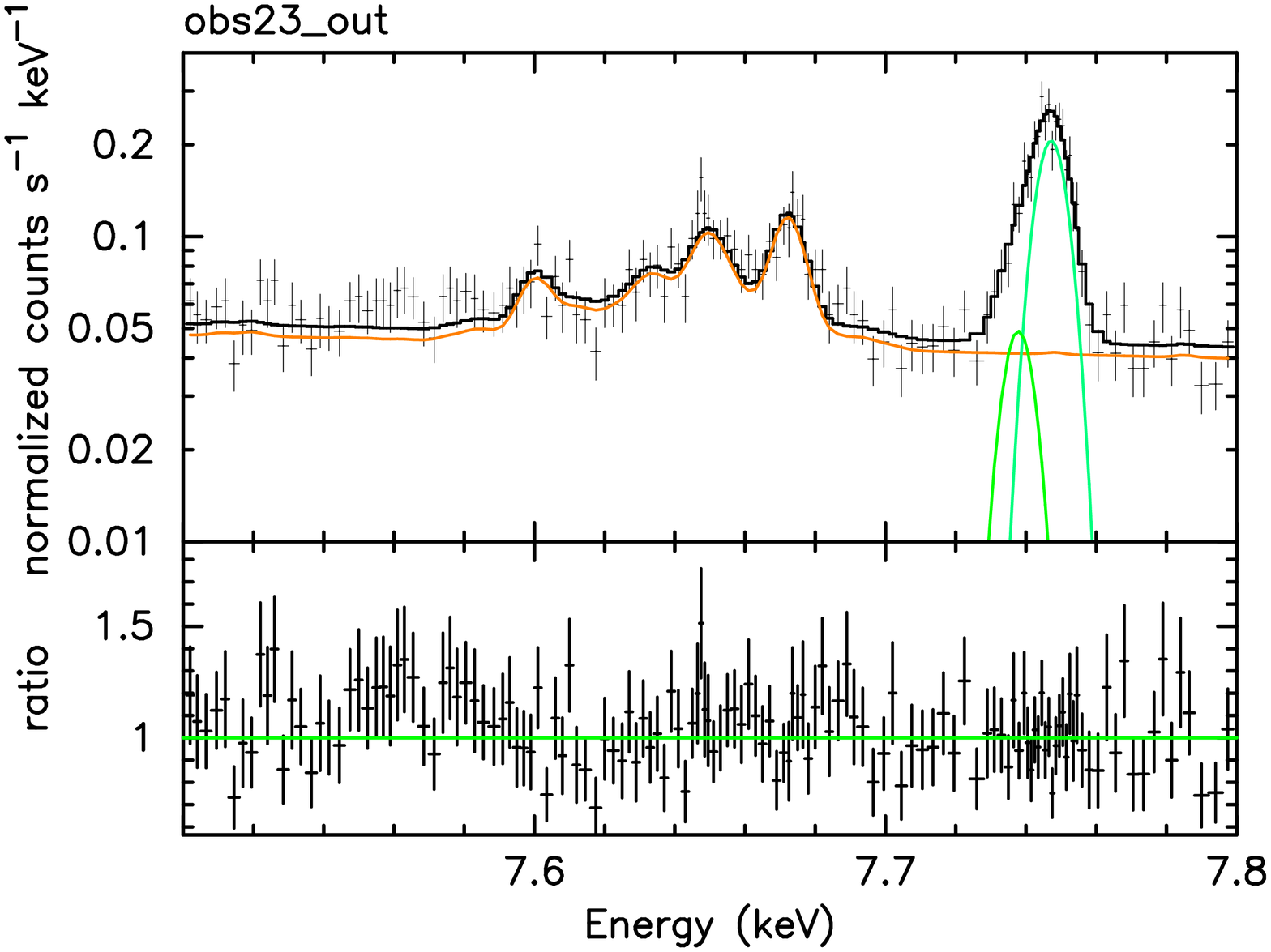}
\end{minipage}\hfill
\begin{minipage}{0.33\textwidth}
\FigureFile(\textwidth,\textwidth){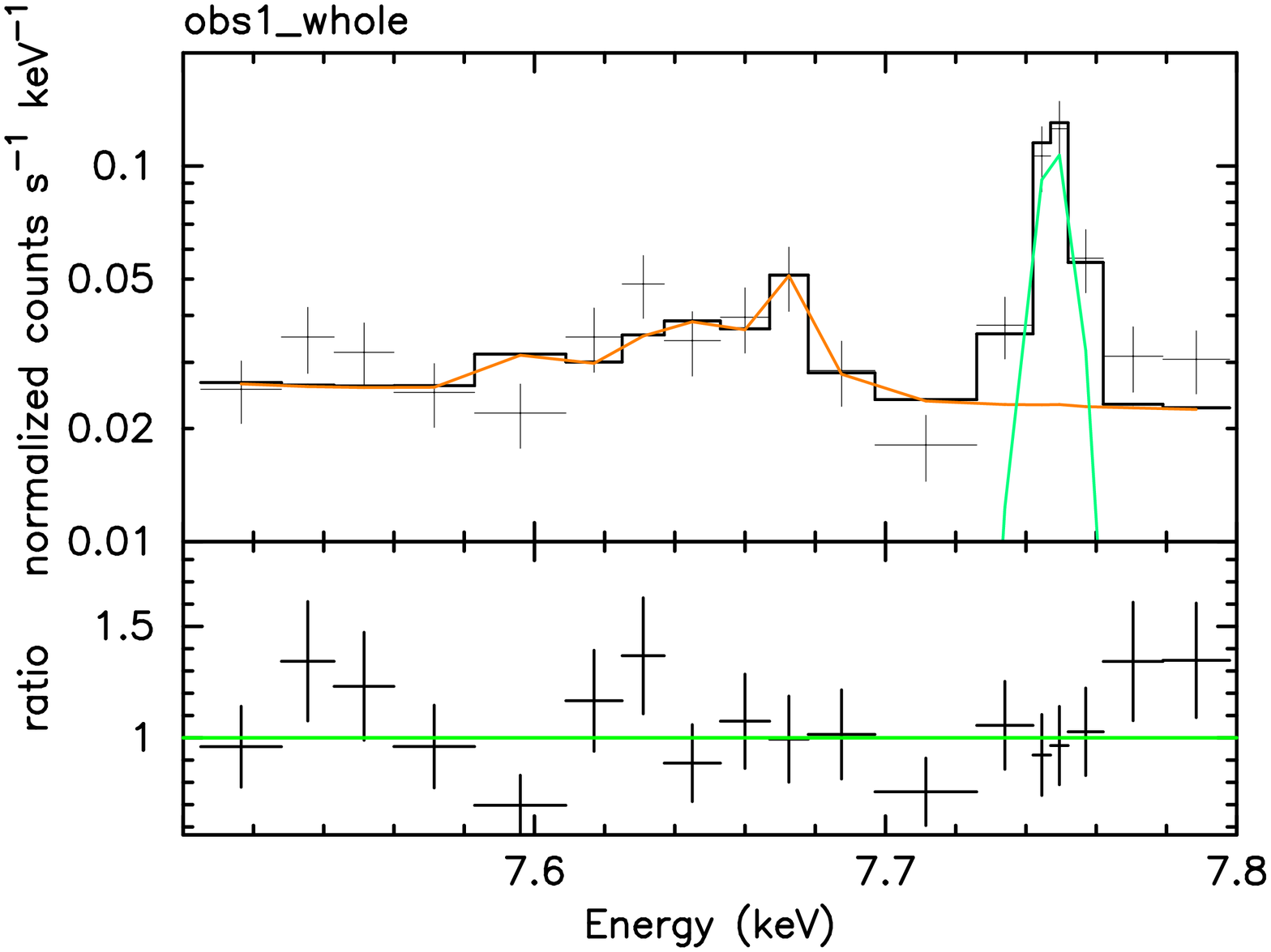}
\end{minipage}

\begin{minipage}{0.33\textwidth}
\FigureFile(\textwidth,\textwidth){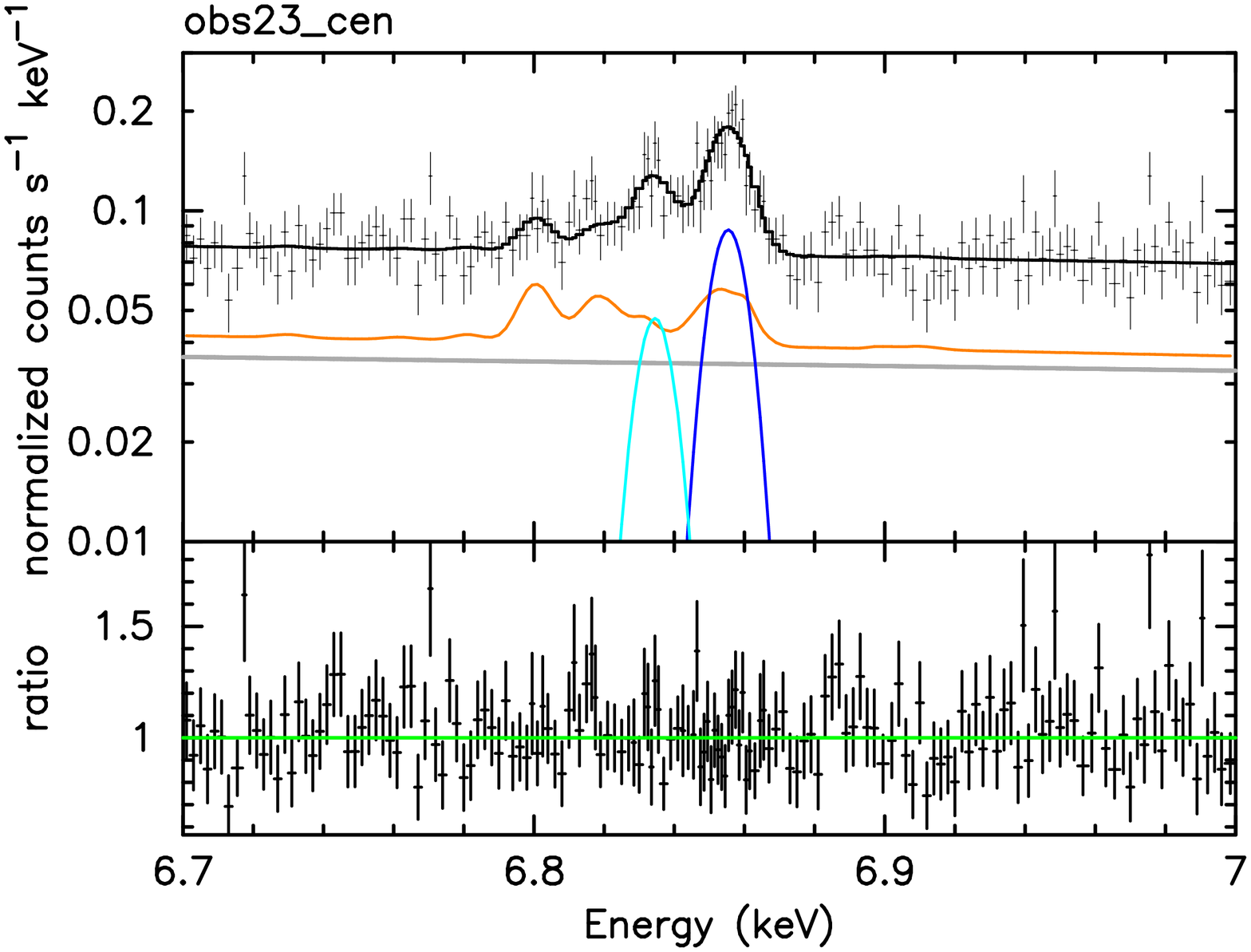}
\end{minipage}\hfill
\begin{minipage}{0.33\textwidth}
\FigureFile(\textwidth,\textwidth){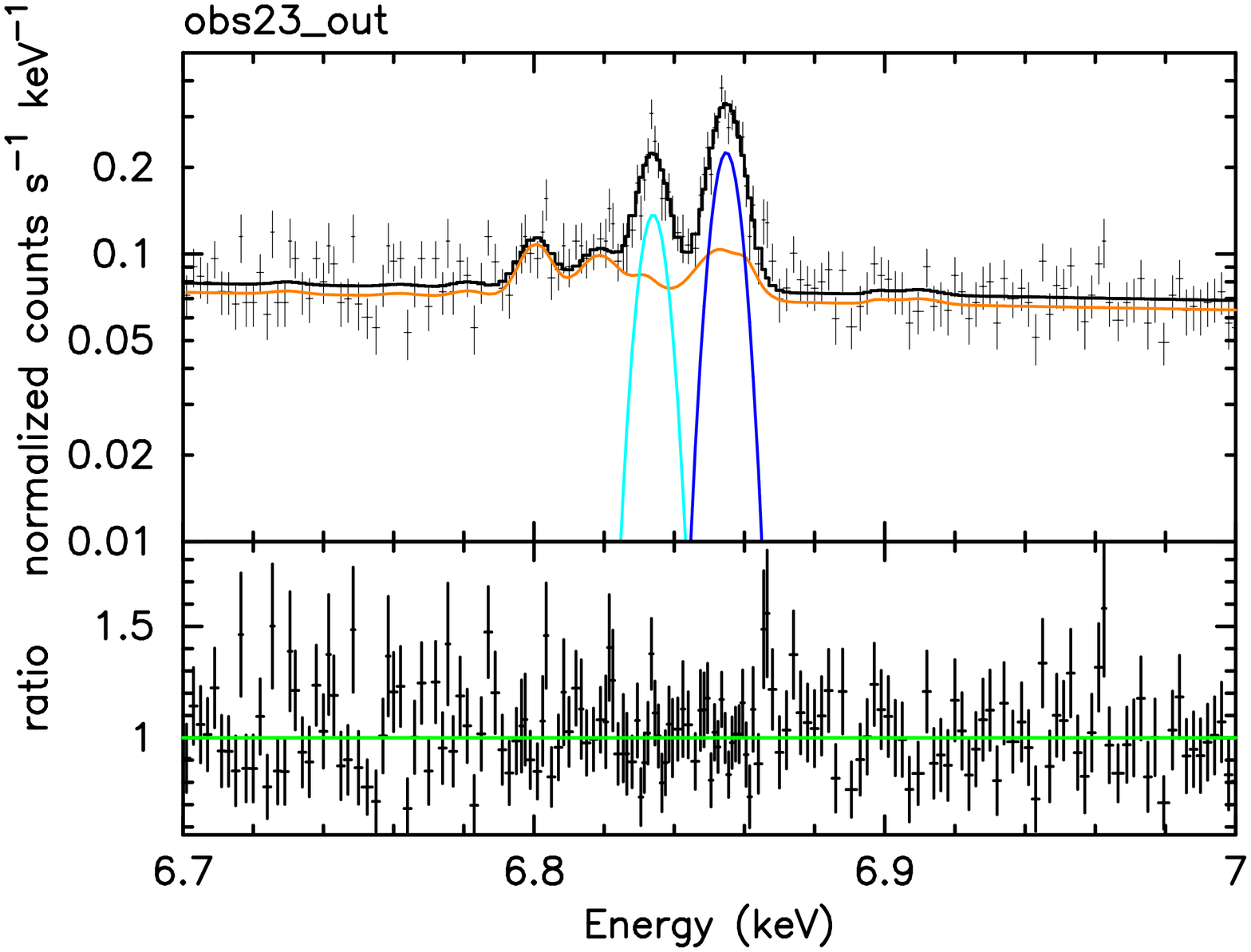}
\end{minipage}\hfill
\begin{minipage}{0.33\textwidth}
\FigureFile(\textwidth,\textwidth){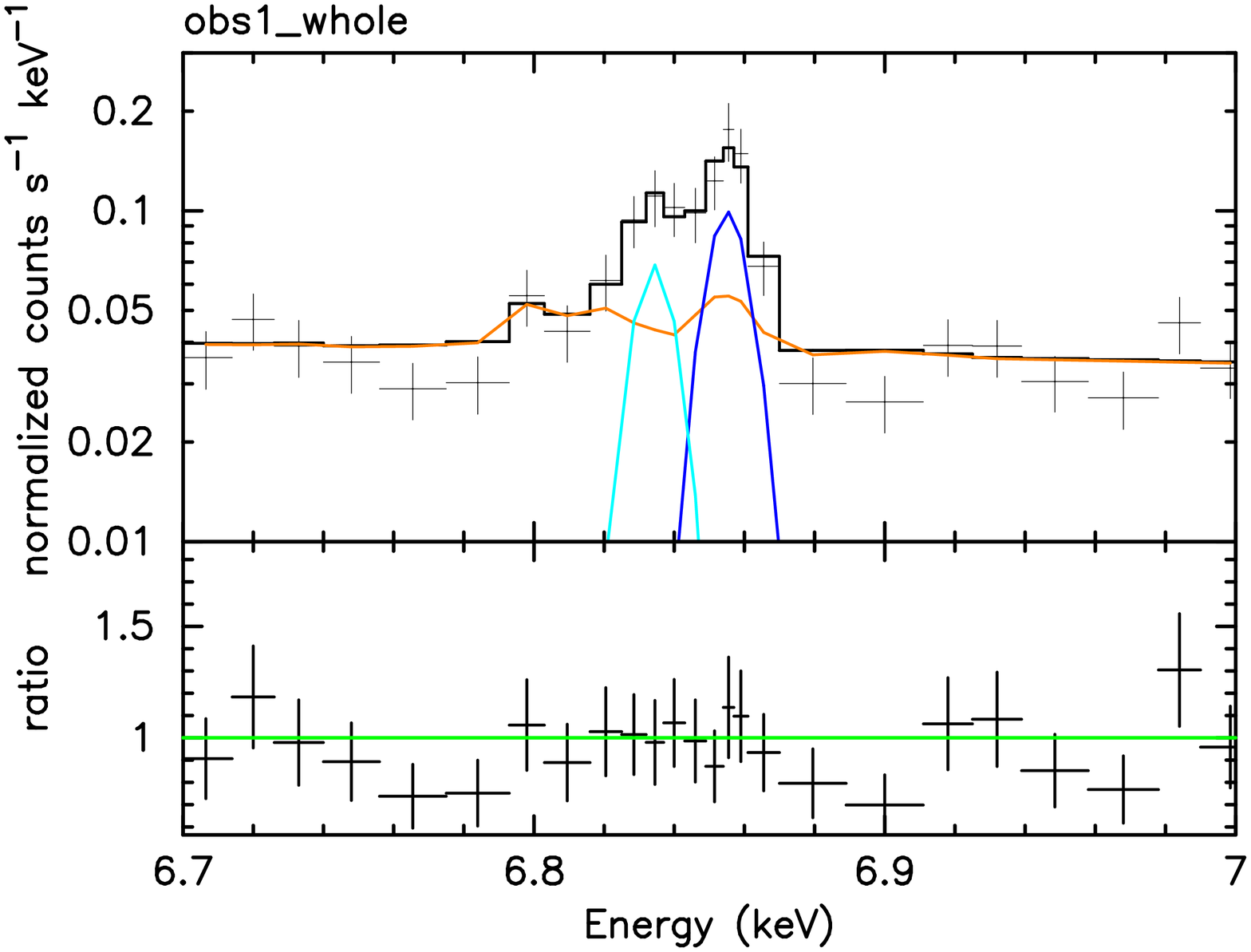}
\end{minipage}

\vspace*{5ex}
\caption{ 
The observed Hitomi spectra extracted from the obs23\_cen, obs23\_out and obs1\_whole regions shown in figure \ref{fig:image}, and binned for display purposes. 
Top panels show the resultant spectral fits in 6.1--7.9 keV band, 
while the second, third and fourth rows show the energy range of the
He$\alpha$ complex, and He$\beta$, and Ly$\alpha$ lines in 
6.4--6.7, 7.5--7.8, and 6.7--7 keV, respectively.
The spectra obtained with the Hitomi SXS are shown in black; 
light gray lines show the emission from the AGN. 
Orange lines indicate the 
``modified'' {\it bvvapec} model, in which the strongest lines have 
been deleted. The Fe\emissiontype{XXV} He$\alpha$ 
forbidden and resonance, He$\beta_{1,2}$, and 
Fe\emissiontype{XXVI} Ly$\alpha_{1,2}$ 
are shown in magenta, 
red, green, light green, blue and cyan lines, respectively. 
The lower panels show the fit residuals in units of ratio.
}
\label{fig:spectra-all}
\end{figure*}

After applying all these corrections, 
the spectra were extracted with the Xselect package in HEAsoft 
for each region as shown in figure \ref{fig:spectra-all}. 
We used only high primary grade event data to generate the spectra.
In order to subtract the non-X-ray background (NXB), we employed 
the day and dark Earth database using the ``sxsnxbgen'' Ftools task. 
We generated a redistribution matrix file (RMF) 
including the escape peak and electron loss continuum effects with the
``{\tt whichrmf=x}'' option in the ``sxsmkrmf'' task to represent the 
spectral shape in the lower and higher energy band.
Because the spatial distributions of the ICM and AGN components are
different, we also made two kinds of Ancillary Response Files (ARFs) for 
the spectrum of each region, $A^{P}$ and $A^{C}$. The response $A^{P}$ 
assumed point-like source emission from NGC~1275 centered on 
(RA, Dec.) = (\timeform{3h19m48.1s}, \timeform{+41D30'42''}); 
while $A^{C}$ is appropriate to the diffuse emission from the ICM, and 
is based on the X-ray image observed with Chandra in the 1.8--9.0 keV 
energy band, with a region of radius 10 arcseconds centered on the AGN 
replaced with the average surrounding brightness by the ``aharfgen'' 
task in HEAsoft.
At the time operations ceased, a full in-orbit calibration of the spatial 
response and effective area had not yet been performed well. 
In this paper, we therefore use ARFs generated based on the ground 
calibration of SXT.
A 'fudge factor' was derived from the ground measurements, to adjust the 
calibration to in-flight performance; however this fudge 
factor has large uncertainties as shown in \citet{crabcal, tz}, and 
the spectral fits with these ``fudged'' ARFs produced 
artificial residual features. We also examined an adjustment of 
``Crab ratio'' using the Crab observation with Hitomi SXS \citep{crabcal}, 
however this adjustment also 
introduced systematic residuals around the Au and Hg 
edges around 12~keV as shown in the ``T'' paper \citep{tz}. 
We therefore decided not to apply such corrections and use
the standard ground calibration-based response. 
Finally, in all spectral fits, the spectra are grouped with 1 eV bin$^{-1}$, 
and 1 count per bin at least, allowing 
the C-statistics method to be used.

We extracted spectra from three regions, obs23\_cen and obs23\_out 
from obs. 2 and 3, and obs1\_whole from obs. 1, with the region 
boundaries coinciding with detector pixel boundaries as shown in 
figure \ref{fig:image}, in order to avoid having to redistribute photons 
between pixels.  
The pointing directions of obs. 2 and 3 are slightly different 
(offset by $\sim0.1$ arcmin), however, this offset is much smaller than 
the size of the SXS point spread function. The obs23\_cen, obs23\_out, 
and obs1\_whole are located on the central 9 pixels around the AGN, 
the outer 26 pixels of obs. 2 and 3, and the whole region of obs. 1, 
respectively, see figure \ref{fig:image}.
The observed spectra in the 6.1--7.9 keV band are shown in figure 
\ref{fig:spectra-all}. Their modeling is discussed below 
in section \ref{sec:spectralfits}.

\section{Observational indications for resonant scattering}
\label{sec:indications}

\begin{figure*}
\includegraphics[width=\textwidth]{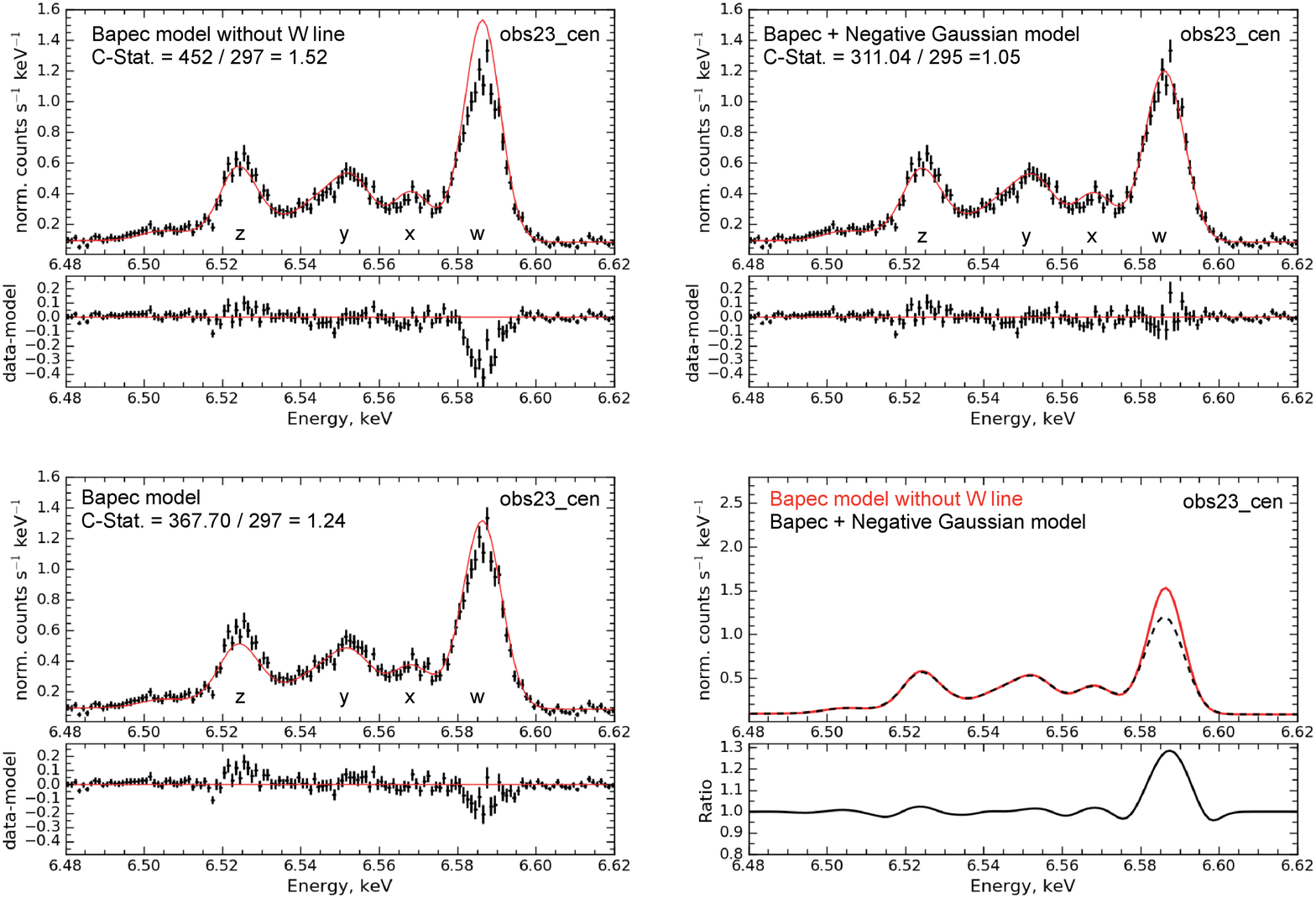}
\caption{Flux suppression in the strongest line of 
He-like Fe\emissiontype{XXV} ($w$) in the Perseus Cluster 
observed in the obs23\_cen region. Black points show the Hitomi data; 
red lines in the corresponding panels show the best-fitting models. 
{\bf Top left:} the spectrum is fitted with the {\it bapec} model, 
excluding the $w$ line from the data; {\bf top right:} the same spectrum 
is fitted with the {\it bapec} model and a Gaussian component centered 
at the energy of the $w$ line, the normalization of the Gaussian model 
is allowed to be negative; {\bf bottom left:} the same spectrum fitted 
with the {\it bapec} model. The comparison of the models from 
the top left (solid) and right (dashed) panels is shown in the 
{\bf bottom right} panel. The suppression of the $w$ line indicates 
the presence of the resonant scattering effect in the Perseus Cluster. 
See Section \ref{sec:illustr} for details.
\label{fig:illustr}
}
\end{figure*}

\begin{figure*}
\includegraphics[width=\textwidth]{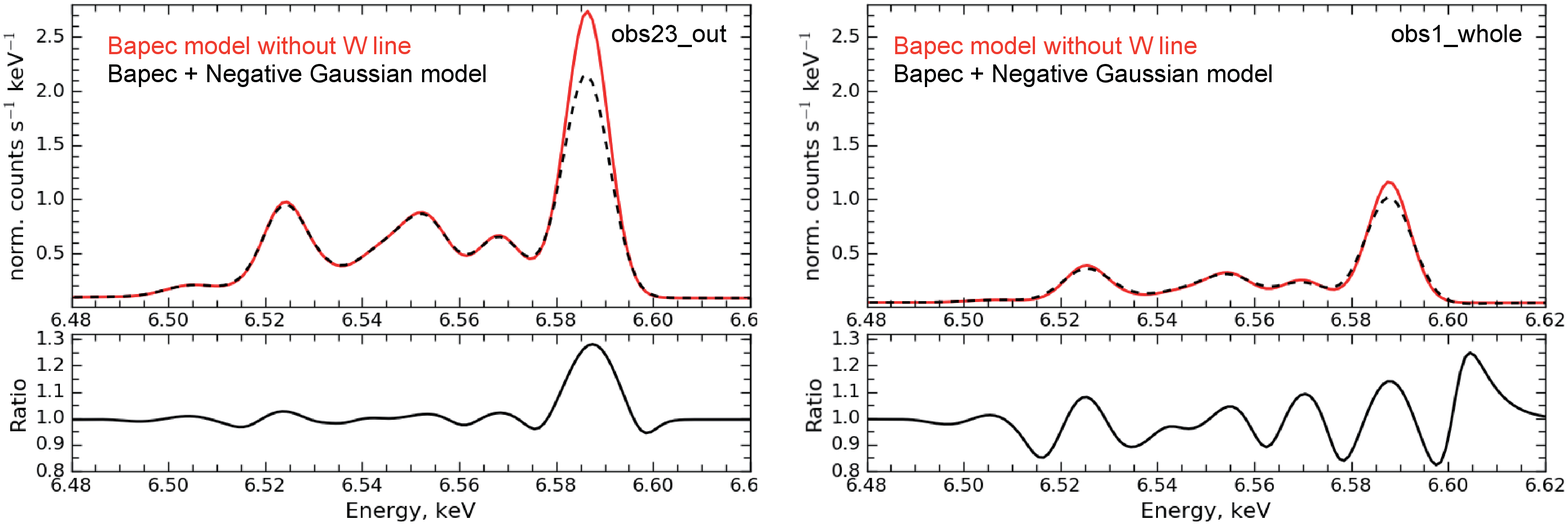}
\caption{The same as the bottom right panel in figure \ref{fig:illustr}, 
but for the spectra observed in obs23\_out (left) and obs1\_whole (right) 
regions. See Section \ref{sec:illustr} for details.
\label{fig:illustr1}
}
\end{figure*}

Theoretical studies of the RS effect in the Perseus Cluster predict 
that, in the absence of gas motions the degree of flux suppression 
in the resonance line of He-like Fe should vary with the projected 
distance from the cluster center:
the line flux will be most suppressed in the innermost region, with 
the suppression decreasing with projected distance out to a radius 
$\sim 100$ kpc.
At larger radii, the line flux is slightly increased relative to the 
value for the optically thin case \citep[e.g.][]{Churazov04}. 
Also, as the result of scattering, the wings of
the line become slightly stronger \citep[see e.g.][]{Zhu13}. 
Below we demonstrate that the Perseus Hitomi data show evidence 
for both of these effects.

\subsection{Flux suppression in the  Fe\emissiontype{XXV} He$\alpha$ 
resonance line}
\label{sec:illustr}

We first consider the spectrum of the He-like Fe\emissiontype{XXV} 
triplet from the innermost region (obs23\_cen in figure \ref{fig:image}), 
where the suppression of the resonance ($w$) line is expected to be 
the strongest. A single-temperature 
{\it bapec}\footnote{The {\it bapec} model describes a plasma in
collisional ionization equilibrium, with arbitrary velocity broadening 
in addition to thermal Doppler broadening, and element abundance ratios 
relative to He fixed to the Solar ratios.}
model for an optically thin plasma can approximately model 
the resonance line flux, but will then underestimate the fluxes of 
the neighboring forbidden ($z$) and intercombination ($y$) lines
(see bottom left panel in figure \ref{fig:illustr} and supplementary 
material in \citet{Hitomi16}). Exclusion of the resonance line from the
modeling provides a better fit for $x,~y$ and $z$ and other weaker lines, 
but clearly overestimates the $w$ flux (top left panel in
figure \ref{fig:illustr}). We then add a Gaussian component to the model 
with the energy of the $w$ line and normalization that is allowed to be 
negative. The best-fitting result of this model is shown in the top right 
panel in figure \ref{fig:illustr}.
The best-fitting normalization of the Gaussian component is indeed negative 
and the model provides a statistically better fit to the data than a pure 
{\it bapec} model\footnote{For all three modeling steps we use 
the same gas temperature, which is taken from the best-fitting model 
to the data without the $w$ line.}. The ratio of
the best-fitting models shown in the top panels shows a suppression of 
the resonance line by a factor of $\sim 1.28$, indicating the presence 
of RS.

The same modeling procedure is applied to spectra from the regions 
at larger distances from the cluster center (obs23\_out and obs1\_whole, 
see figure \ref{fig:image}). When fitting the obs23\_out spectrum, 
the {\it bapec+negative Gaussian} model provides a statistically better 
fit than the pure {\it bapec} model. The $w$ flux in the obs23\_out region 
is suppressed by factor of $\sim 1.28$ (left panel in 
figure \ref{fig:illustr1}). In the most distant from the cluster center 
region, obs1\_whole, the {\it bapec+negative Gaussian} model does not 
provide a statistically better fit of the data than a {\it bapec} model. 
The measured line suppression is small, less than 1.15 (right panel 
in figure \ref{fig:illustr1}). These simple experiments illustrate 
the possible presence of the RS effect in the Perseus core.

\subsection{The broadening of the Fe\emissiontype{XXV} He$\alpha$ resonance 
line}

In addition to flux suppression in the resonance line, the RS broadens 
the wings of the line. Even though the effect is significantly smaller 
than the line suppression,
we have checked for indications of line broadening in the $w$ line 
compared to other lines in the triplet.
We fit the observed data excluding the $w$ line with a single-temperature 
{\it bapec} model, from which the $w$ line has been removed. 
Freezing the best-fitting parameters of this model, we fit the whole 
triplet, with the $w$ line restored, adding an additional Gaussian 
component with the central energy of the $w$ line. Such modeling allows 
us to measure the broadening of the $w$ line independently from the 
broadening of other lines in the triplet. Accounting for statistical 
uncertainties, the turbulent broadening of the $w$ line varies 
between 171--183 km s$^{-1}$, while the broadening of the $x,~y$ and 
$z$ lines are smaller, 145--165 km s$^{-1}$. 
A similar difference is observed in the obs23\_out region. 
Namely, the $w$ line turbulent broadening in this region is 
159--167 km s$^{-1}$, while in all other lines it is 136--150 km s$^{-1}$.

\section{Observed line ratios}
\label{sec:spectralfits}

\begin{table*}[t]
  \caption{Summary of the best-fit properties of temperatures, Fe abundance, turbulent velocity ($\sigma_{\rm v}$), C-statistics, line ratios, and line widths ($\sigma_{\rm v+th}$). }
\label{tab:ratios}
\begin{center}
\begin{tabular}{lrrrrrrr} \hline
Region ID & $kT^{\ast}$ & Fe$^{\ast}$ & $\sigma_{\rm v}^{\ast}$ & C-stat/d.o.f$^{\ast}$ & C-stat/d.o.f$^{\dagger}$ & & \\
  & keV & solar & km s$^{-1}$ & 1.8--20 keV & 6.1--7.9 keV & & \\\hline
obs23\_cen  & $3.92 \pm 0.03$ & $0.65 \pm 0.01$ & $155 \pm 7$ & 10609/11151 & 1793/1784  &  & \\
obs23\_out  & $4.05 \pm 0.01$ & $0.65 \pm 0.01$ & $141 \pm 5$ & 14559/11744 & 1964/1784  &  & \\
obs1\_whole & $5.06 \pm 0.07$ & $0.53 \pm 0.02$ & $159 \pm 17$ & 6333/6930  & 1283/1494  & & \\
\hline \hline
Region ID & $w/z^{\dagger}$ & $w$/He$\beta^{\dagger}$ & $w$/Ly$\alpha_1^{\dagger}$ & $w$/Ly$\alpha_2^{\dagger}$ & $z$/He$\beta^{\dagger}$ & $z$/Ly$\alpha_1^{\dagger}$ & $z$/Ly$\alpha_2^{\dagger}$ \\ 
Line ratio & & & & & & & \\ 
\hline
obs23\_cen  & $2.45 \pm 0.11$ & $5.98 \pm 0.57$ & $9.79 \pm 0.98$ & $18.17 \pm 2.76$ & $2.45 \pm 0.25$ & $4.00 \pm 0.42$ & $7.42 \pm 0.36$ \\
obs23\_out  & $2.59 \pm 0.08$ & $6.23 \pm 0.57$ & $9.36 \pm 0.57$ & $15.41 \pm 1.24$ & $2.40 \pm 0.23$ & $3.61 \pm 0.24$ & $5.95 \pm 0.20$ \\
obs1\_whole & $3.27 \pm 0.34$ & $6.35 \pm 0.95$ & $6.87 \pm 1.11$ & $9.80 \pm 1.96$  & $1.94 \pm 0.33$ & $2.10 \pm 0.38$ & $3.00 \pm 0.35$ \\
\hline
Region ID & $w^{\dagger}$ & $z^{\dagger}$ & Ly$\alpha^{\dagger}$ & He$\beta^{\dagger}$ & & &  \\ 
Line width ($\sigma_{\rm v+th}$) &  eV & eV & eV & eV & & & \\ 
\hline
obs23\_cen  & $4.49 \pm 0.11$ & $3.57 \pm 0.21$ & $5.29 \pm 0.55$ & $3.45 \pm 0.50$ & & &  \\
obs23\_out  & $4.20 \pm 0.08$ & $3.54 \pm 0.15$ & $3.46 \pm 0.25$ & $4.24 \pm 0.42$ & & &  \\
obs1\_whole & $4.43 \pm 0.24$ & $3.58 \pm 0.50$ & $6.09 \pm 0.89$ & $4.81 \pm 0.80$ & & &  \\
\hline
\hline\\[-1ex]
\multicolumn{8}{l}{\parbox{0.9\textwidth}{\footnotesize
\footnotemark[$\ast$]
Fits in the broad 1.8--20.0 keV band with the AGN and modified {\it bvvapec} models, from which the resonance line is excluded and a Gaussian component is added instead. $\sigma_{\rm v}$ is a turbulent velocity in {\it bvvapec} model without the resonance line. The numbers in this table are slightly smaller than those in the ``V'' paper \citep{velocity}, see section \ref{sec:spectralfits} for the details. 
}}\\
\multicolumn{8}{l}{\parbox{0.9\textwidth}{\footnotesize
\footnotemark[$\dagger$]
Fits in the narrow, 6.1--7.9 keV, band with the AGN and modified {\it bvvapec} models, from which we exclude the He-$\alpha$ resonance and forbidden, He-$\beta$1\&2, and Ly-$\alpha$1\&2 lines.
}}\\
\end{tabular}
\end{center}
\end{table*}

\begin{figure*}[!th]
\begin{minipage}{0.33\textwidth}
\FigureFile(\textwidth,\textwidth){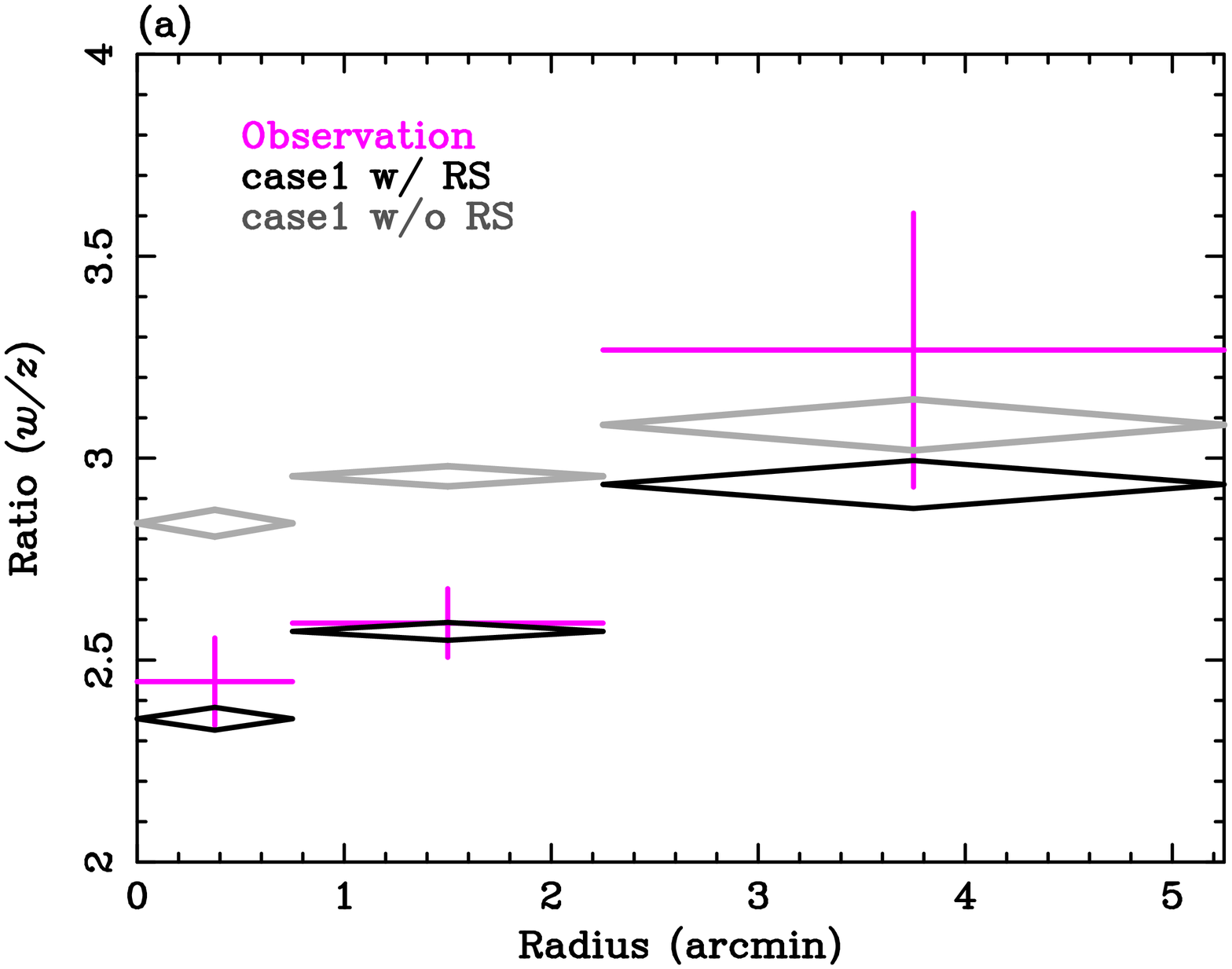}
\end{minipage}\hfill
\begin{minipage}{0.33\textwidth}
\FigureFile(\textwidth,\textwidth){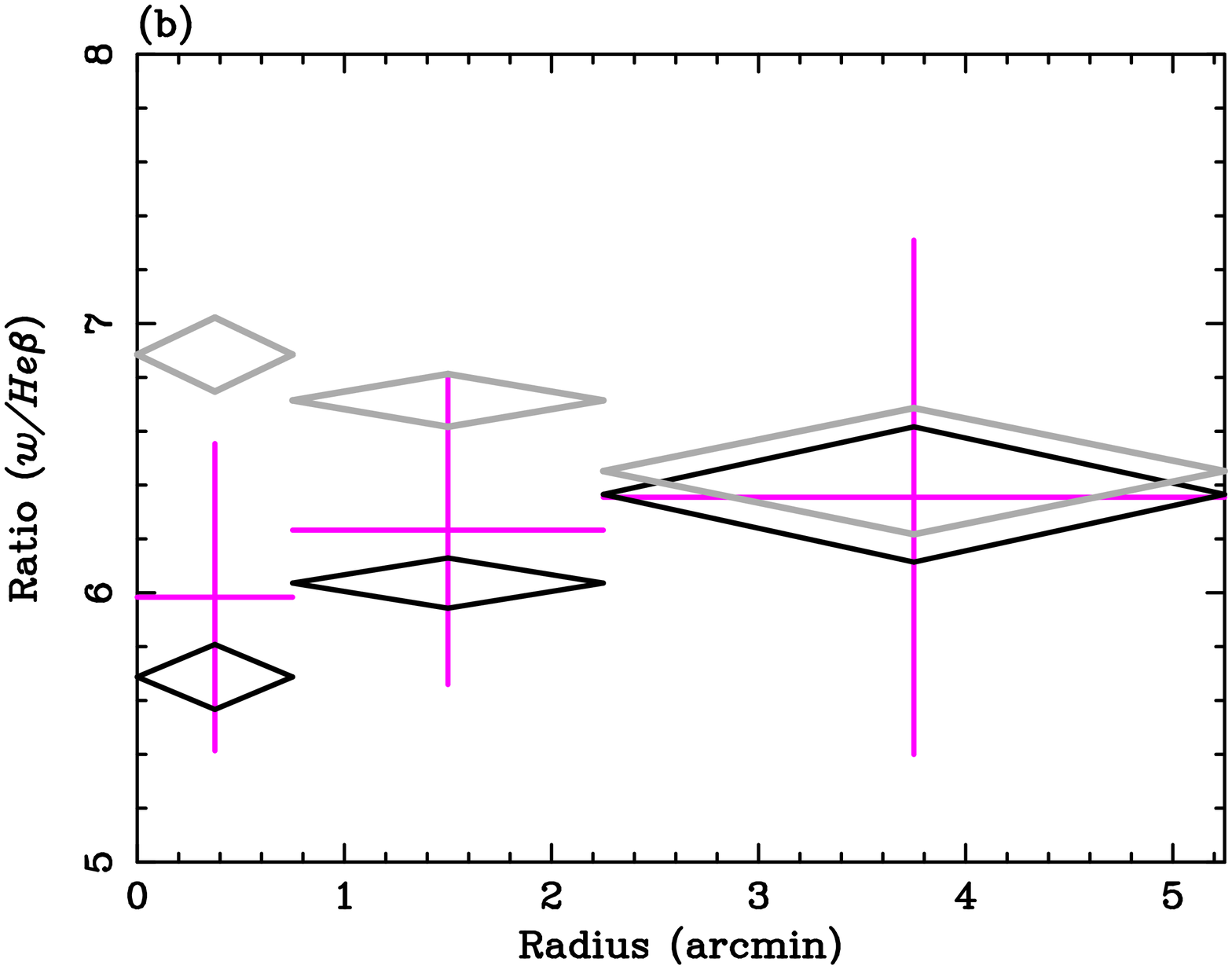}
\end{minipage}\hfill
\begin{minipage}{0.33\textwidth}
\FigureFile(\textwidth,\textwidth){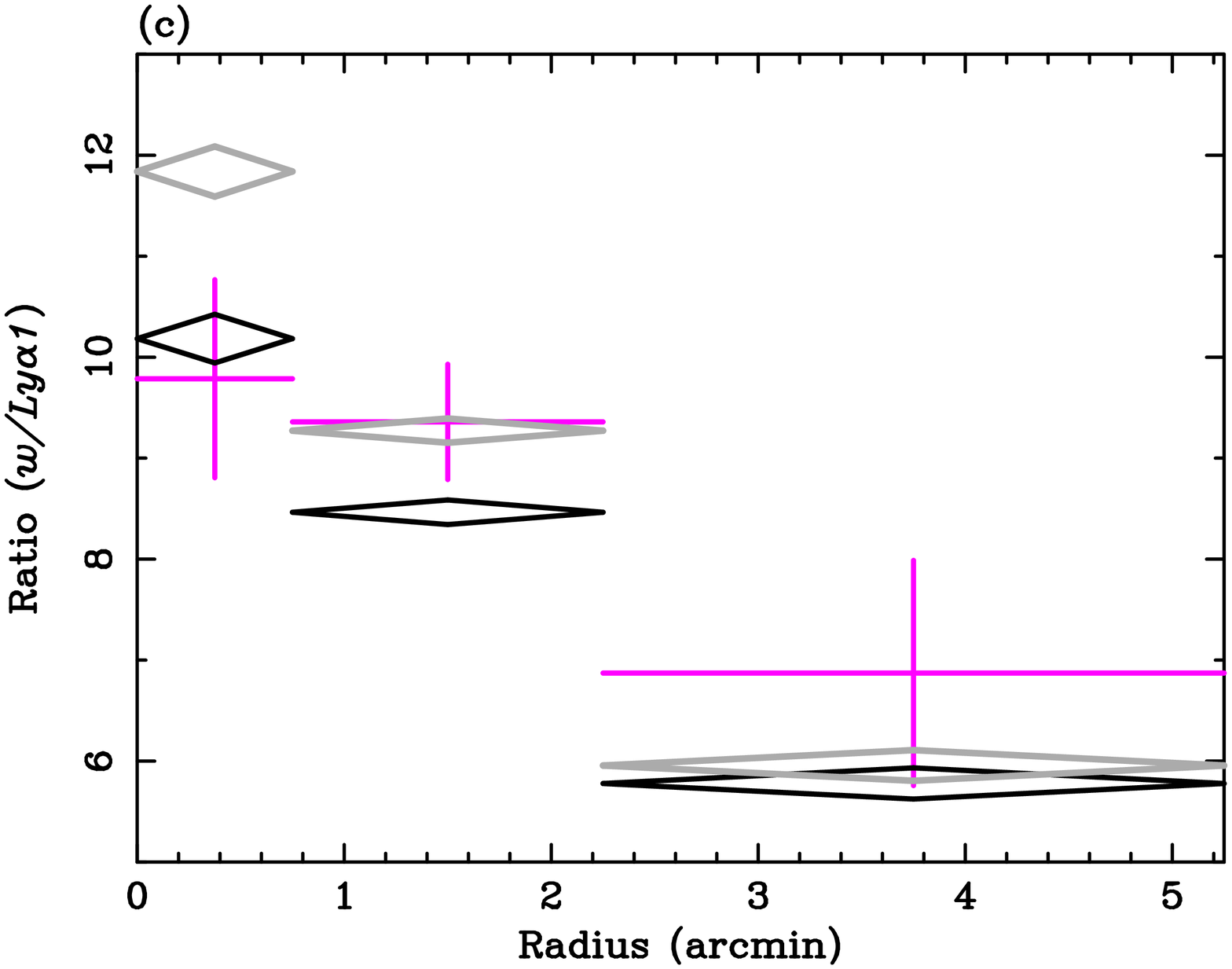}
\end{minipage}

\begin{minipage}{0.33\textwidth}
\FigureFile(\textwidth,\textwidth){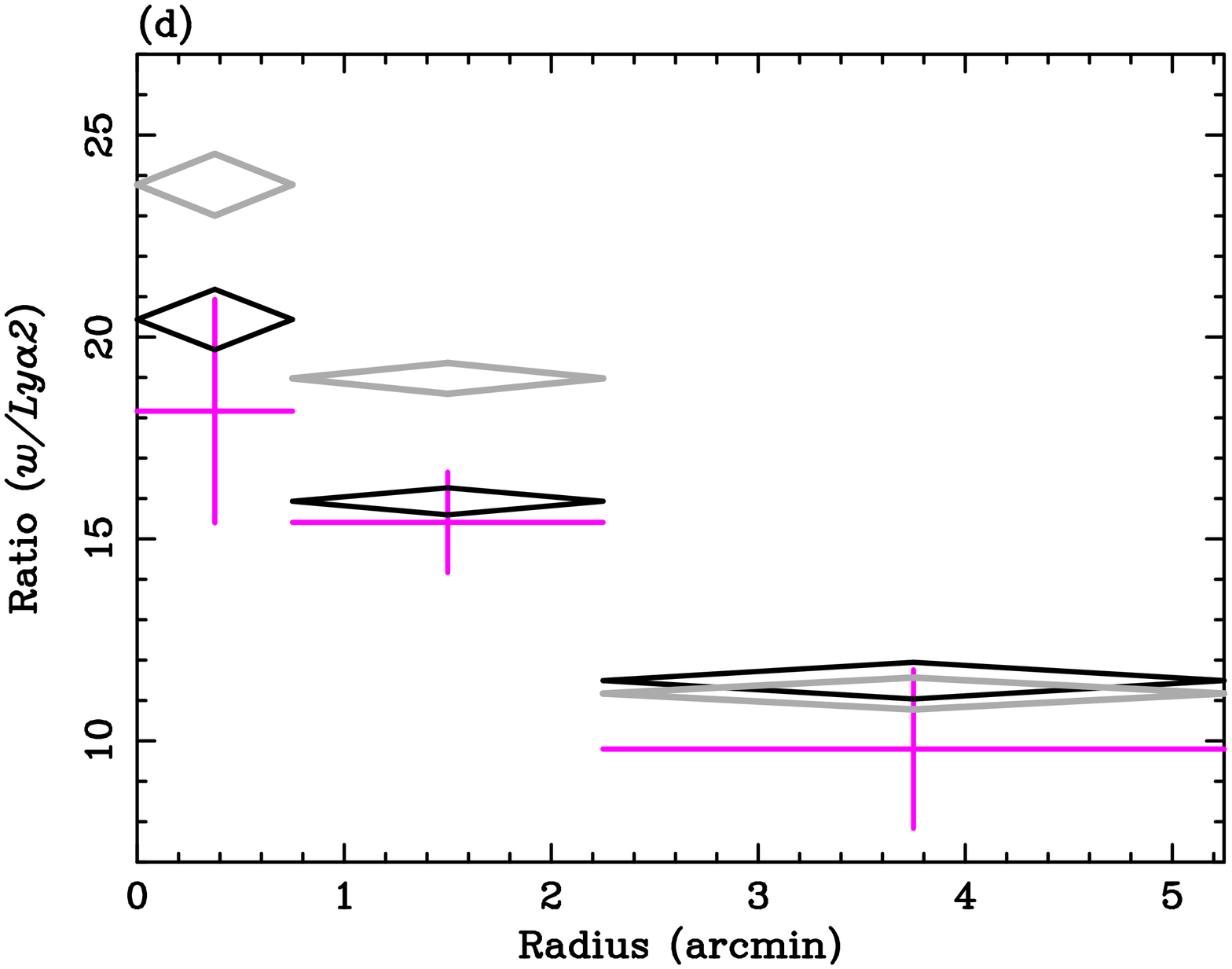}
\end{minipage}
\begin{minipage}{0.33\textwidth}
\FigureFile(\textwidth,\textwidth){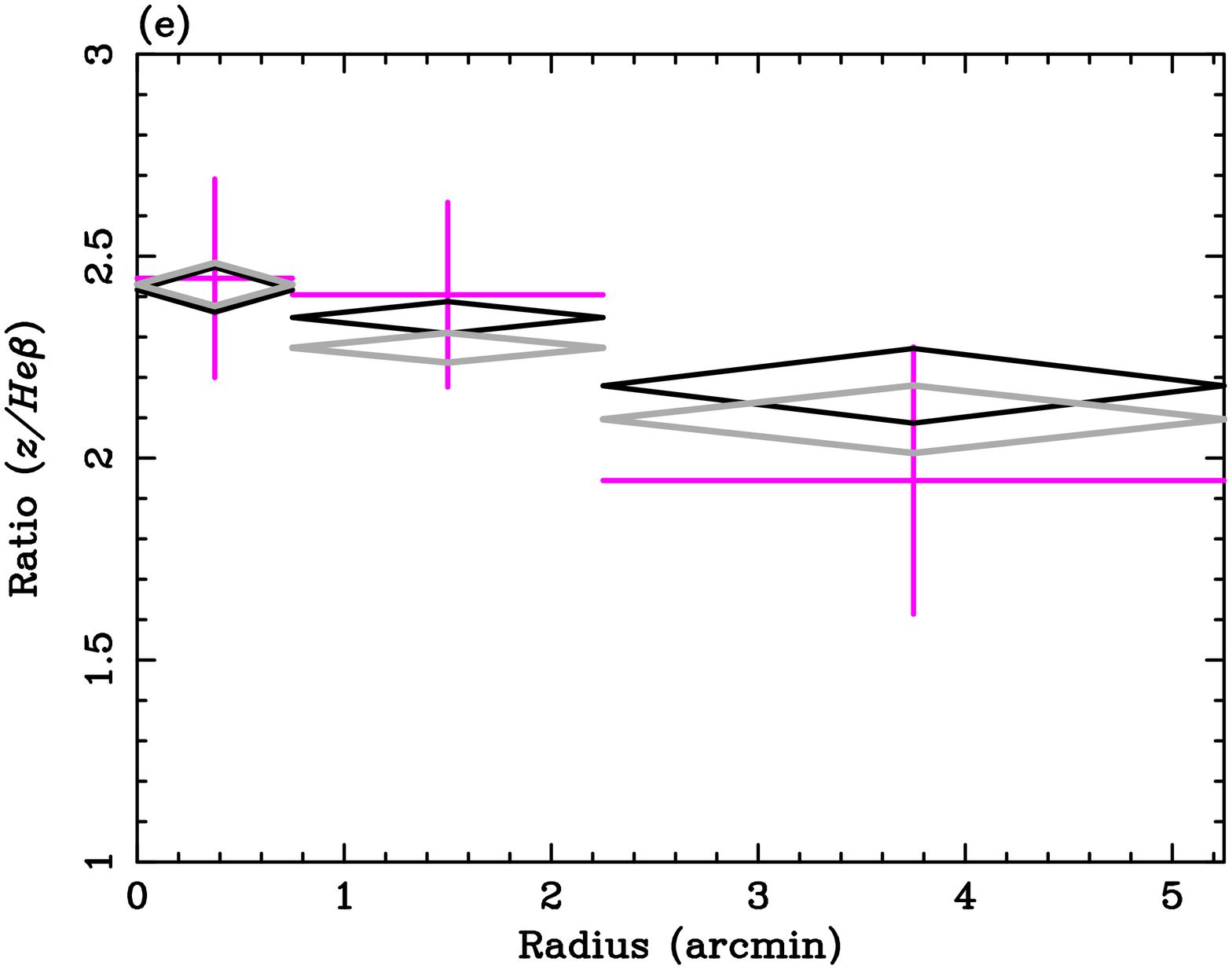}
\end{minipage}\hfill
\begin{minipage}{0.33\textwidth}
\end{minipage}

\begin{minipage}{0.33\textwidth}
\FigureFile(\textwidth,\textwidth){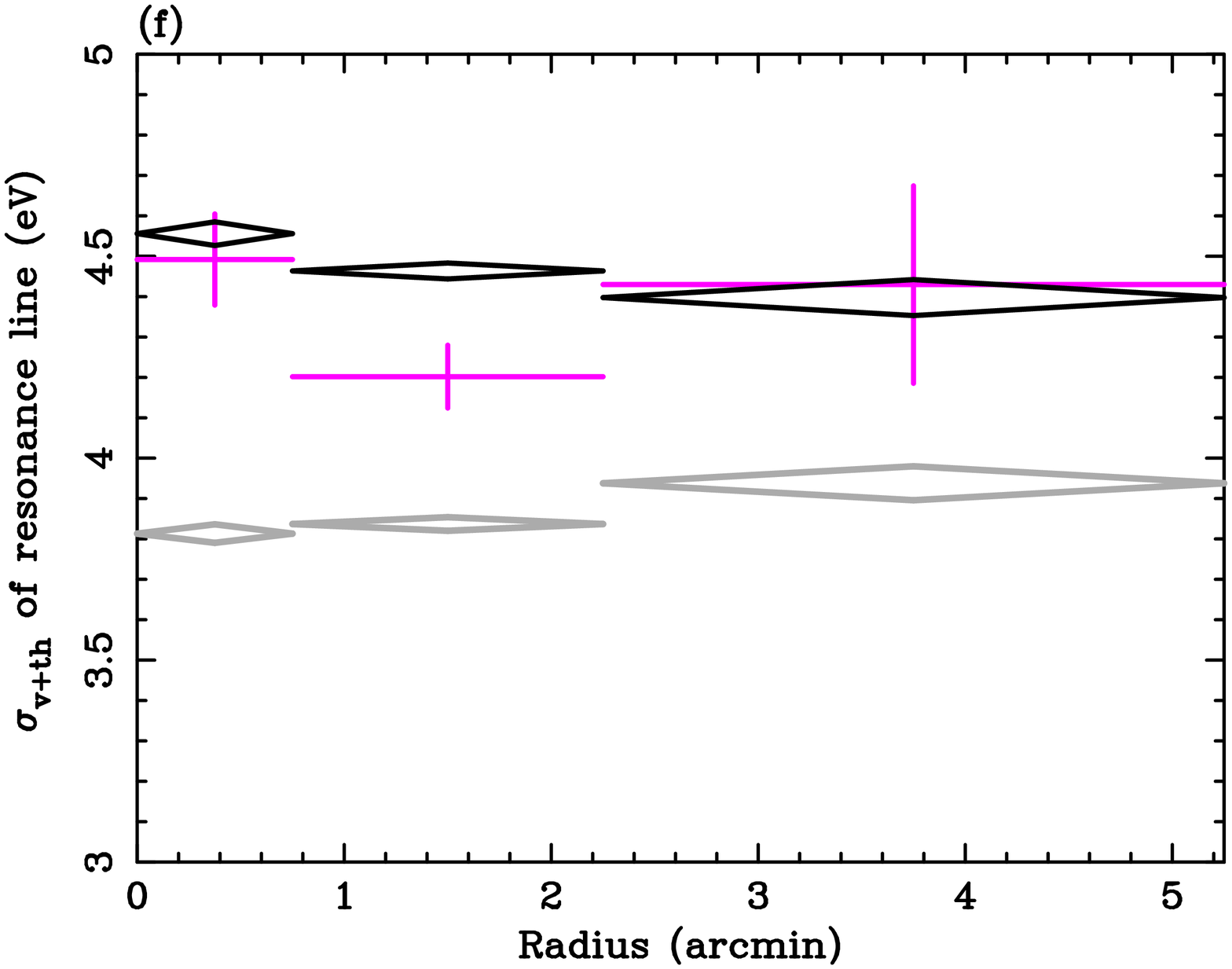}
\end{minipage}\hfill
\begin{minipage}{0.33\textwidth}
\FigureFile(\textwidth,\textwidth){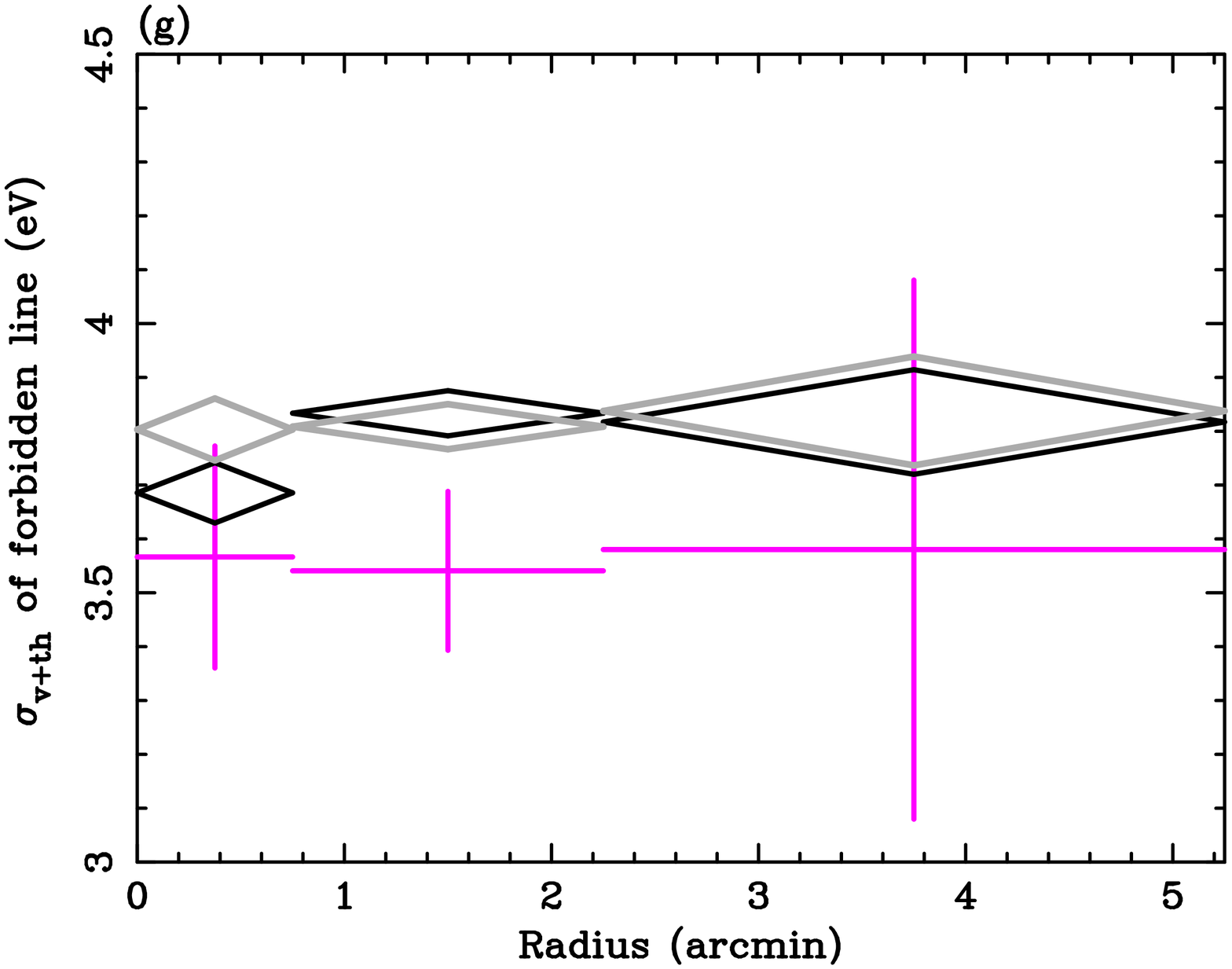}
\end{minipage}\hfill
\begin{minipage}{0.33\textwidth}
\FigureFile(\textwidth,\textwidth){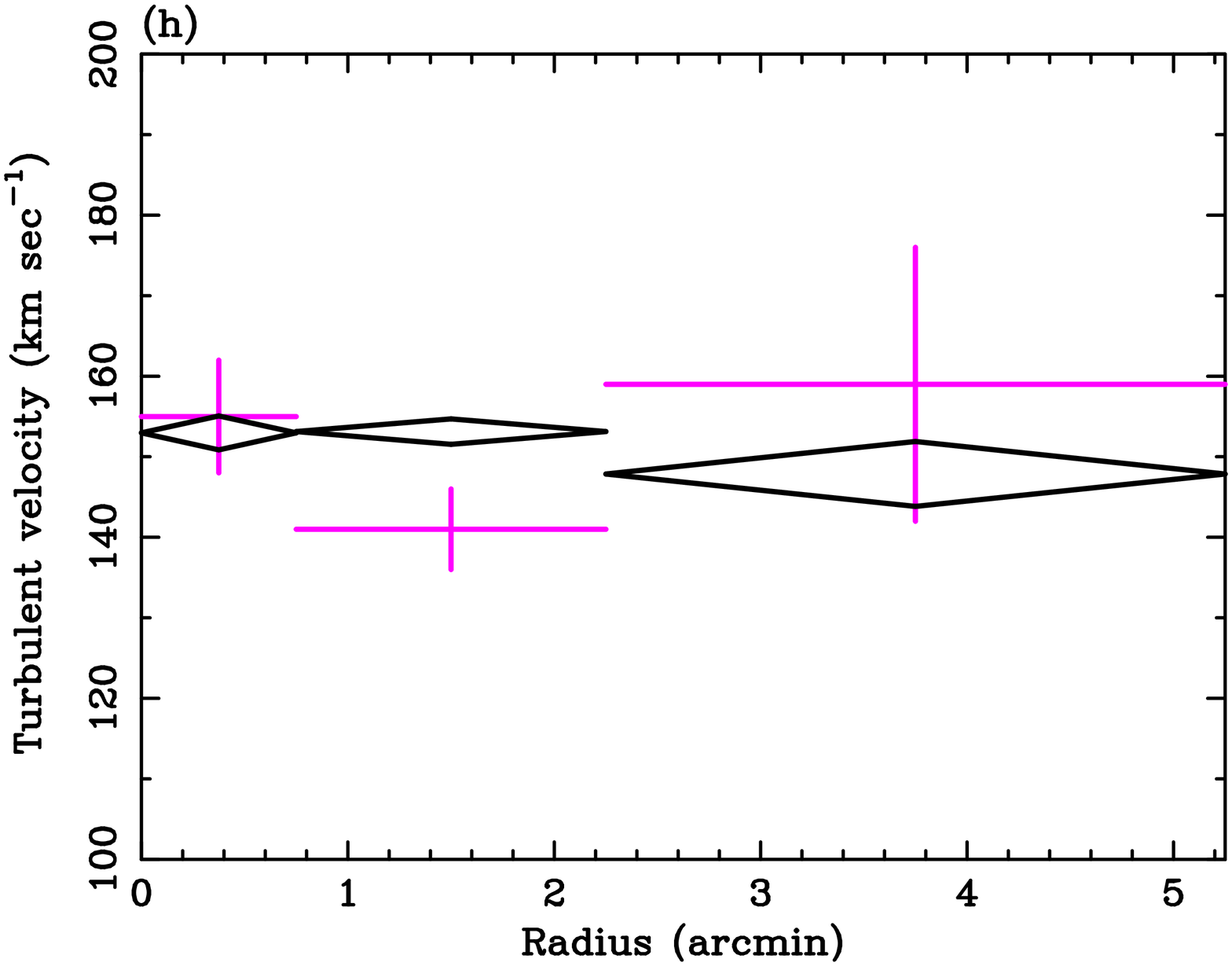}
\end{minipage}

\vspace*{2ex}
\caption{
(a)--(e): Comparisons of the observed and predicted 
ratios of the Fe He$\alpha$ resonance (w), He$\alpha$ forbidden (z), 
He$\beta$, and Ly$\alpha_{1,~2}$ lines. 
Observations are shown as magenta crosses and the simulations
with RS as black diamonds and the same without RS as gray diamonds for 
the assumption of the constant $\sigma_{\rm v}$ of 150 km s$^{-1}$ 
(case 1).
(f)--(g): Comparisons of the widths of the resonance (w) and 
forbidden (z) lines between observation (magenta crosses) and 
simulations with RS (black diamonds) and without RS (gray diamonds). 
(h): Comparisons of the derived turbulent 
velocities from the spectral fits between observation (magenta 
cross) in broad band fits and the simulations.
}
\label{fig:comp-results}
\end{figure*}

\begin{figure*}[h]
\begin{center}
\begin{minipage}{0.25\textwidth}
\FigureFile(\textwidth,\textwidth){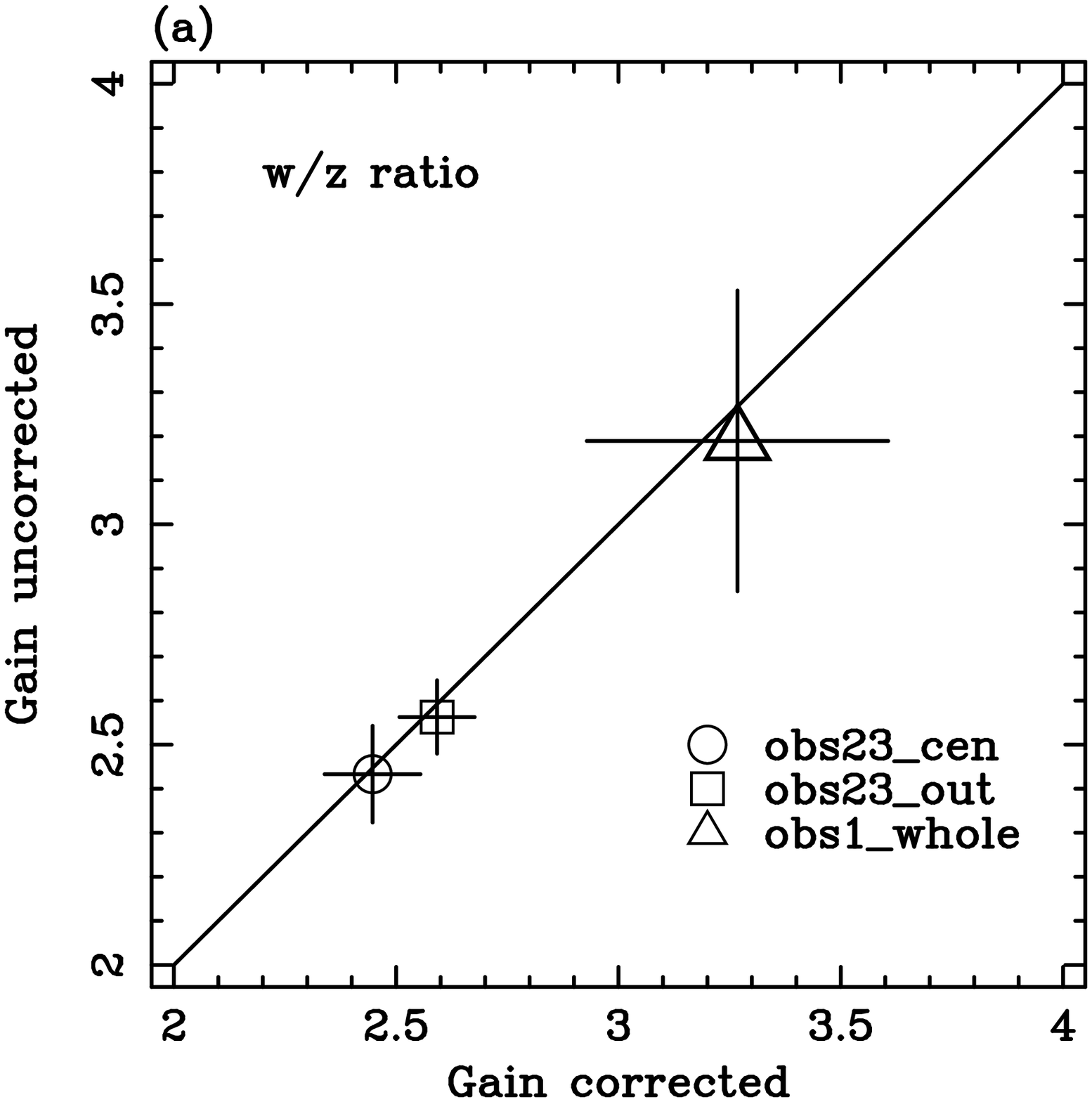}
\end{minipage}\hfill
\begin{minipage}{0.25\textwidth}
\FigureFile(\textwidth,\textwidth){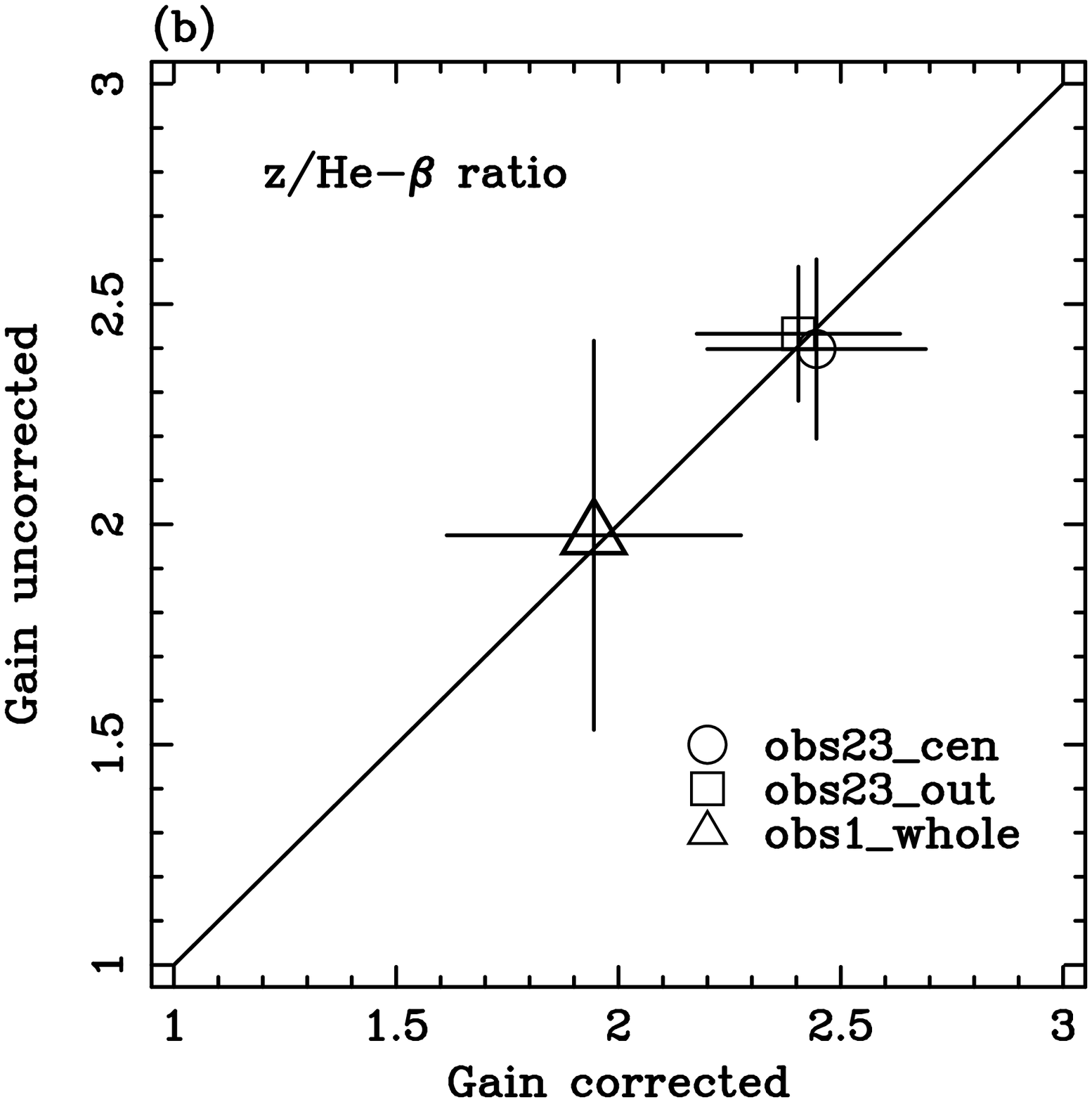}
\end{minipage}\hfill
\begin{minipage}{0.25\textwidth}
\FigureFile(\textwidth,\textwidth){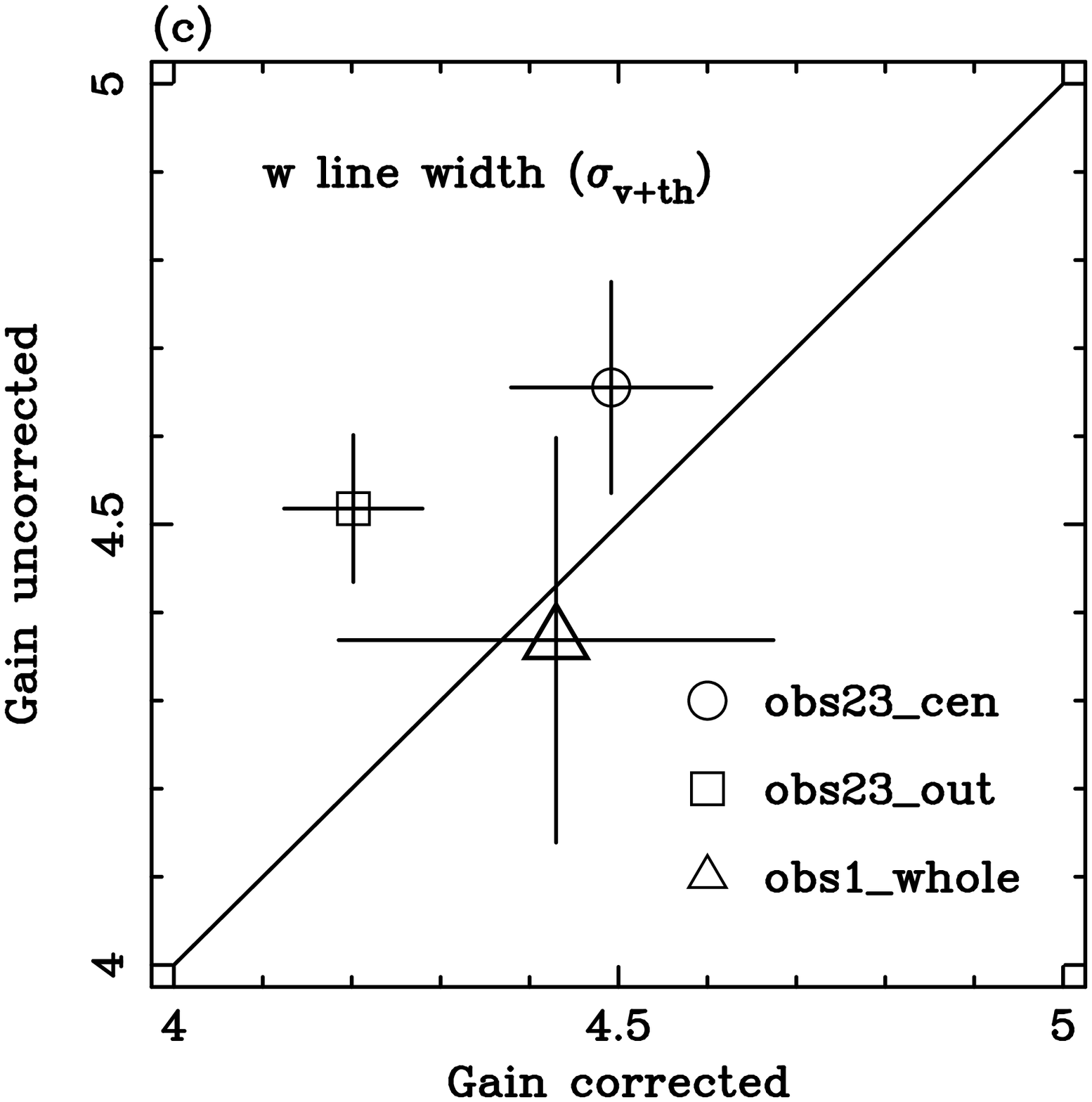}
\end{minipage}\hfill
\begin{minipage}{0.25\textwidth}
\FigureFile(\textwidth,\textwidth){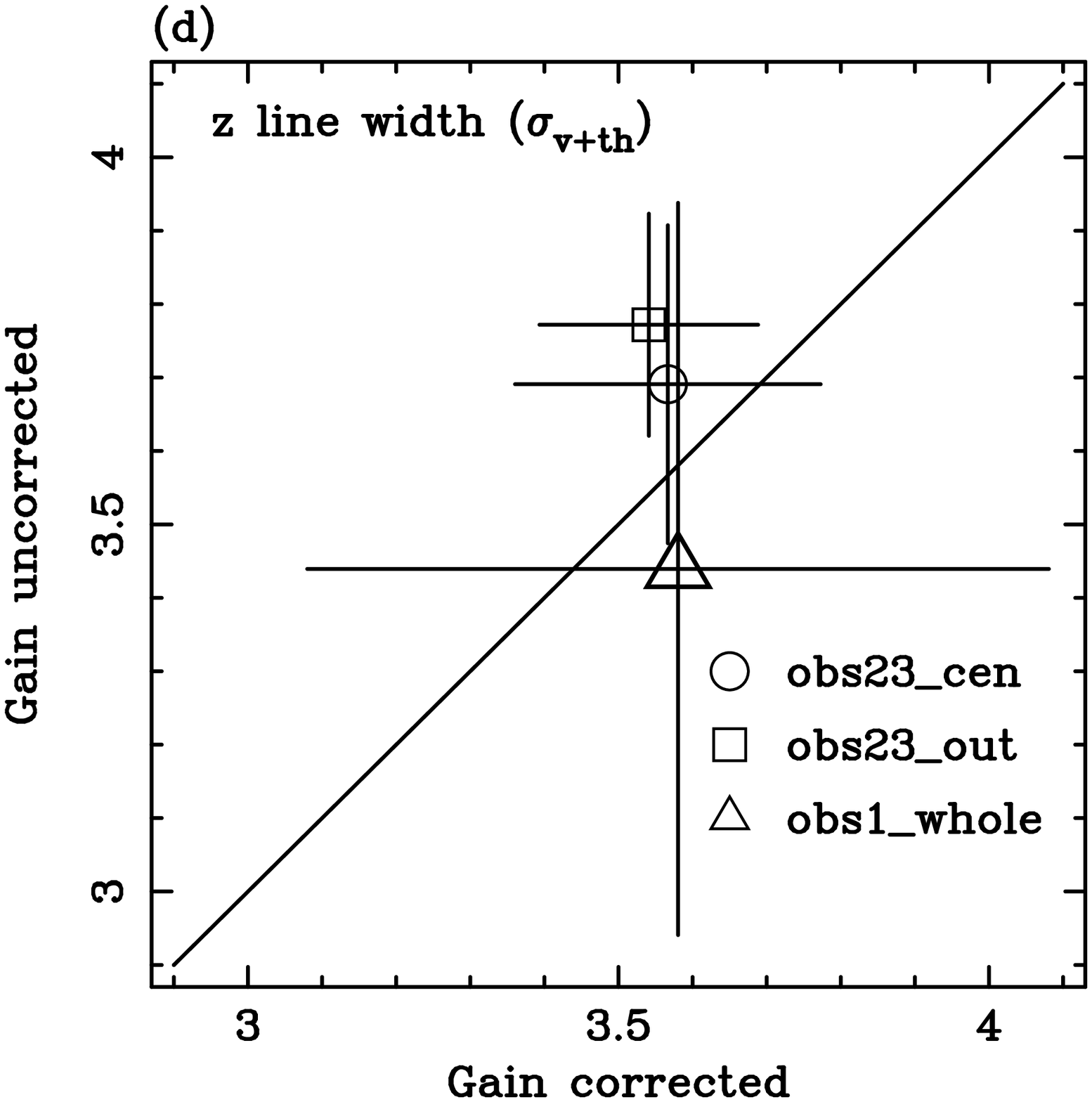}
\end{minipage}
\end{center}
\caption{ 
Scattering plots between the gain corrected and uncorrected data for (a) 
the w/z line ratio, (b) the z/He$\beta$ line ratio, 
(c) w line width ($\sigma_{\rm v+th}$) and (d) z line width 
($\sigma_{\rm v+th}$).
Open circles, squares, and triangles correspond to the measurements in the obs23\_cen, obs23\_out, and obs1\_whole regions, respectively.
}
\label{fig:scatplot}
\end{figure*}

We fitted the spectra with a combination of emission models representing 
the AGN and the ICM, each with its own response, in Xspec. The AGN is 
represented by a power-law with redshifted absorption, with additional 
(redshifted) Fe K$\alpha_{1,~2}$ fluorescent emission lines; the ICM is 
modeled with a redshifted collisional ionization equilibrium plasma, with 
adjustable elemental abundances, and additional Gaussian emission lines 
if necessary. The two components share a common foreground neutral Galactic 
absorption. Formally, we have
AGN model: $TBabs_{\rm GAL} \times (pegpwrlw_{\rm AGN} 
+ zgauss_{\rm AGN,~Fe K\alpha_1}+zgauss_{\rm AGN,~Fe~K\alpha_2})$, and 
ICM model: $TBabs_{\rm GAL} \times ( bvvapec_{\rm ICM}+zgauss_{\rm Fe~lines})$.
The AGN parameters are fixed at the numbers in the NGC~1275 paper 
\citep{NGC1275}.
In this paper, we modify the {\it bvvapec} model, 
setting the emissivities of the strong Fe lines to zero.

Firstly, we derived the ICM temperature, Fe abundance, 
turbulent velocity, and normalizations from the spectral fits 
in 1.8--20.0 keV\@ with a single temperature model for each region. 
In the broad band fit, to determine those parameters, 
we adopted the modified {\it bvvapec} model, from which the 
Fe\emissiontype{XXV} He$\alpha$ resonance line is excluded, and the 
corresponding Gaussian model is added. The resultant parameters and 
C-statistics from the spectral fits for each region are shown 
in table \ref{tab:ratios}. The ``projected'' temperature increases 
slightly with radius, while the Fe abundance drops by $\sim0.1$ solar 
from the center to the obs1\_whole region. 
The measured temperature and Fe abundance gradients 
agree with the results from the ``T'' and ``Z'' 
papers \citep{tz, abundance}.
As for the turbulent velocity, $\sigma_{\rm v}$, the derived values 
are almost constant with radius.
These $\sigma_{\rm v}$ in the previous Hitomi paper 
\citep{Hitomi16} and ``V'' paper \citep{velocity} are slightly different.
We note that the data reduction, calibrations and plasma codes 
are more improved than the previous Hitomi paper. 
And, the numbers shown in table 4 for PSF 
uncorrected in the ``V'' paper from their narrow band fits in 6.4--6.7 keV 
are slightly smaller than those from the broadband fits shown 
in table \ref{tab:ratios} in this paper. 
The difference comes from the broader line width of the Ly$\alpha$ lines 
(see table \ref{tab:ratios} and the ``V'' paper).

Fixing the ICM temperature, Fe abundance, and $\sigma_{\rm v}$ 
at the values from the broad band fits, we exclude the 
Fe\emissiontype{XXV} He$\alpha$ forbidden ($z$) and resonance ($w$), 
the He$\beta$ and the Fe\emissiontype{XXVI} Ly$\alpha$ lines from the 
{\it bvvapec} model and include Gaussian line models instead with 
the central energies of these lines. 
The best-fitting normalizations of the Gaussian components give total fluxes 
of these lines. Here, the line widths of the Ly$\alpha_2$ and He$\beta_2$ 
are linked to Ly$\alpha_1$ and He$\beta_1$ lines, respectively, and 
other parameters except for the redshift are varied in the spectral fits.
This fitting model is very useful since {\it bvvapec} describes 
the weak satellite lines, while the added Gaussian lines allow us 
to measure the fluxes of the strongest emission lines 
in a model independent way, taking blending with weaker emission 
lines into account. 

The observed spectra are well-described by the model, 
except around the 6.55 keV feature,
as shown in figure \ref{fig:spectra-all}. 
The resulting line ratios and widths (thermal and 
turbulent broadenings, $\sigma_{\rm v + th}$) are summarized 
in table \ref{tab:ratios} and figure \ref{fig:comp-results}. 
The Ly$\alpha_1$ and Ly$\alpha_2$ lines are clearly resolved in 
obs23\_cen and obs23\_out regions, while the He$\beta$ lines are not, 
due to their close central energies.
Note that the emission lines are represented well by the corresponding 
Gaussian models, as confirmed by the study of 
possible non-Gaussianity in the ``V'' paper.
The derived radial profile of the $w/z$ ratio increases with the distance 
from the center, while the $z/He\beta$ ratio is almost the same everywhere.
The measured $w$ line widths in the obs23\_cen and 
obs23\_out regions are broader than the $z$ ones 
at the $\sim2~\sigma$ level. 
The comparison of the measured line ratios and line broadening 
with the results of numerical simulations of the RS effect 
is discussed in section \ref{sec:rs-sim}.

Systematic uncertainties, such as (a) the ICM modeling of a single or 
two temperature structure, (b) gain correction, (c) the point spread 
function (PSF) deblending, and (d) plasma codes (AtomDB version 3.0.8 or 3.0.9) 
should be considered in the spectral analysis. 
Estimates of their effects are examined below.
As a result, these uncertainties almost do not affect our results and conclusions.

As for the ICM modeling, the Fe lines in 6--8 keV are well modeled 
with a single temperature model 
with the exception of the resonance ($w$) line shown in 
figure \ref{fig:spectra-all} and table \ref{tab:ratios}.
On the other hand, as described in the ``T'' paper, 
a two temperature (2T) model improves the spectral fits 
when AtomDB version is 3.0.9. 
The $w/z$ ratios measured from the 1T and 2T models agree within 
the statistical error with either AtomDB version 3.0.8 or 3.0.9\@.
In this paper, since we examined the spectral fits for the observations 
and simulations in the same manner, i.e., the same model formula,
 as described in section \ref{subsec:sim_tus} and compare the resultant 
fit parameters for each other, the choice of the 1T or 2T models does 
not affect our conclusions, as long as the continuum spectra are 
well-represented by the models.
Note that the gas temperature measured from the line ratios obtained 
from the 2T model 
in the ``T'' paper agree well with the deprojected temperature profile 
from the Chandra data.

We repeated the spectral analysis using
gain-uncorrected Hitomi data to estimate the uncertainty.
Figure \ref{fig:scatplot} shows comparison plots 
of the resultant fits for the $w/z$ and $z/He\beta$ line ratios and 
line widths, $\sigma_{\rm v + th}$, of the $w$ and $z$ lines 
between the gain corrected and uncorrected data. 
The line ratios from both data sets are consistent within the statistical 
errors. At the same time, the width of the $w$ line in obs23\_cen decreases, 
as expected, by about 5\% when the gain correction is applied.
We did not correct the spectral fit for the PSF effects. 
The azimuth-averaged values in regions 1--4 for the PSF uncorrected numbers 
in the ``V'' paper which roughly correspond to the obs23\_out region 
are almost consistent with our results within statistical errors.
The PSF effect is accounted for in the simulations in this paper described 
in section \ref{sec:rs-sim}.
The residuals around 6.55 keV in the obs23\_out region 
are likely associated with uncertainties in the plasma model for 
the Fe\emissiontype{XXIV} Li-like line 
(see also figure 8 in ``Atomic'' paper). These come from the underestimation 
of the Li-like lines in AtomDB version 3.0.8. The updated version 3.0.9 
are corrected for the problem as shown in Appendix \ref{sec:v309}.
This feature has negligible impact on our results for line ratios and widths, 
however, since the Li-like lines are separated from the $w$ and $z$ lines.

\section{Radiative transfer simulations}
\label{sec:rs-sim}

The line suppression due to the RS effect is sensitive to the velocity 
of gas motions: the larger the velocity of gas motions the lower 
the probability of scattering and the closer the 
line ratios to those for
an optically thin plasma. In order to interpret 
the observed line suppression and infer the velocity of gas motions, 
we performed radiative transfer Monte Carlo (MC) 
simulations of the RS in 
the Perseus Cluster. We followed two independent approaches: 
(i) using the {\tt Geant4} and HEAsim tools and assuming 
a velocity field consistent with 
the direct velocity measurements as presented in the ``V'' paper 
\citep{velocity}, and 
(ii) using a proprietary code written specifically for MC 
simulations of radiative transfer in the cluster ICM (ICMMC).
Both approaches are based on the emission models for an optically thin 
plasma taken from AtomDB version 3.0.8,
and take into account projection 
effects (gas density, temperature, abundance of heavy elements) and the 
spatial response of the telescope. The latter is treated 
differently in both approaches. The results based on both simulations 
broadly agree. Details of the two approaches are discussed below.

\subsection{Model of the Perseus Cluster}
\label{subsec:input-model}

\begin{figure}
 \begin{center}
  \includegraphics[width=8cm]{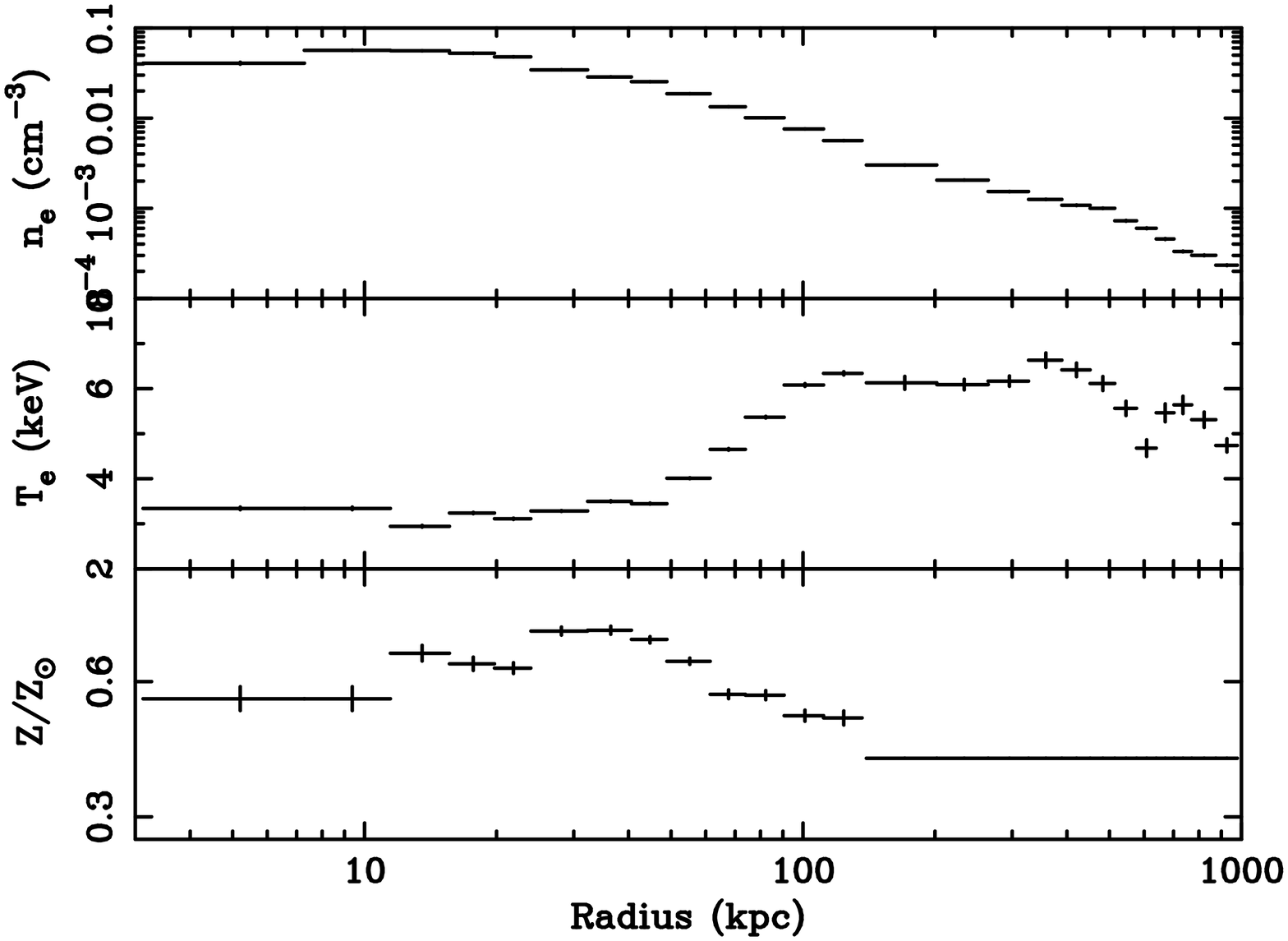}
 \end{center}
\caption{
Model of the Perseus Cluster used for the Monte Carlo simulations 
of radiative transfer in strong emission lines. Top: deprojected 
electron number density; middle: deprojected gas electron temperature; 
bottom: deprojected abundance of heavy elements relative to Solar abundance 
from \citet{Lodders09}. Chandra data are used in the inner 
$\sim 150$ kpc region. These profiles of the temperature 
and electron density
are merged with Suzaku deprojected data at large radii, $r>\sim 150$ kpc, taken 
from \citet{Urban14}, and the abundance is adopted to be the averaged number in \citet{Werner13,Matsushita13,Urban14}.
}
\label{fig:input}
\end{figure}

For the MC simulations, we adopt a spherically symmetric model of the 
Perseus Cluster. We used archival Chandra data to measure 
the profiles of gas density, 
temperature and abundance of heavy elements. Excluding point sources 
and the central AGN, projected spectra are obtained in radial annuli,
centered on the central galaxy, NGC~1275.
These are deprojected following the procedure described 
by \citet{Churazov2003}. The spectra are fitted with an {\it apec}
model in a broad energy band, 0.5--8.5 keV\@, accounting for Galactic 
foreground absorption by a column density of 
$N_{\rm H} = 1.38 \times 10^{21}$ cm$^{-2}$, and treating the abundance 
of heavy elements as a free parameter, using
the solar abundance table by \citet{Lodders09}. The Chandra 
deprojected profile within $\sim 150$ kpc is shown in figure 
\ref{fig:input}. There is a density drop in the innermost region 
(the first point from the center) likely associated with the bubbles of 
relativistic plasma that push up the X-ray gas. Due to this density drop 
and the presence of multi-temperature plasma, the deprojected temperature 
and the heavy element abundances are not 
determined reliably in this region. 
Therefore, we assume constant temperature and abundance profiles 
in the inner $\sim 10$ kpc region. The Chandra deprojected profile 
is then merged with the Suzaku deprojected data \citep{Urban14} 
at large radii, $r>150$ kpc. 
As for the abundances in $r=150$--1000 kpc, since the observed abundances 
($\sim$0.3 solar) from Suzaku in \citet{Urban14, Werner13} are relatively 
smaller than those ($\sim$0.5 solar) from XMM in \citet{Matsushita13}, 
we adopted the averaged number of $\sim$0.4 solar as the input parameter.
Figure \ref{fig:input} shows the combined radial profiles.

\subsection{Optical depth}
\label{subsec:optical_depth}

\begin{figure*}[tb]
\begin{minipage}{0.45\textwidth}
\includegraphics[width=\textwidth]{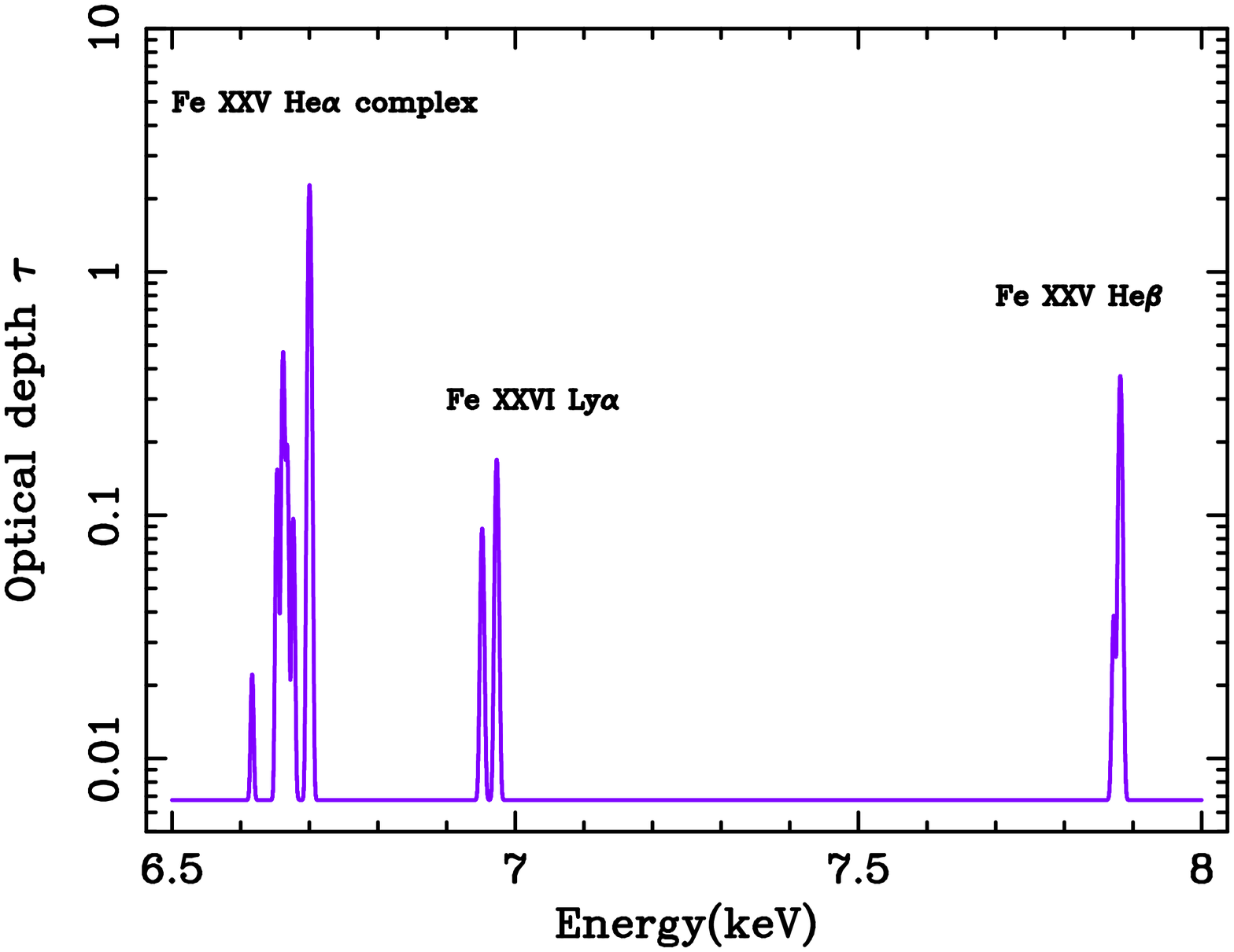}
\end{minipage}
\hfill
\begin{minipage}{0.45\textwidth}
\includegraphics[width=\textwidth]{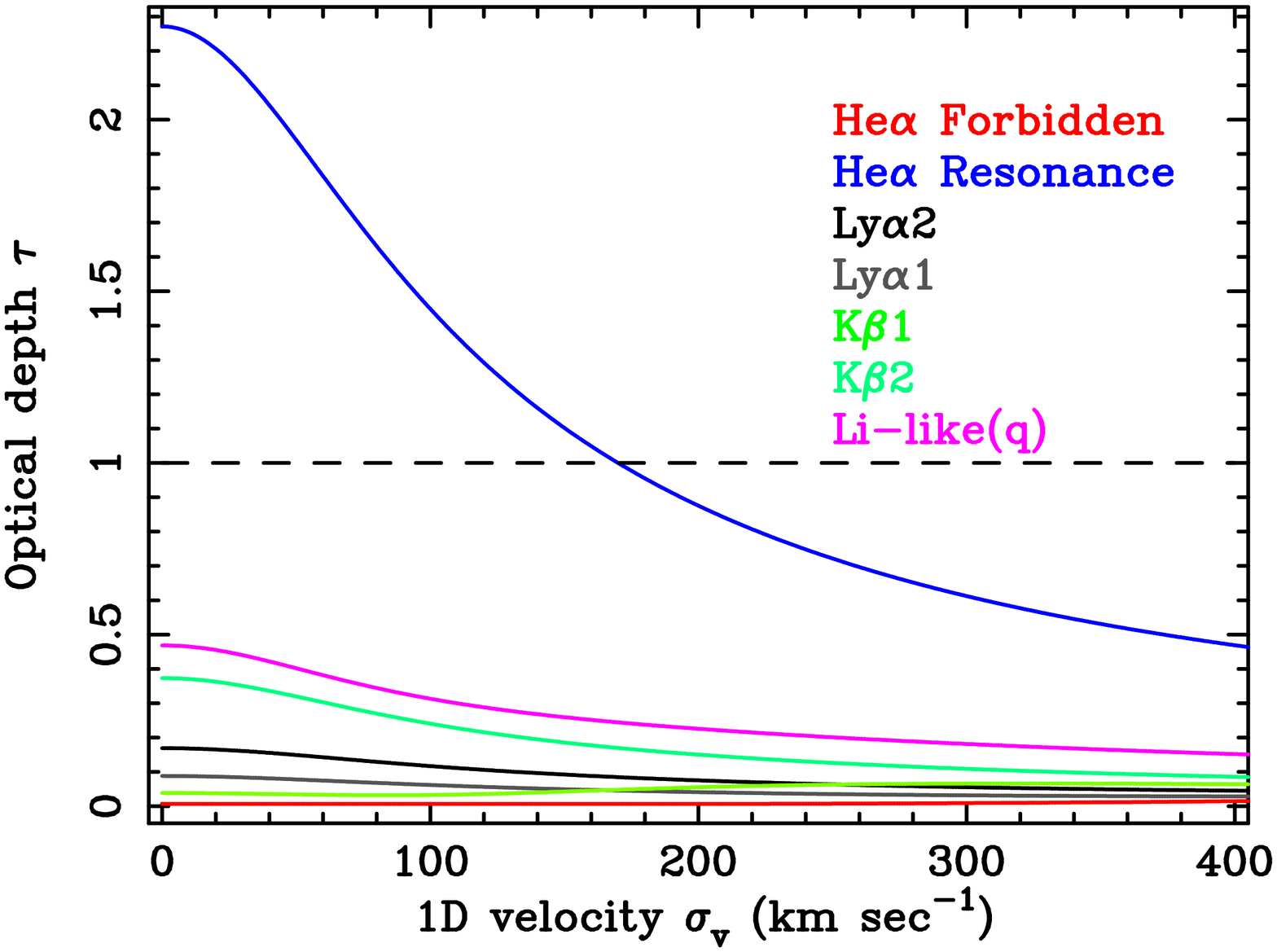}
\end{minipage}
\vspace*{2ex}
\caption{
Left: Optical depth in lines and continuum as a function of photon energy 
calculated assuming zero turbulent velocity and integrated over a $r=0-40'$ 
region (see also table \ref{tb:felines}). 
Right: Optical depth profile of the Fe\emissiontype{XXIV,~XXV,~XXVI} lines 
versus the velocity of gas motions in units of km s$^{-1}$.
}
\label{fig:opticaldepth}
\end{figure*}

\begin{table*}[tb]
\caption{Rest frame Fe line properties in the 6--8 keV band that 
have optical depth $>\sim0.01$. Optical depths are integrated over a 
$r=0-40'$ region with $\sigma_{\rm v}=0$ km s$^{-1}$. 
Energies and oscillator strengths are from AtomDB version 3.0.8.}
\label{tb:felines}
\begin{center}
  \begin{tabular}{l c c c c c l} \hline \hline
    \makebox[40pt] [l]{Ion} & \makebox[40pt] {Energy} & \makebox[40pt] {Lower Level$^\ast$} & \makebox[40pt] {Upper Level$^\ast$} & \makebox[70pt] {Oscillator strength} & \makebox[70pt] {Optical depth $\tau$} & \makebox[70pt] [l]{Comments$^\ast$}\\
    & (eV)&  &  & & $\sigma_{\rm v}=0$ km s$^{-1}$ & \\ \hline
    Fe\emissiontype{XXIV}& 6616.73 & $1s^2 2s_{1/2}\ ^2S_{1/2}$ & $1s_{1/2}2s_{1/2} 2p_{1/2}\ ^4P_{3/2}$  & 3.26$\times 10^{-2}$ & 2.22$\times10^{-2}$ & $u$ \\
    Fe\emissiontype{XXV} & 6636.58 & $1s^2\ ^1S_0$             & $1s2s\ ^3S_1$                        & 3.03$\times 10^{-7}$ & 6.75$\times 10^{-3}$ & He$\alpha$, $z$ \\
    Fe\emissiontype{XXIV}& 6653.30 & $1s^2 2s_{1/2}\ ^2S_{1/2}$ & $1s_{1/2}2s_{1/2} 2p_{1/2}\ ^2P_{1/2}$  & 3.13$\times 10^{-1}$ & 1.54$\times 10^{-2}$ & $r$ \\
    Fe\emissiontype{XXIV}& 6661.88 & $1s^2 2s_{1/2}\ ^2S_{1/2}$ & $1s_{1/2}2s_{1/2} 2p_{3/2}\ ^2P_{3/2}$  & 9.78$\times 10^{-1}$ & 4.69$\times 10^{-1}$ & $q$ \\
    Fe\emissiontype{XXV} & 6667.55 & $1s^2\ ^1S_0$             & $1s_{1/2}2p_{1/2}\ ^3P_{1}$            & 5.79$\times 10^{-2}$ & 1.92$\times 10^{-1}$ & He$\alpha$, $y$ \\
    Fe\emissiontype{XXIV}& 6676.59 & $1s^2 2s_{1/2}\ ^2S_{1/2}$ & $1s_{1/2}2s_{1/2} 2p_{3/2}\ ^2P_{1/2}$  & 1.92$\times 10^{-1}$ & 9.67$\times 10^{-2}$ & $t$ \\
    Fe\emissiontype{XXV} & 6682.30 &  $1s^2\ ^1S_0$            &  $1s_{1/2}2p_{3/2}\ ^3P_{2}$           & 1.70$\times 10^{-5}$ & 7.26$\times 10^{-3}$ & He$\alpha$, $x$ \\
    Fe\emissiontype{XXV} & 6700.40 &  $1s^2\ ^1S_0$            &  $1s_{1/2}2p_{3/2}\ ^1P_{1}$           & 7.19$\times 10^{-1}$ & 2.27 & He$\alpha$, $w$\\
    Fe\emissiontype{XXVI}& 6951.86 &  $1s$                     & $2p_{1/2}$                           & 1.36$\times 10^{-1}$ & 8.81$\times 10^{-2}$ & Ly$\alpha_2$\\
    Fe\emissiontype{XXVI}& 6973.07 &  $1s$                     & $2p_{3/2}$                           & 2.73$\times 10^{-1}$ & 1.69$\times 10^{-1}$ & Ly$\alpha_1$\\
    Fe\emissiontype{XXV} & 7872.01 & $1s^2\ ^1S_0$             & $1s 3p\ ^3P_1$                       & 1.18$\times 10^{-2}$ & 3.87$\times 10^{-2}$ & He$\beta_2$, intercomb.\\
    Fe\emissiontype{XXV} & 7881.52 & $1s^2\ ^1S_0$             & $1s 3p\ ^1P_1$                       & 1.37$\times 10^{-1}$ & 3.73$\times 10^{-1}$ & He$\beta_1$, resonance\\
\hline \hline\\[-1ex]
\multicolumn{7}{l}{\parbox{0.9\textwidth}{\footnotesize
\footnotemark[$\ast$]
Letter designations for the transitions as per \citet{Gabriel72}
}}\\
  \end{tabular}
\end{center}
\end{table*}

Using the equations shown in \citet{Zhu13}, the optical depth 
is calculated from the center of the cluster out to a radius corresponding 
to an angular size on the sky of $40'\sim 830$ kpc,
corresponding to 2/3 times $r_{500}$ \citep{Urban14}. 
The left panel of figure \ref{fig:opticaldepth} 
shows the optical depth for each line (see also table \ref{tb:felines}) 
for the case of zero $\sigma_{\rm v}$
calculated using the cluster model described 
in section \ref{subsec:input-model}.
RS is expected to be important in the central regions of 
the Perseus Cluster, where the optical depth is larger than 1. 
The Fe\emissiontype{XXV} He$\alpha$ $w$ has the largest optical depth 
$\sim 2.3$, while the Fe\emissiontype{XXV} He$\alpha$ $z$ line is 
essentially optically thin and not affected by the RS.

The optical depth is inversely proportional to the Doppler line width, 
which depends on the thermal broadening and turbulent gas motions. 
Therefore, the stronger the turbulence, the smaller the optical depth
(see figure \ref{fig:opticaldepth}, right panel). However, even if 
the gas is moving with a characteristic velocity as large as 
$\sim$150--200 km s$^{-1}$, 
as measured directly through the line broadening, 
we still expect RS to affect
the $w$ line (the optical depth is $\sim 1$). 
All other lines considered in this work are effectively optically thin.

\subsection{Monte Carlo simulations with {\tt Geant4}}
\label{subsec:sim_tus}

\begin{table*}[t]
\caption{Assumed velocity field of the one-component velocity 
($\sigma_{\rm v}$) in our simulation with {\tt Geant4}.}
\label{tb:radial_turbulent}
\begin{center}
\begin{tabular}{lrrrrrr} \hline
case ID & \multicolumn{6}{c}{$\sigma_{\rm v,~1D}$ (km s$^{-1}$)} \\\hline
 & $r<0.'5$ & $0.'5 < r < 1'$ & $1'<r<2'$ & $2'<r<5'$ & $r>5'$ & Temp. model$^{\ast}$ \\\hline
case 1  & 150 & 150 & 150 & 150 & 150  & nominal \\
case 1a & 150 & 150 & 150 & 150 & 150  & nominal$+10$\% \\
case 2  & 200 & 200 & 150 & 150 & 100  & nominal \\
case 3  & 200 & 200 & 150 & 100 & 100  & nominal \\
case 4  & 200 & 150 & 150 & 150 & 150  & nominal \\
case 5  & 200 & 200 & 150 & 100 & 300  & nominal \\
\hline\\[-1ex]
\multicolumn{7}{l}{\parbox{0.6\textwidth}{\footnotesize
\footnotemark[$\ast$]
Assumed ``nominal'' temperature model as shown in figure \ref{fig:input}. 
We estimate the temperature uncertainties changing the temperature 
by $+10$\% which is corresponding to the azimuthal dependence of 
the temperature profile from Chandra and XMM. 
}}\\\end{tabular}
\end{center}
\end{table*}

\begin{figure*}[tb]
\begin{minipage}{0.45\textwidth}
\includegraphics[width=\textwidth]{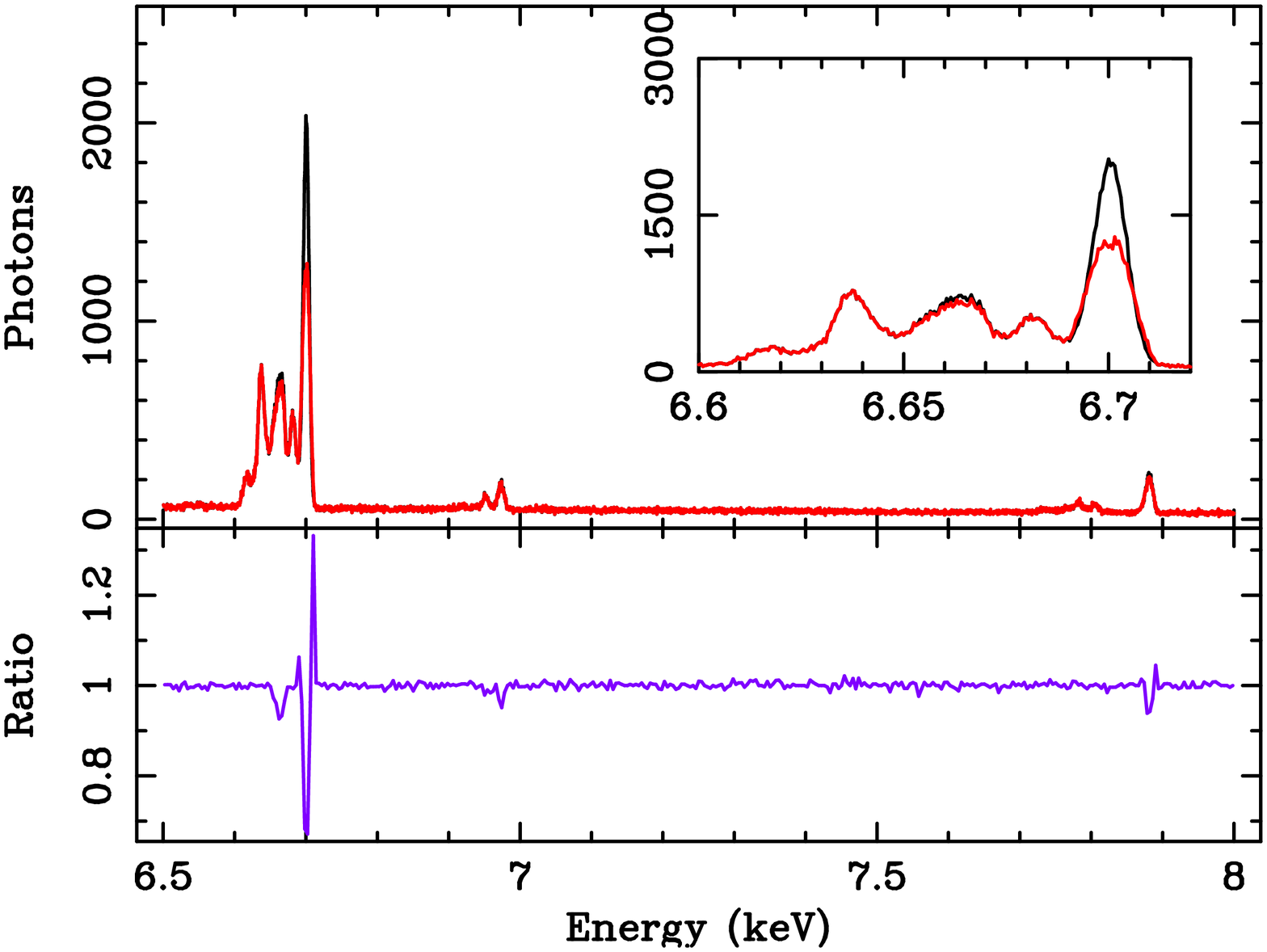}
\end{minipage}
\hfill
\begin{minipage}{0.45\textwidth}
\includegraphics[width=\textwidth]{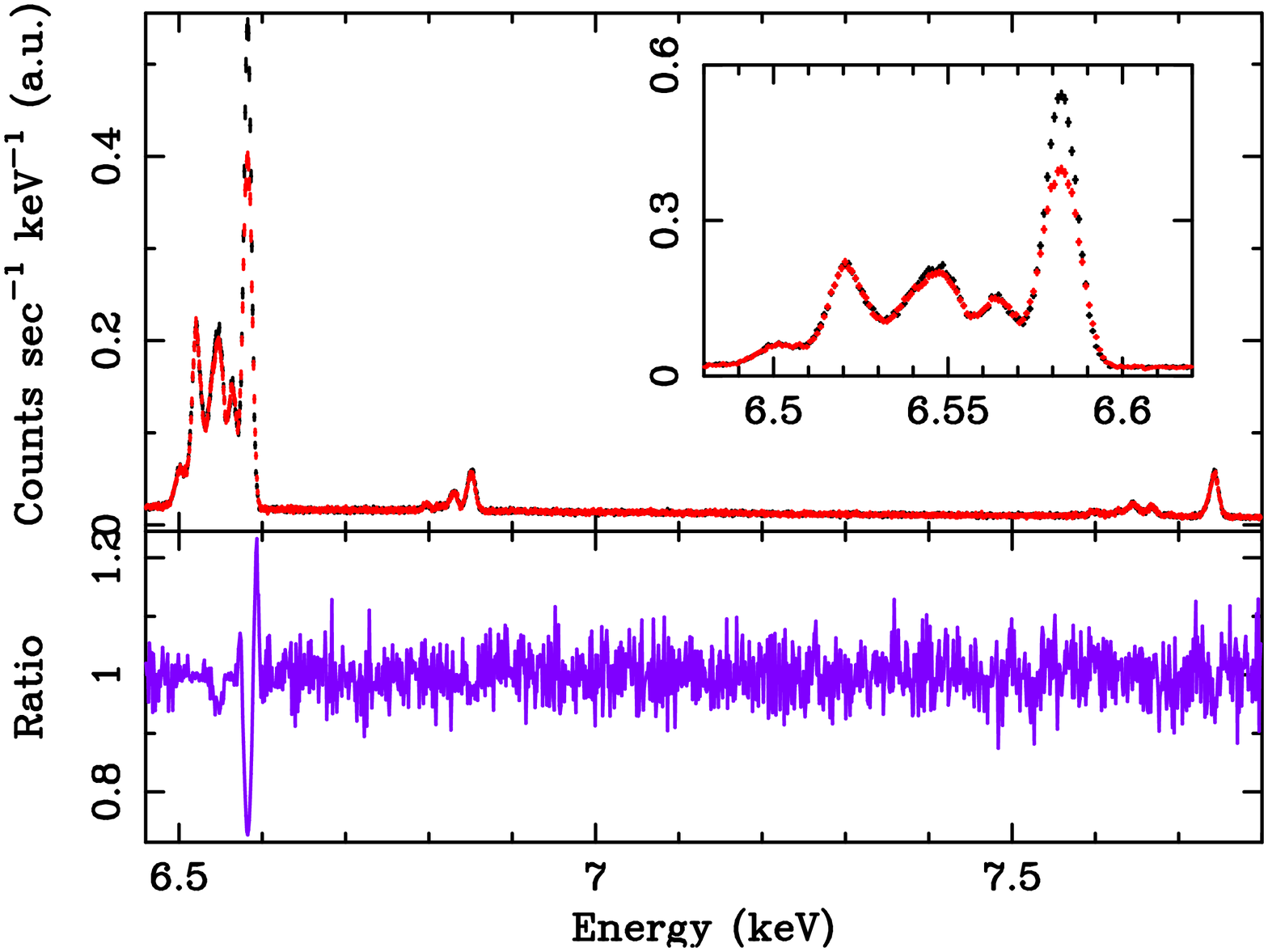}
\end{minipage}
\vspace*{2ex}
\caption{Left: Photon lists generated by the {\tt Geant4} simulator 
assuming a uniform $\sigma_{\rm v}$ profile of 150 km s$^{-1}$ (case 1)
for the inner, $0.'5$ (radius), region in Perseus (top panel). 
Red and black lines correspond to simulations with and without RS, 
respectively. Right: mock spectra for the obs23\_cen region 
with the HEAsim tool 
with the photon lists generated by the {\tt Geant4} simulator. 
Small panels show a zoom-in
around the Fe\emissiontype{XXV} He$\alpha$ complex. 
The suppression of the $w$ line in the simulated spectra is
consistent with the previous results by \citet{Zhu13}.
}
\label{fig:mock} 
\end{figure*}

The RS simulation was performed with the main reaction processes shown 
in \citet{Zhu13}, using the input Perseus model shown 
in figure \ref{fig:input}.
Assuming spherical symmetry, we calculated multiple scatterings of photons 
in the Perseus core; 
the {\tt Geant4} tool kit produces a list of simulated photons incident 
on the Hitomi SXS. 
In the {\tt Geant4} frame work, we assume 400 spherical shells in a 
$r=$0--40$'$ region, and scaled to be 1 kpc$=$1 cm to preserve 
the scattering probability under the low density environment in the ICM.
The seed photons in the simulator are generated 
according to the thermal emissivity associated with our adopted 
spatial distributions of density, temperature, and abundance, and we 
assume the photons are emitted isotropically. Scattering probabilities are 
calculated using the mean free path of each photon in each shell, 
assuming thermally and turbulently broadened Fe line absorption, 
as well as Thomson scattering, 
including a proper energy transfer and scattering direction of the incident
photons after RS in the cluster and ion velocity field, 
which are uniquely implemented in {\tt Geant4}.
The Fe line emissivity and oscillator strength 
are taken from AtomDB version 3.0.8. 
In the simulation, we include scattering by the set of 
the Fe\emissiontype{XXIV,~XXV,~XXVI} lines shown in table 
\ref{tb:felines}. Other ions were neglected since their optical depths  
are negligibly small.
To run the simulation, we adopted an input spectral model 
of optically thin plasma generated 
with {\it bapec}. The emission model includes all emission lines, 
including the weak satellite lines. 

We examined three assumptions for the velocity ($\sigma_{\rm v}$) field 
based on the line-of-sight velocity dispersion shown in ``V'' paper:
a uniform $\sigma_{\rm v}$ of 150 km s$^{-1}$ (case 1) as a reference 
for comparison with the simulations shown in \citet{Zhu13}, 
a peak $\sigma_{\rm v}$ toward the AGN and a nearly flat field elsewhere 
(cases 2--4), 
and a case in which the $\sigma_{\rm v}$ rises outside of the field observed 
by Hitomi (case 5). 
The parameters for each simulation are listed in
table \ref{tb:radial_turbulent}. 
Figure \ref{fig:mock} (left panel) shows the simulated incident spectrum 
from the inner $0.'5$ radius of the cluster for a uniform $\sigma_{\rm v}$ 
of 150 km s$^{-1}$ (case 1 in table \ref{tb:radial_turbulent}). 
The bottom panel of the figure shows the ratio 
of the photon lists for the models with and without RS.
The $w$ line flux
is obviously suppressed by the RS effect. Note that the suppressed 
$w$ line shape is not represented by a Gaussian model which has the 
same $\sigma$ as the $w$ line without RS. 
The suppression of the $w$ line in our simulation agrees 
with previous results by \citet{Zhu13}. 
On the other hand, the predicted line broadenings 
due to the distortion with the {\tt Geant4} simulator are slightly 
wider than those from ICMMC. However, the difference is quite negligible 
after smoothed by the Hitomi responses as described in the next 
paragraph.

After generating the projected photon lists with the {\tt Geant4} simulator, 
we processed them with the HEAsim software in Ftools to make mock event files 
for the Hitomi SXS FOV, taking into account the Hitomi responses. The HEAsim 
software calculates the redistribution of the input photons, including 
the Hitomi mirror and detector responses such as the effective area, the PSF, 
and the energy resolution. Here, we used the responses in the HEAsim tools
and normalized the flux to the observed value, taking into account events 
out of the SXS FOV. We assumed a 1 Ms exposure time for each simulation. 
Black and red spectra in the right panel of 
figure \ref{fig:mock} show the mock spectra for obs23\_cen with and 
without RS, respectively, for the 150 km s$^{-1}$ 
uniform $\sigma_{\rm v}$ (case 1) model. 
One can clearly see the flux suppression in the $w$ line 
when RS is taken into account. 
As shown in the bottom panels in figure \ref{fig:mock}, the resonance 
line shape are clearly distorted by the line broadening as well as 
the line suppression in the mock spectrum.
Note that the mock spectra have finite numbers of photons
since the mock spectra are normalized to a given, finite flux.

To estimate the potential impact of systematic 
uncertainties in the input model, we also performed
simulations with the temperature and abundance profiles of 
the input model changed 
by $+10$\% (case~1a), and $\pm10$\%, respectively.
Also, we explored the effects of the moving core 
within 1$'$, with 150 km s$^{-1}$ in redshift 
relatively against the surrounding gas, as pointed out 
in ``V'' paper \citep{velocity}.

\subsection{Monte Carlo simulations with the ICMMC code}
\label{subsec:sim_ICMMC}

In order to interpret the observed resonance line suppression and infer 
velocities of gas motions, we also applied a different approach, 
based on Monte Carlo simulations of radiative transfer in 
hot gas described in \citet{Zhu10} \citep[see also][]{Sazonov02,Churazov04}. 
Here, instead of simulating the whole spectrum and fitting it 
with plasma models to obtain line ratios, we performed calculations 
in specific lines. Such simulations directly provide fluxes 
in the considered lines for models with and without RS; their ratios, 
corrected for the PSF, are then compared with the observed values. 
This approach has been successfully applied to the analysis of RS and 
velocity measurements in massive elliptical galaxies and galaxy groups  
\citep{Werner09,dePlaa12,Ogo17}. In these previous works the detailed 
treatment of individual interactions in the simulations is described.

Since the Hitomi measurements of line broadening and variations 
of line centroids do not show strong radial velocity gradients 
in the Perseus Cluster, and the properties of the velocity field 
outside the inner $\sim$ 100 kpc are unknown, we conservatively assume 
that the velocity of gas motions is approximately the same 
within the considered regions. The simulations are done for a grid 
of characteristic velocity amplitudes, the results of which are then 
compared with the observed line ratios (see section \ref{sec:comp_icmmc}).

\section{Comparisons of the observed line ratios and the simulations}
\label{sec:comparison}

\subsection{GEANT4 simulations}

\begin{figure*}[h]
\begin{minipage}{0.33\textwidth}
\FigureFile(\textwidth,\textwidth){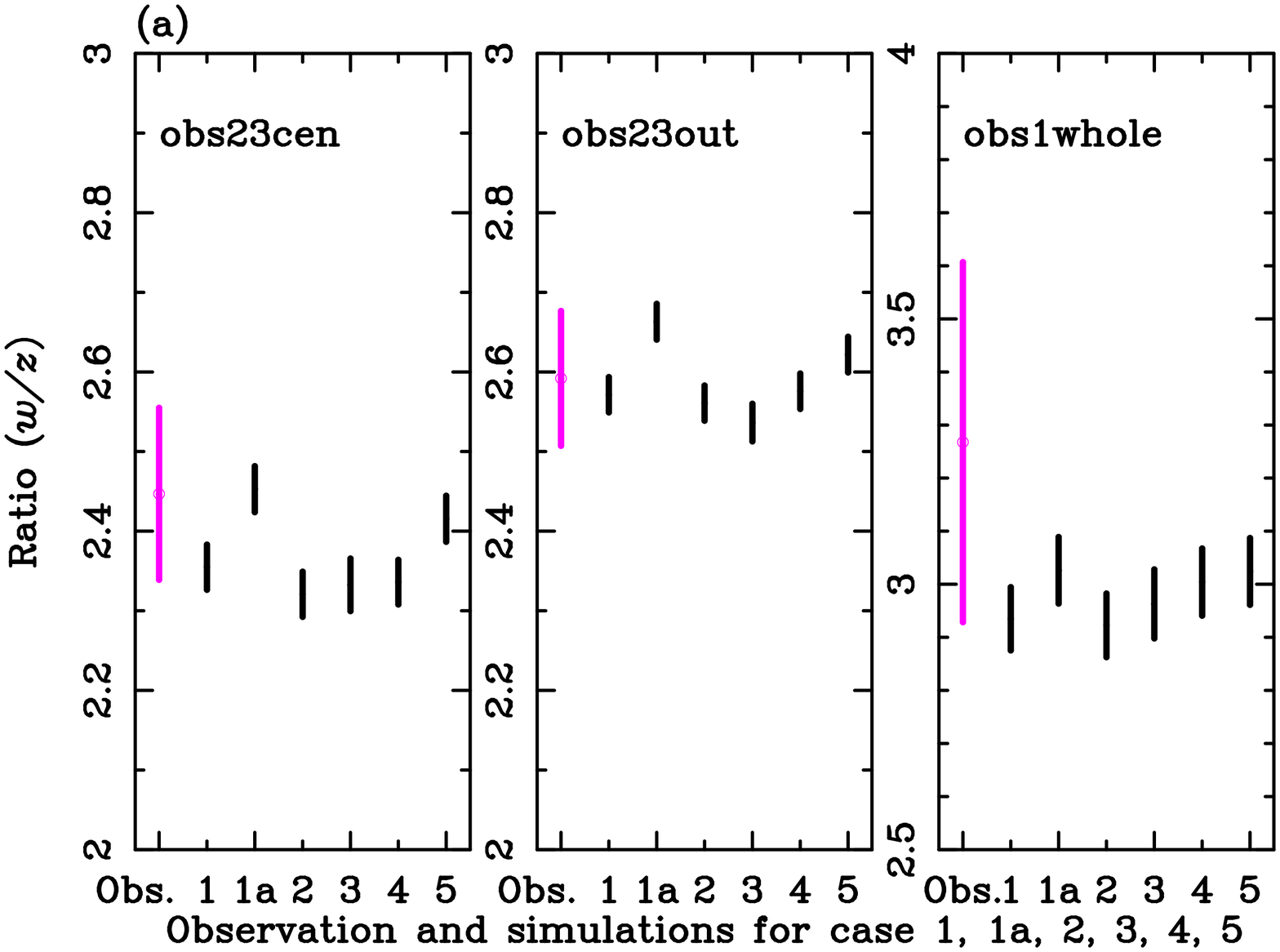}
\end{minipage}\hfill
\begin{minipage}{0.33\textwidth}
\FigureFile(\textwidth,\textwidth){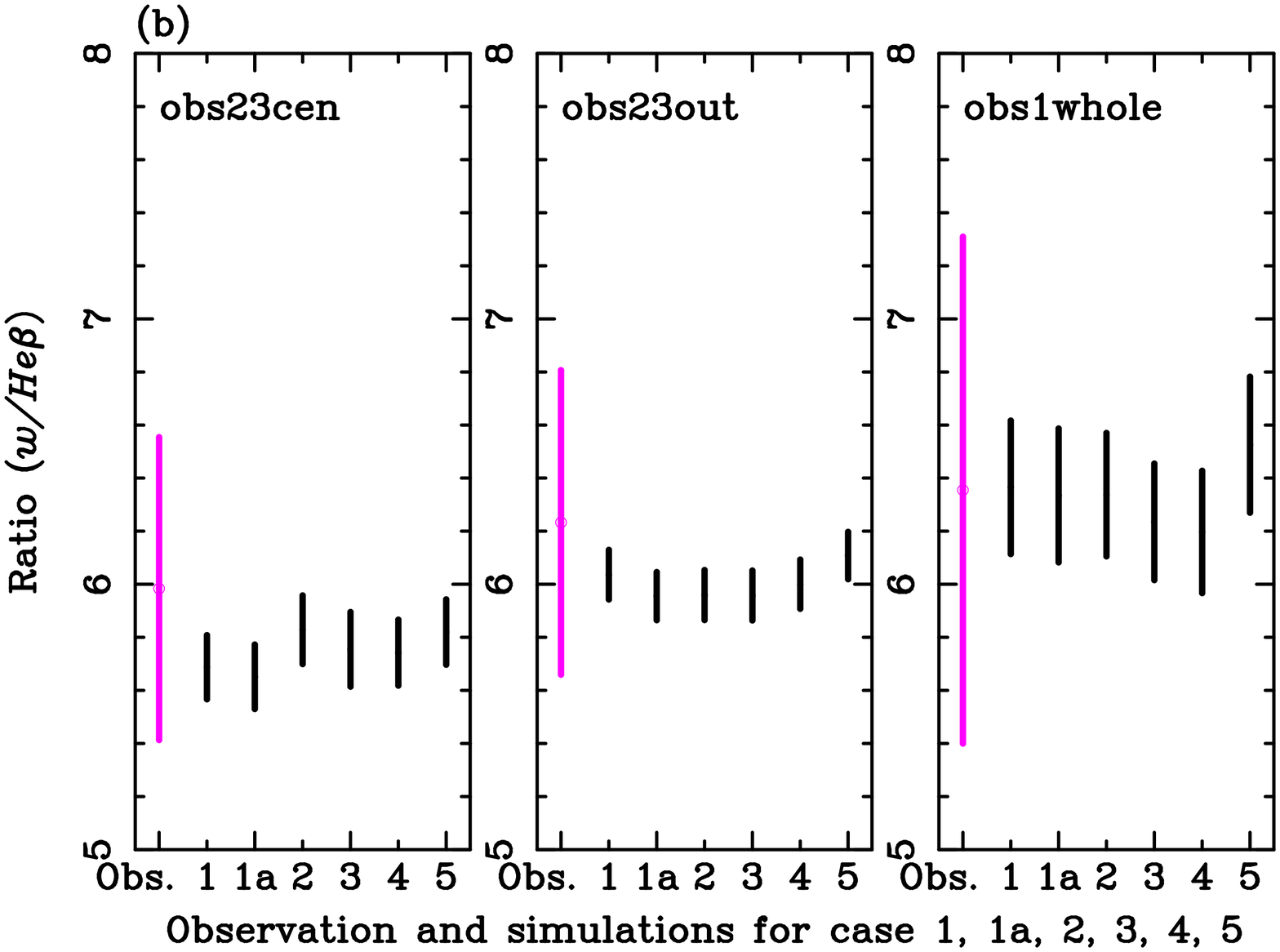}
\end{minipage}\hfill
\begin{minipage}{0.33\textwidth}
\FigureFile(\textwidth,\textwidth){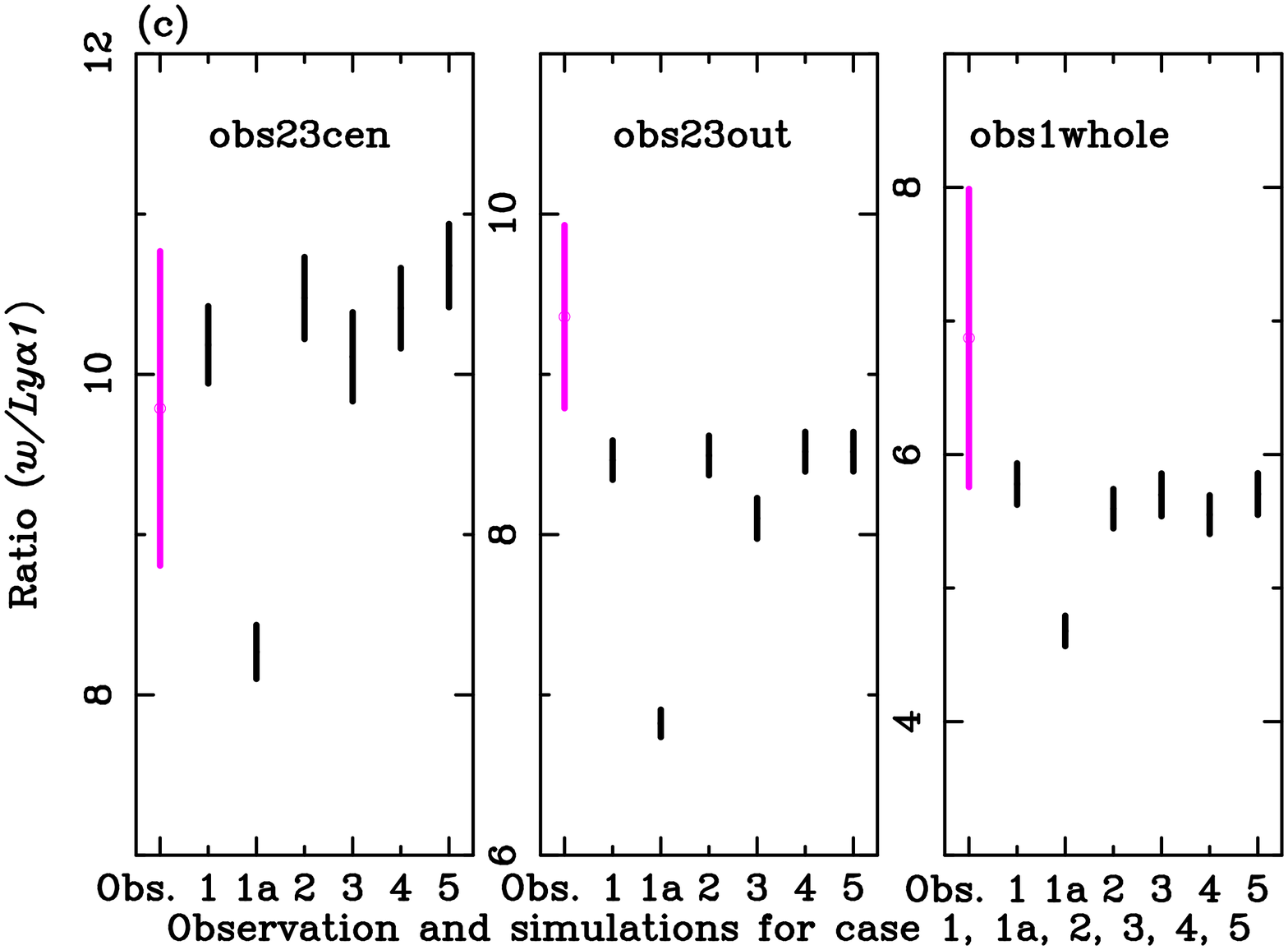}
\end{minipage}

\begin{minipage}{0.33\textwidth}
\FigureFile(\textwidth,\textwidth){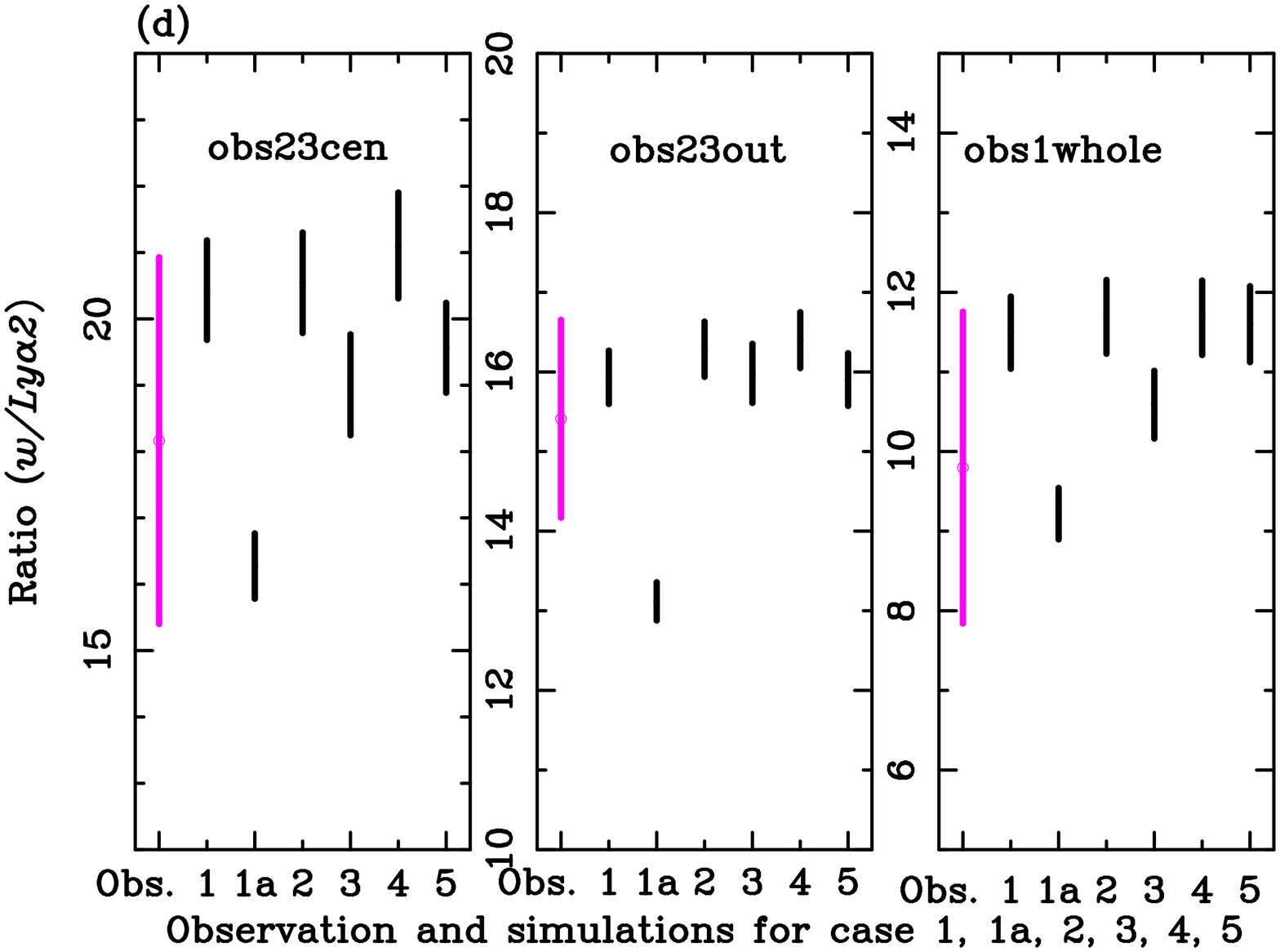}
\end{minipage}
\begin{minipage}{0.33\textwidth}
\FigureFile(\textwidth,\textwidth){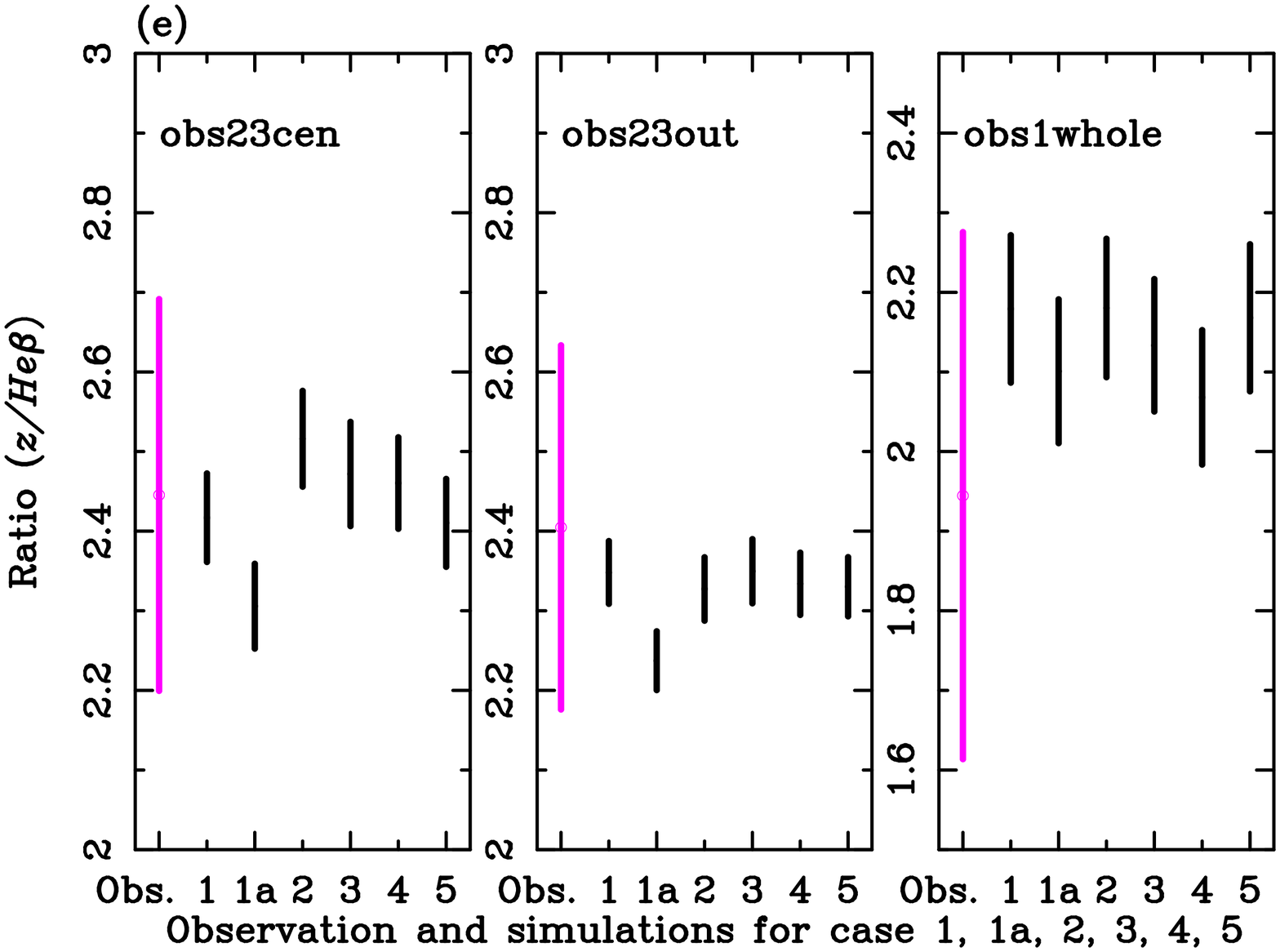}
\end{minipage}\hfill
\begin{minipage}{0.33\textwidth}
\end{minipage}

\begin{minipage}{0.33\textwidth}
\FigureFile(\textwidth,\textwidth){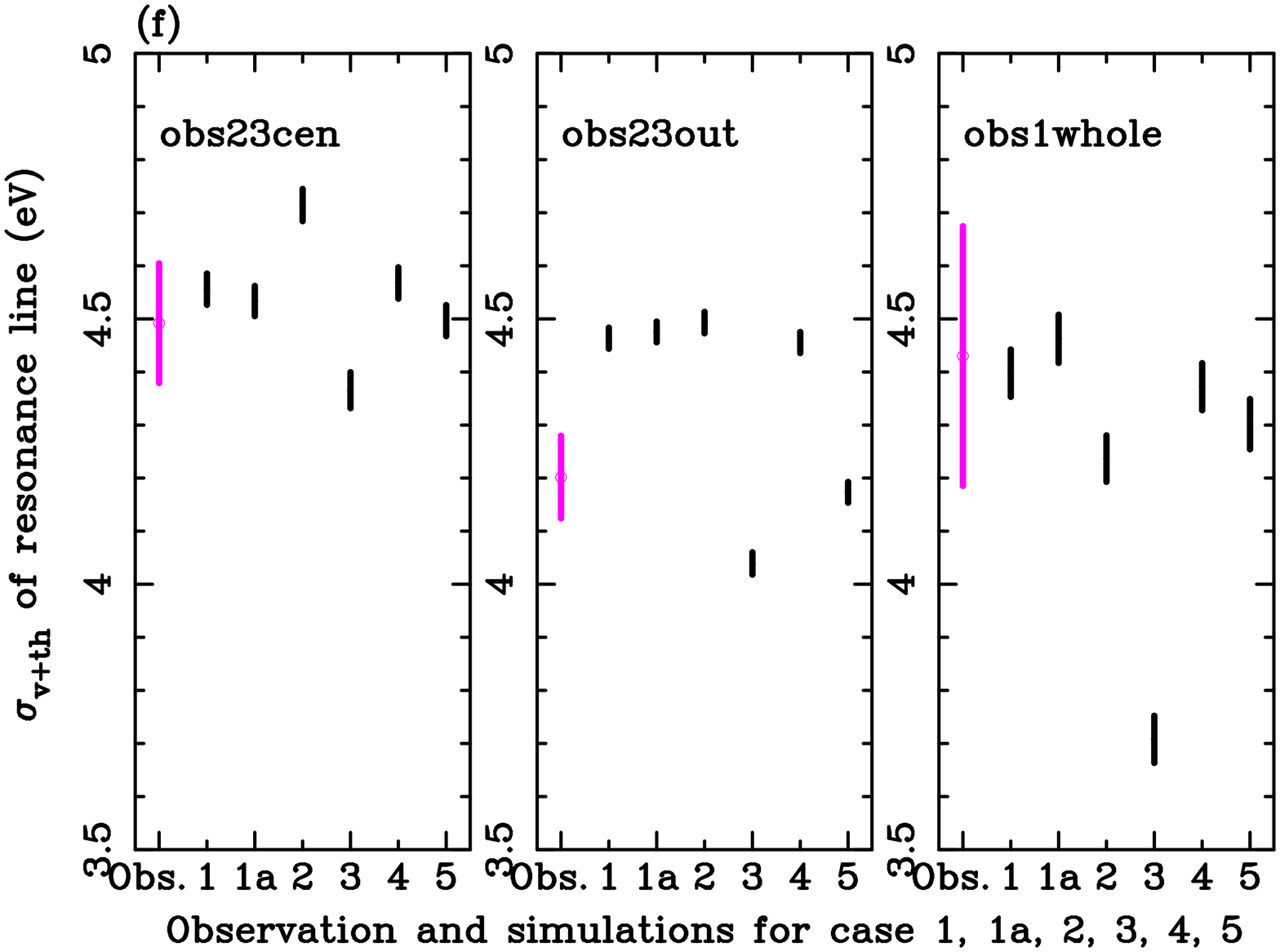}
\end{minipage}\hfill
\begin{minipage}{0.33\textwidth}
\FigureFile(\textwidth,\textwidth){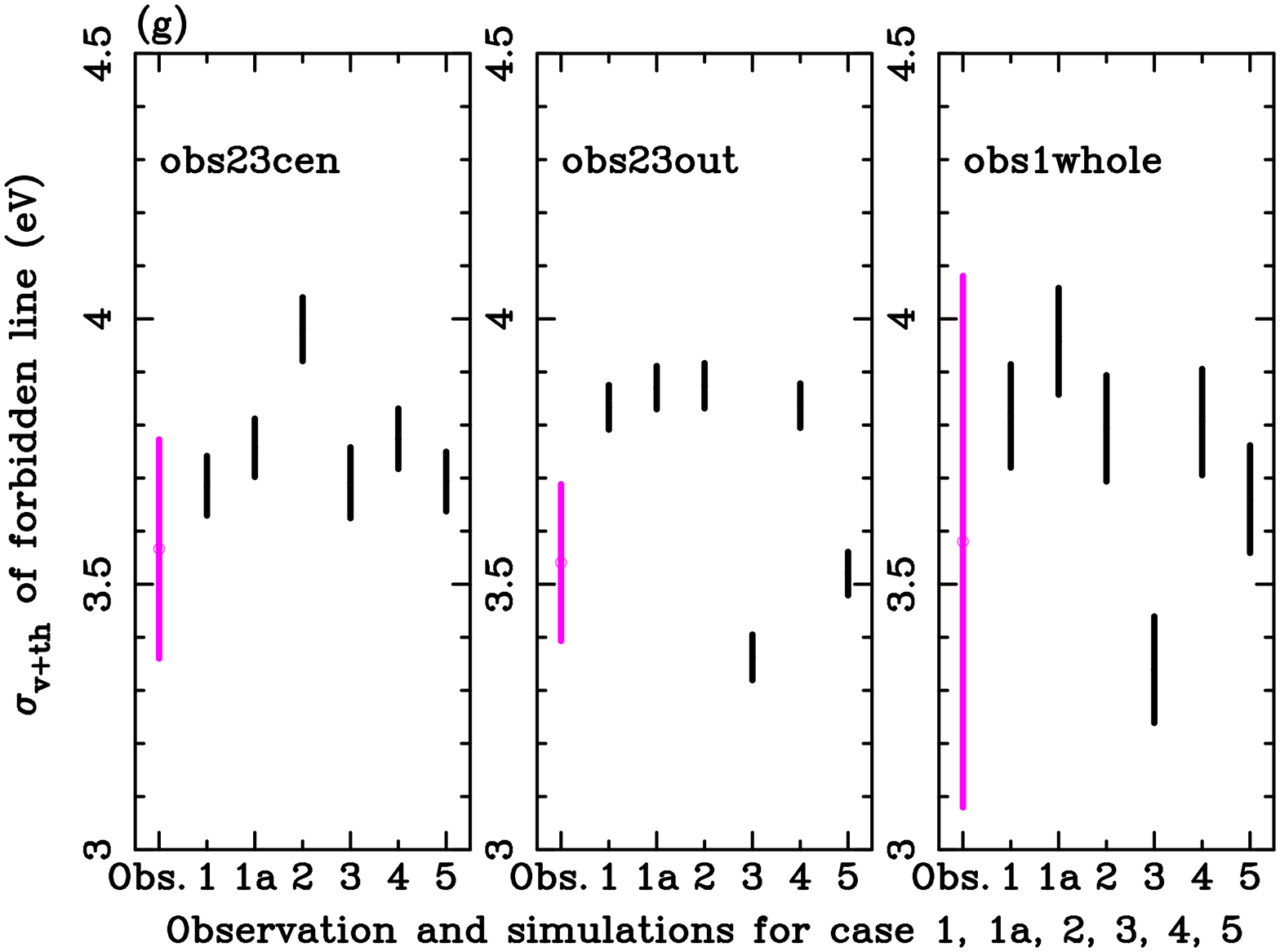}
\end{minipage}\hfill
\begin{minipage}{0.33\textwidth}
\FigureFile(\textwidth,\textwidth){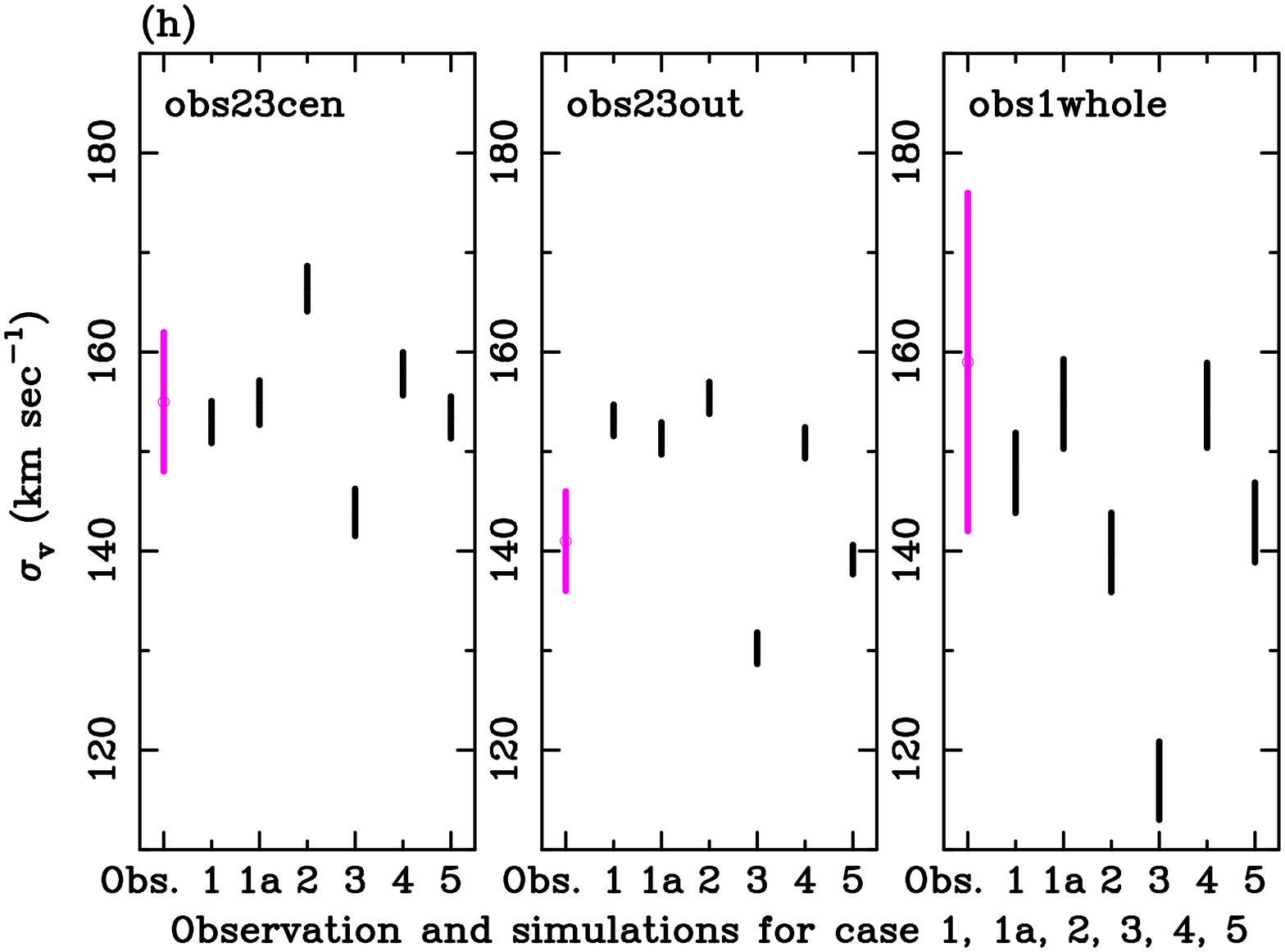}
\end{minipage}

\vspace*{2ex}
\caption{
(a)--(e): Comparisons of line ratios of the Fe He$\alpha$ resonance (w) 
to forbidden (z), He$\beta$, and Ly$\alpha_{1,~2}$, 
and the Fe He$\alpha$ w to He$\beta$ between 
observation (magenta, noted as Obs.) and simulations
with RS for case 1, 1a, 2, 3, 4, and 5.
(f)--(g): Same as line widths of the w and z.
(h): Same as the derived $\sigma_{\rm v}$ from the mock spectra.
}
\label{fig:comp-models}
\end{figure*}

We compared the spatial distribution of the observed line ratios with 
the simulations described in section \ref{subsec:sim_tus}. 
In order to compare the line ratios and widths,
we fitted the simulated spectra with the same 
spectral model and responses for the ICM discussed in section 
\ref{sec:spectralfits}, i.e. the ``modified'' {\it bvvapec} 
({\it bvvapec} with the strongest lines deleted) plus Gaussian models. 
The mock spectra are well represented by this model.
To understand the impact of limited photon statistics in the modeling,
we divided the simulated event list into ten 100 ks parts, each of which 
had similar statistics in Fe\emissiontype{XXV} 
He$\alpha$ to the observed Hitomi data.

Figure \ref{fig:comp-results} shows comparisons of the 
observed and predicted line ratios and widths, and $\sigma_{\rm v}$ 
for case 1 (flat $\sigma_{\rm v}$ field), with and without RS. 
The observed ratios of the Fe\emissiontype{XXV} He$\alpha$ $w/z$ are 
consistent with those from simulation with RS.
The simulated ratios without RS,
shown by light gray diamonds, are clearly far away from the 
observed ones in the inner regions.
Figure \ref{fig:comp-models} shows the comparisons for all models 
listed assuming a plausible velocity field based on 
the ``V'' paper \citep{velocity} in table \ref{tb:radial_turbulent}.
For all the regions, simulations of the $w/z$ ratio for all the cases 
are broadly consistent with the observations 
as shown in figures \ref{fig:comp-results} and \ref{fig:comp-models}.
The observed widths of the $w$ and $z$ lines for the central 
region obs23\_cen and the obs1\_whole are well 
represented by the simulation with RS for cases 1, 4, and 5.
The simulation for case 4 which is close to the line-of-sight 
velocity dispersion field in the ``V'' paper agrees well with the observed 
line ratios and widths.
The simulated line widths with RS for case 2
look slightly broader than the observed one, while simulations 
with lower $\sigma_{\rm v}$ ($< 100$ km s$^{-1}$) 
in $r>2'$, such as case 3, are poorly described in the outer regions. 
Consequently, our simulation assuming plausible velocity field 
based on the ``V'' paper is consistent with
the observation, while the constant distribution and the relatively 
large $\sigma_{\rm v}$ of $\sim300$ km s$^{-1}$ at large radius 
would not be rejected from our simulation.
The simulations show that the predicted 
line ratios and widths are affected by the assumed velocity field 
rather than the RS effects.
For the obs23\_out region which includes the north-west 'ghost' bubble 
as shown in the ``V'' paper, the line widths from simulations are
broader than observations due to the azimuthal dependence.

As for the $w/He\beta$, $w/Ly\alpha_1$, and $w/Ly\alpha_2$ lines, 
the simulated ratios with the RS effects also broadly agree with 
the observed ones within the statistical errors, except for $w/Ly\alpha_1$ 
in the obs23\_out and obs1\_whole regions. 
The Ly$\alpha$ line ratios are sensitive to 
the azimuthal dependence and hotter component of projected temperature.
On the other hand, the observed $z/He\beta$ ratios, 
whose lines have low optical depth than the 
other lines as shown in figure \ref{fig:opticaldepth},
are also consistent with the simulated ratios.

The temperatures derived from the simulated spectra are lower 
than the observed ones for all the regions. 
It should be noted that the $w/z$ line ratio does not change much 
even if the temperature and $\sigma_{\rm v+th}$ change.
In fact, changing the temperature in simulations by $+10$\%  
for case 1a, which corresponds to the azimuthal scattering, 
does not change the results within the observed statistical errors.
The derived $\sigma_{\rm v}$ for cases 1, 4 and 5 agree well 
with the observations in the innermost region, while those in 
the obs23\_out region are lower than the simulated ones. 
We also estimated the uncertainties by changing the Fe abundance 
by $\pm10$\%. The resultant line ratios do not change by more 
than $\sim3$\%.

In this simulation, we assumed spherical symmetry in the cluster core. 
If bulk motion existed along the line of sight in the cluster core, 
the line widths should be broader along line of sight.
The ``V'' paper \citep{velocity} actually shows a large scale bulk 
velocity gradient of $\sim100$ km s$^{-1}$.
As shown in section \ref{sec:spectralfits}, we adopted the gain correction, 
which gave $\sim5$\% broader line widths than the uncorrected data,  
but the ``V'' paper did not.
In order to estimate the uncertainties, we performed simulations with 
the assumption of the core moving within $0.'5$ radius 
with 150 km s$^{-1}$ relative to the surrounding gas based on case 1.
The resultant $w/z$ line ratios in obs23\_cen did not change within the 
statistical errors for case1. Therefore, we confirmed that the RS effect 
is not very sensitive to bulk motion in the Perseus cluster core 
as was earlier shown as well by \citet{Zhu11}.
On the other hand, the derived $w$ and $z$ line widths are 
broader by 2 and 1\% than those from case 1, and also 4 and 3\% wider 
than the observation. These discrepancies are smaller than the 
difference between the gain-corrected and uncorrected data as shown 
in figure \ref{fig:scatplot}.


\subsection{ICMMC simulations}
\label{sec:comp_icmmc}

\begin{figure*}
\begin{minipage}{0.5\textwidth}
\includegraphics[trim=0 3 30 0,width=\textwidth]{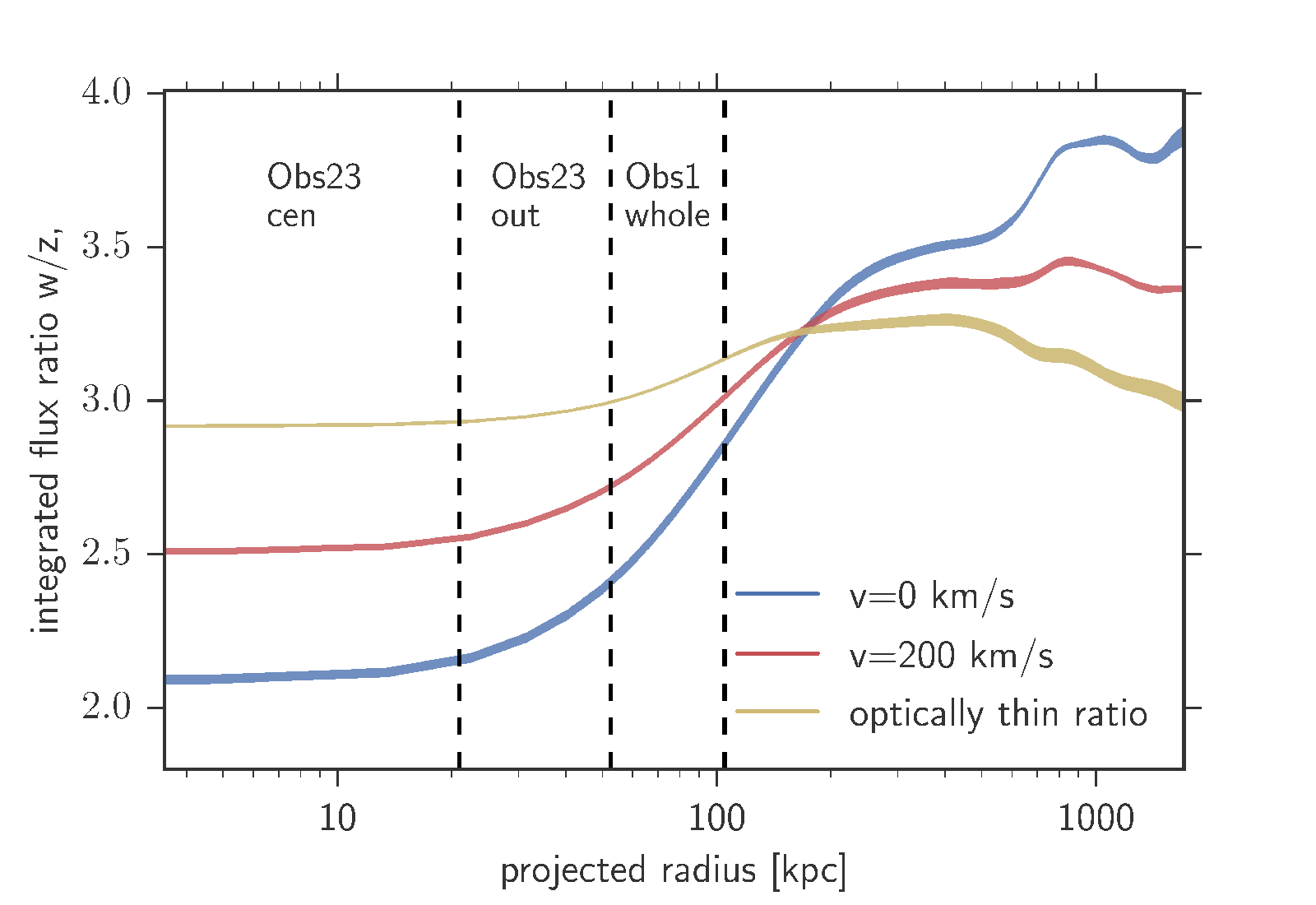}
\end{minipage}\hfill
\begin{minipage}{0.5\textwidth}
\includegraphics[trim=0 3 30 0,width=\textwidth]{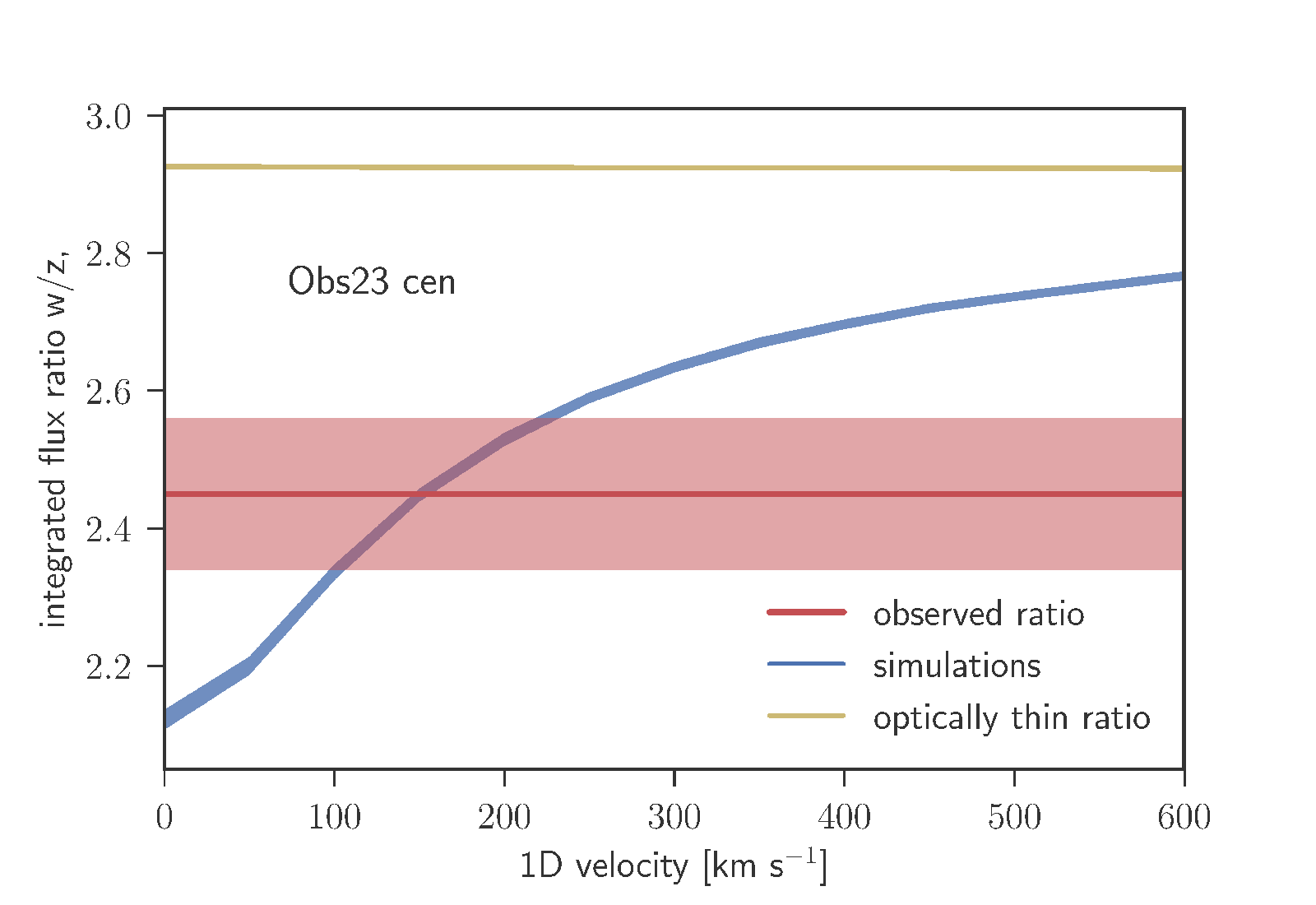}
\end{minipage}
\begin{minipage}{0.5\textwidth}
\includegraphics[trim=0 0 30 3,width=\textwidth]{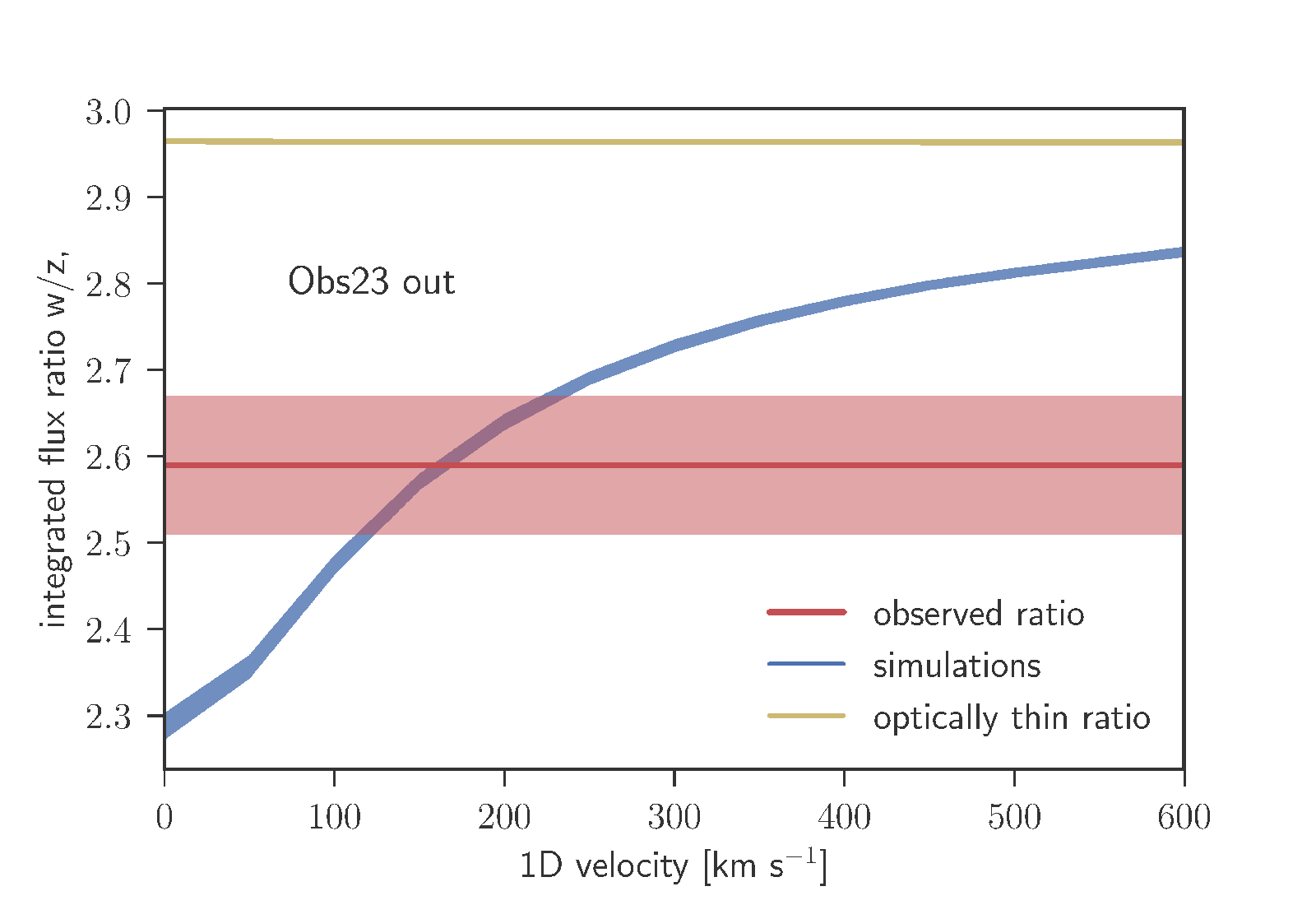}
\end{minipage}\hfill
\begin{minipage}{0.5\textwidth}
\includegraphics[trim=0 0 30 3,width=\textwidth]{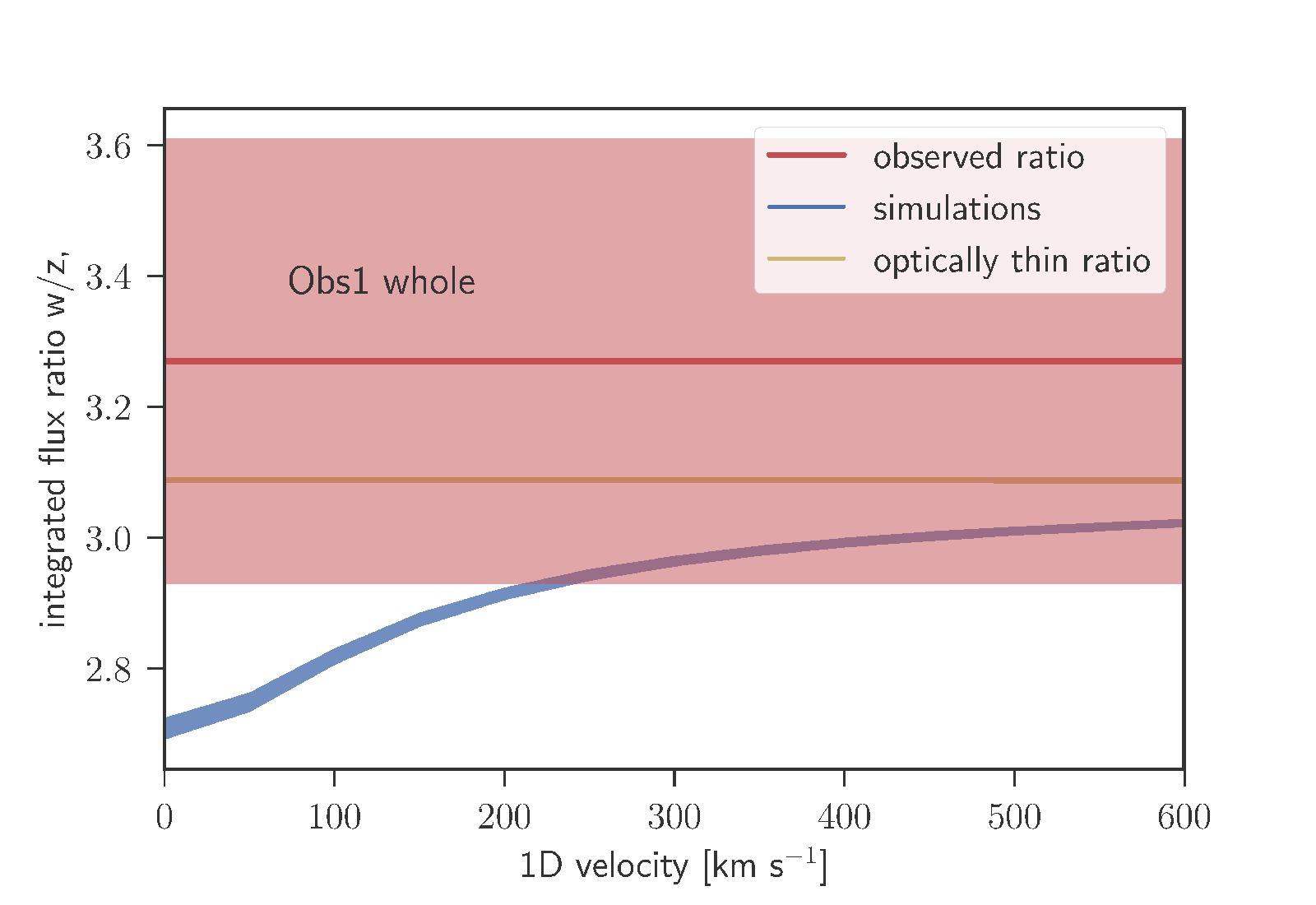}
\end{minipage}
\caption{
Top, left: the $w/z$ ratio as a function of
projected distance from the cluster center calculated using ICMMC 
simulations of RS assuming zero (blue)
and 200 km s$^{-1}$ (red) velocity of gas motions along the line of sight. 
The optically thin line ratio is shown in yellow. 
All ratios are combined with the Hitomi PSF. The width of each
curve reflects the $1\sigma$ statistical uncertainty on the deprojected
thermodynamic profiles (see figure \ref{fig:input}). 
The approximate locations of spectral
extraction regions (obs23\_cen, obs23\_out, and obs1\_whole) are shown
with dashed lines. Top, right and bottom: integrated flux ratio over
spectral extraction regions (see figure \ref{fig:image}) as a function
of line-of-sight velocity (blue); integrated optically thin line ratio vs.
velocity (yellow) and the observed line ratio from the Hitomi data (red). 
The overlap between the red regions and blue curves provides constrains on the
velocity of gas motions. 
}
\label{fig:app_fig1}
\end{figure*}

Using the thermodynamic model of the Perseus Cluster shown in figure
\ref{fig:input} and the APEC (based on AtomDB version 3.0.8) plasma model, 
we calculated the line ratios
($w/z$, $w/(He{\beta_1}+He{\beta_2})$, $w/Ly{\alpha_1}$ and 
$w/Ly{\alpha_2}$) as a function of projected distance
from the cluster center, assuming different levels of isotropic
turbulence and accounting for RS. The results of simulations are then 
combined with the 2D PSF maps of Hitomi provided by the ``V'' paper 
\citep{velocity}. Results for the $w/z$ ratio are shown in the top left 
panel in figure \ref{fig:app_fig1}. The uncertainties on the line ratios 
are a result of series of simulations, in which we vary the assumed 
temperature, density, and abundance profiles within the uncertainties 
for the cluster model shown in figure \ref{fig:input}. These flux ratios
are integrated over the observed regions (obs23\_cen, obs23\_out,
and obs1\_whole, see figure \ref{fig:image}) and compared 
with the observed line ratios.  
The rest of panels in figure \ref{fig:app_fig1} show the $w/z$ ratio 
spatially integrated within the observed regions as a function of the
one-component (1D) velocity. As expected, the larger the velocity, 
the closer the line ratio to the optically thin case. The observed line 
ratios are plotted in red. The overlap between the observed and theoretical
line ratios allows us to constrain the velocity of gas motions. Note that 
the RS effect is the smallest in the obs1\_whole region. 
Also, the statistical uncertainty on the measured $w/z$ ratio is
the largest in this region. Therefore, a longer Hitomi-like observation 
will be required for a positive velocity measurement using the RS effect 
for the region obs1\_whole.

\begin{figure}
\begin{minipage}{0.5\textwidth}
\includegraphics[trim=0 0 10 0,width=\textwidth]{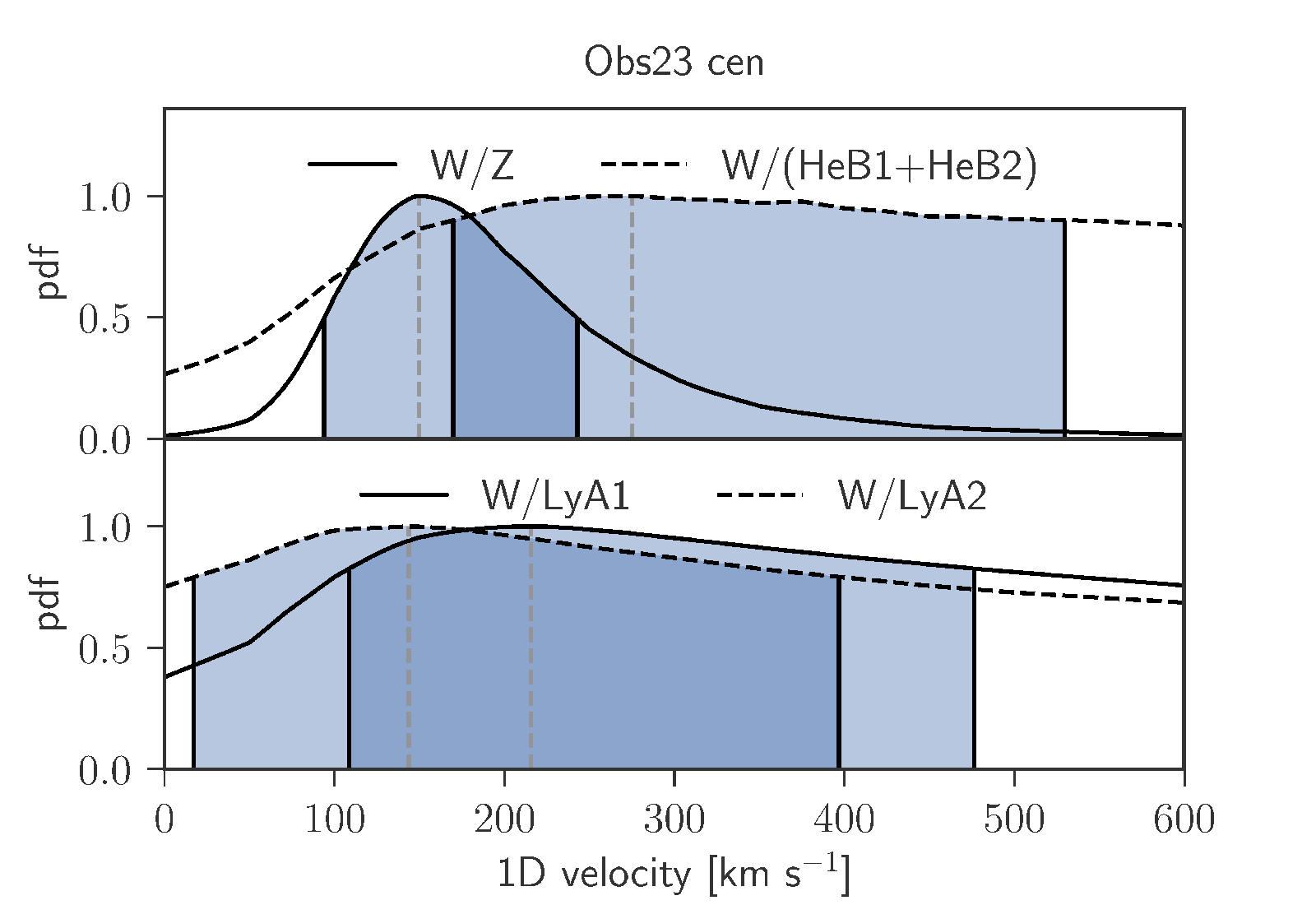}
\end{minipage}\hfill
\begin{minipage}{0.5\textwidth}
\includegraphics[trim=0 0 10 0,width=\textwidth]{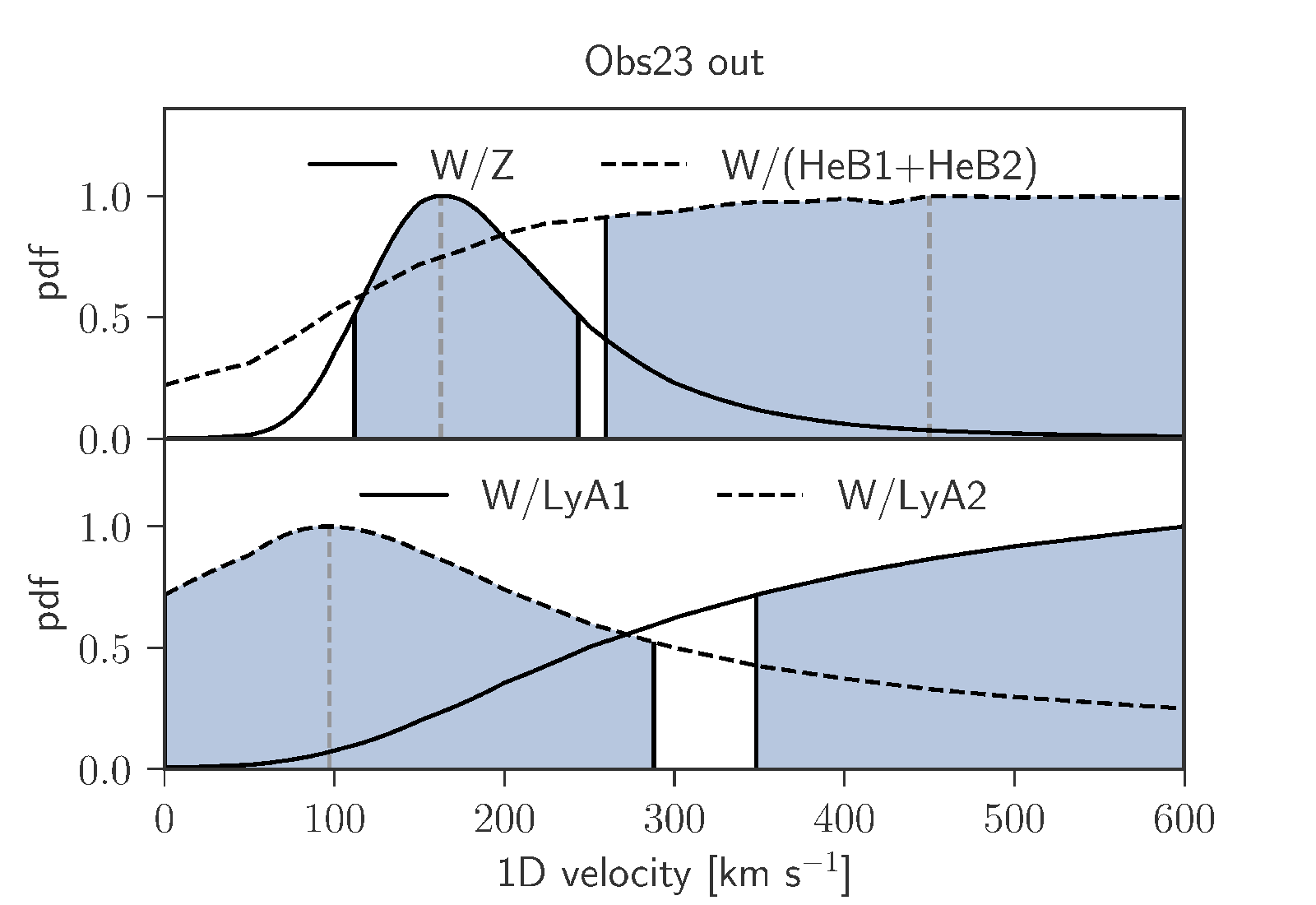}
\end{minipage}\hfill
\begin{minipage}{0.5\textwidth}
\includegraphics[trim=0 0 10 0,width=\textwidth]{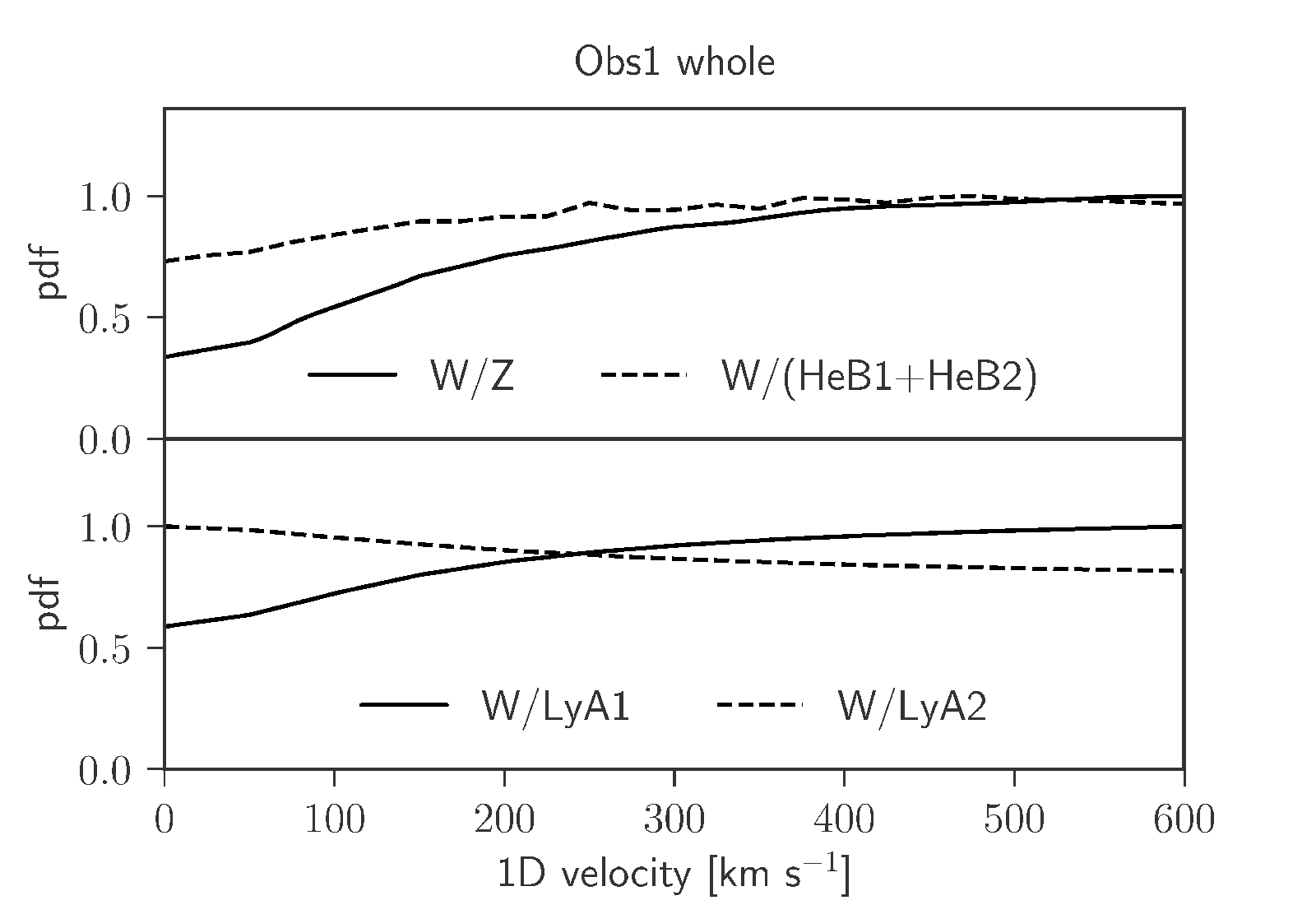}
\end{minipage}\hfill
\caption{Velocity probability distributions from the convolution of
the observed line ratios and those predicted from numerical simulations of
radiative transfer in Perseus combined with the Hitomi PSF. 
The three considered regions are shown in the top left,
top right and bottom panels. Confidence intervals, measured under the
assumption that the maximum Mach number is unity, are shown in blue. See the
legend for different line ratios.
\label{fig:app_fig2}
}
\end{figure}

For each line ratio, the results of the simulations 
(blue regions in figure \ref{fig:app_fig1}) are then combined with
the measured error distributions in the observed line ratios (red regions), 
and probability distributions
for turbulent velocities are obtained, as shown 
in figure \ref{fig:app_fig2} \citep[see][for the method]{Ogo17}.
Assuming that the maximum Mach number, $M_{\rm max}=\sqrt{3}V_{\rm los}/c_s$,
of gas motions is unity, $1\sigma$ confidence intervals on 
the velocity measurements are obtained
(blue regions in figure \ref{fig:app_fig2}). The velocity distributions
obtained from the $w/z$ line ratio are peaked, and the confidence intervals
are relatively insensitive to the choice of $M_{\rm max}$ in both obs23\_cen 
and obs23\_out regions. The measured turbulent velocities are 
$150^{+80}_{-56}$ km s$^{-1}$ and $162^{+78}_{-50}$ km s$^{-1}$ 
in these regions, respectively, consistent with the
direct velocity measurements through line broadening
\citep{Hitomi16}. In contrast the velocity distributions inferred 
from the $w/(He{\beta_1}+He{\beta_2})$ (hereafter, $w/He{\beta}$) 
line ratio is quite uncertain and depends on
the assumed maximal Mach number. Longer, Hitomi-like observations will 
improve the results for this ratio.
The ratios $w/Ly{\alpha_1}$ and $w/Ly{\alpha_2}$ provide velocities 
$220_{-111}^{+260}$ km s$^{-1}$ and $144_{-127}^{+256}$ km s$^{-1}$, 
respectively, in the obs23\_cen region. In the outer region, 
$w/Ly{\alpha_2}$ gives velocity $97_{-97}^{+193}$ km s$^{-1}$, 
consistent with the direct velocity measurements, while the $w/Ly{\alpha_1}$ 
ratio gives 2 $\sigma$ lower limit 178 km s$^{-1}$\@. 
The latter result is very sensitive to the choice 
of $M_{\rm max}$. Note
that the $Ly{\alpha}$ lines of He-like Fe have the peaks of their
emissivity times ionic fraction at gas temperatures $\sim 10$ keV,
while the same quantity for the $w$ line of He-like Fe peaks around 
$\sim 5$ keV. Since our thermodynamic model for
Perseus is calculated from the $0.5-8.5$ keV band spectra, the
contribution of high-temperature gas could be underrepresented
in our fiducial cluster model,
which would affect the emissivity of the $Ly{\alpha}$ lines. Therefore,
the $w/Ly{\alpha}$ line ratios are the least reliable of the ratios 
considered here. The bottom panel in figure \ref{fig:app_fig2}
shows the velocity distributions measured in obs1\_whole region, 
for which the constraints are weak (see also 
figure \ref{fig:app_fig1}).

Our interpretation of the observed line suppressions due to RS relies 
on the assumption of spherical symmetry and the choice of the reference 
emissivity model for an optically thin plasma. Below, we check how these 
assumptions affect the result as well as the effect of the Hitomi PSF.

\begin{figure*}
\begin{minipage}{0.5\textwidth}
\includegraphics[trim=0 40 20 0,width=\textwidth]{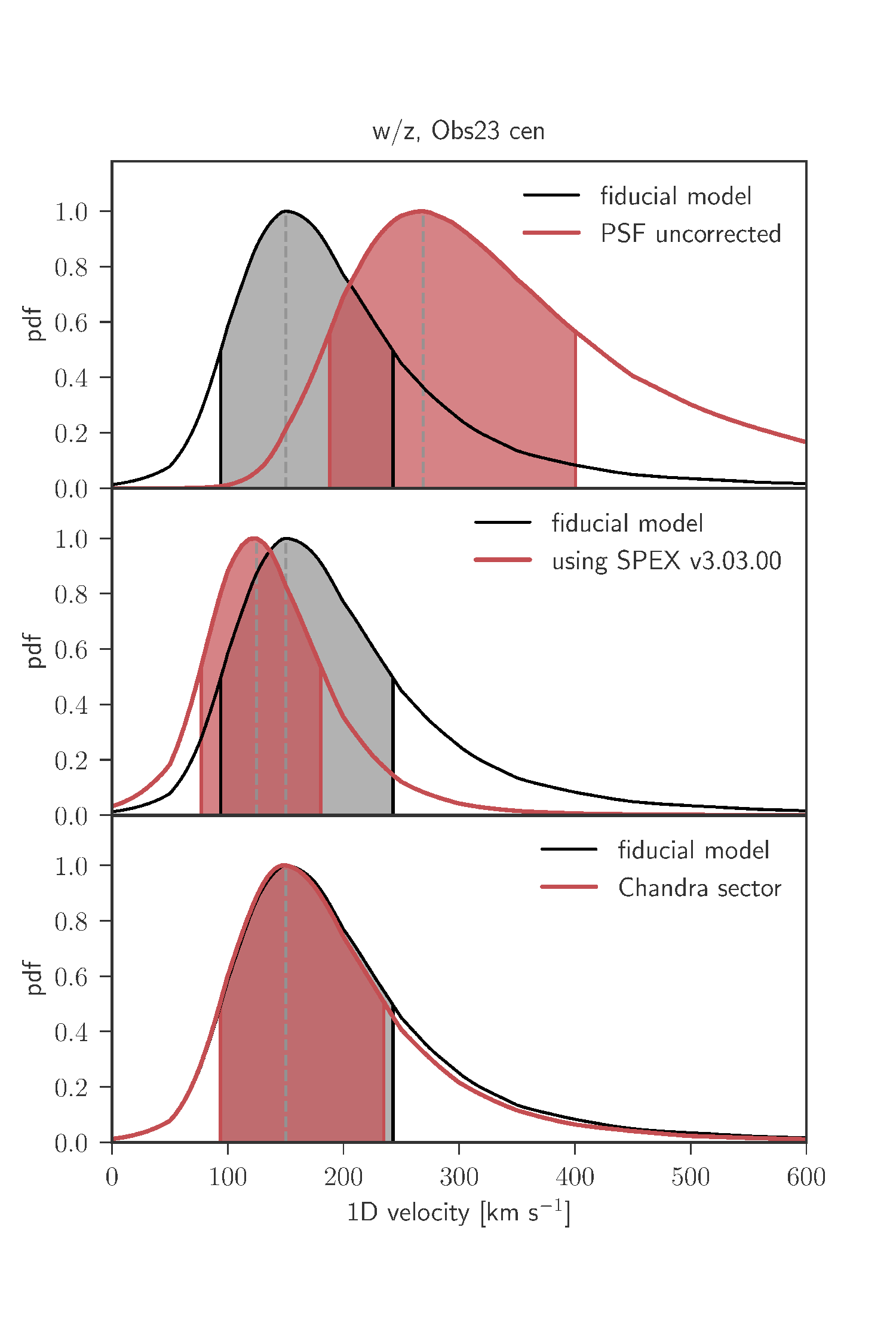}
\end{minipage}
\begin{minipage}{0.5\textwidth}
\includegraphics[trim=0 40 20 0,width=\textwidth]{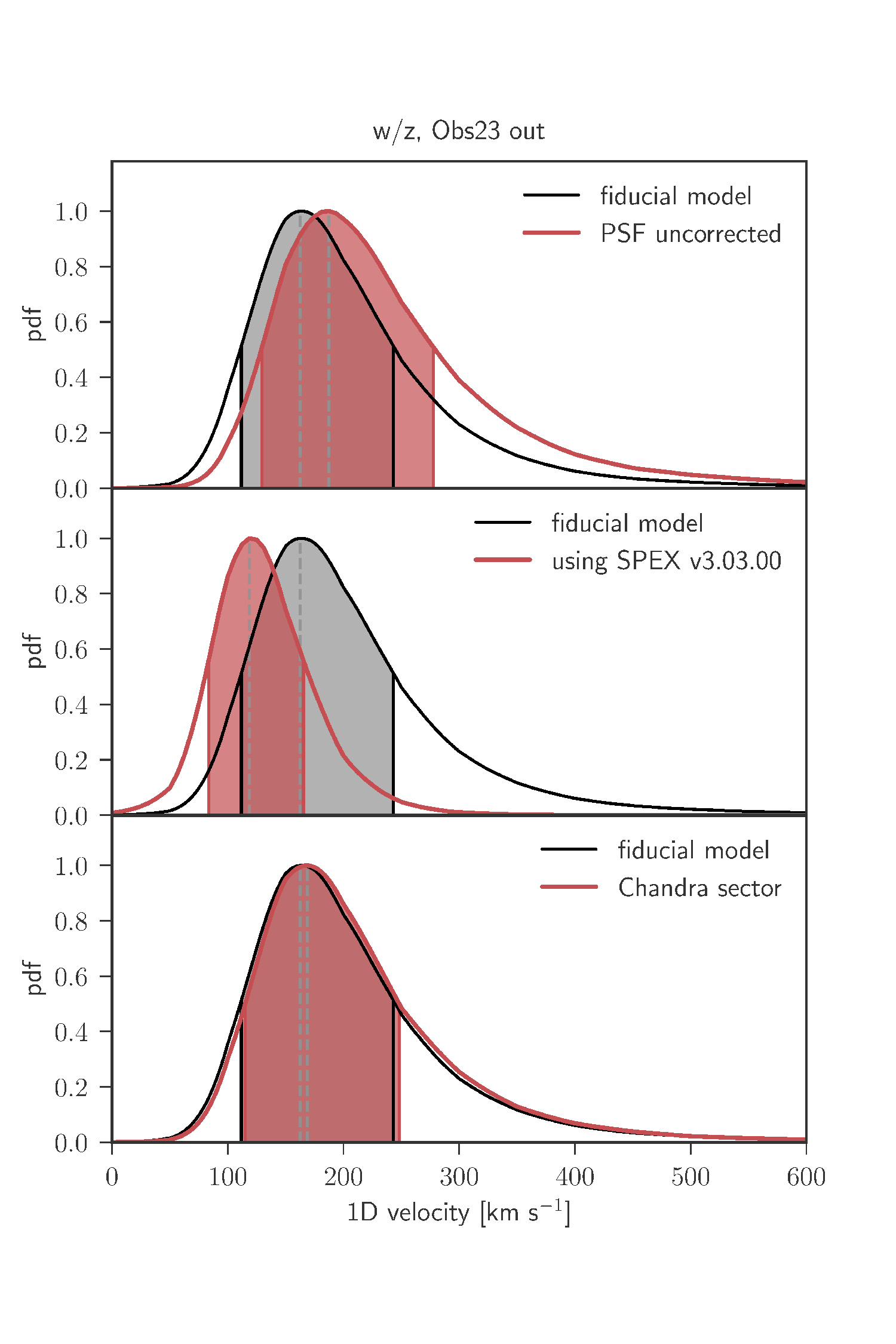}
\end{minipage}
\caption{The sensitivity of the measured velocity of gas motions
in Perseus from the $w/z$ line ratio to (from the top) the Hitomi PSF, 
the choice of optically thin plasma model and assumption of
spherical symmetry for the regions
obs23\_cen on the left and obs23\_out on the right.
\label{fig:app_fig3}
}
\end{figure*}

\begin{figure*}
\begin{minipage}{0.33\textwidth}
\includegraphics[trim=0 0 0 0,width=\textwidth]{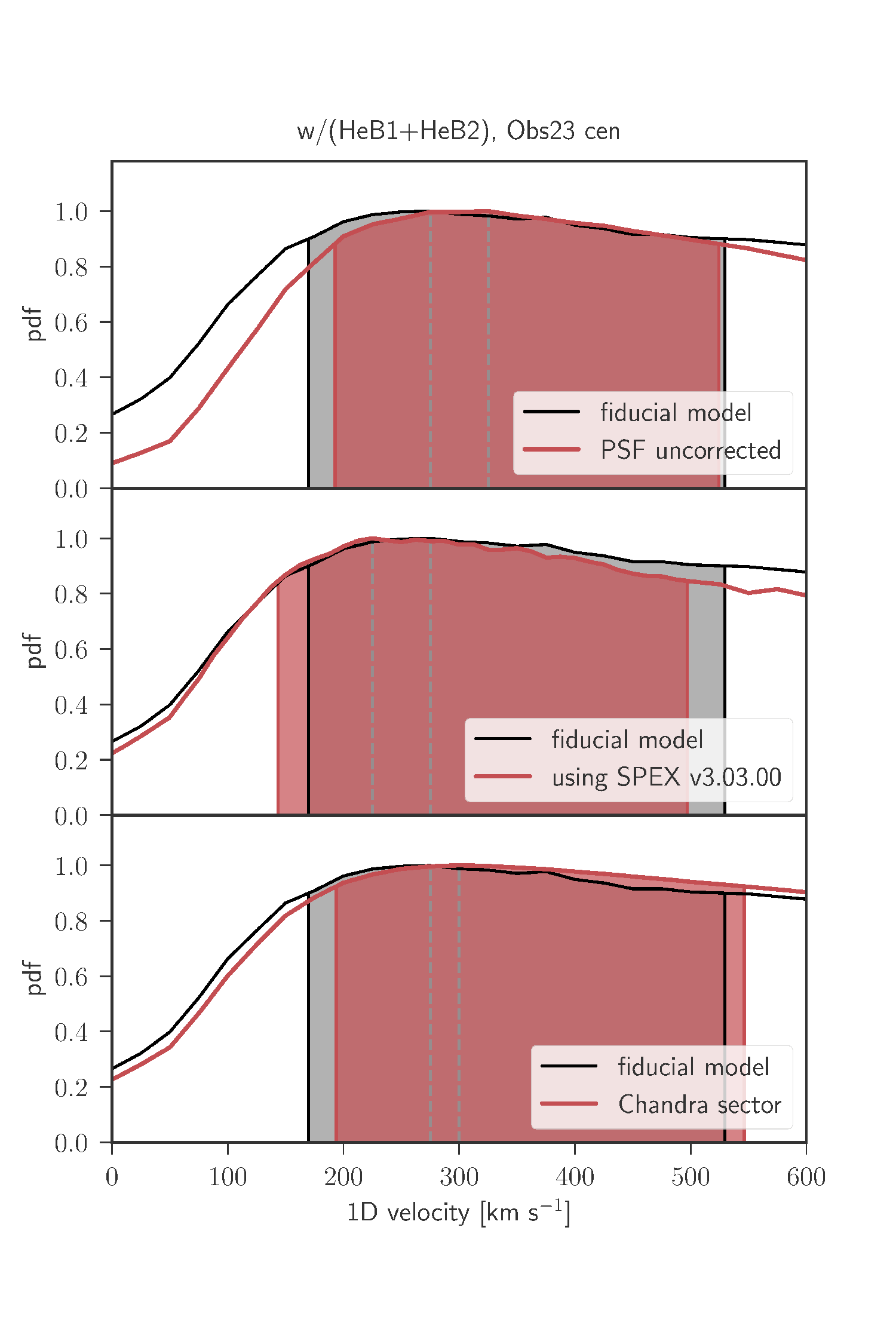}
\end{minipage}\hfill
\begin{minipage}{0.33\textwidth}
\includegraphics[trim=0 0 0 0,width=\textwidth]{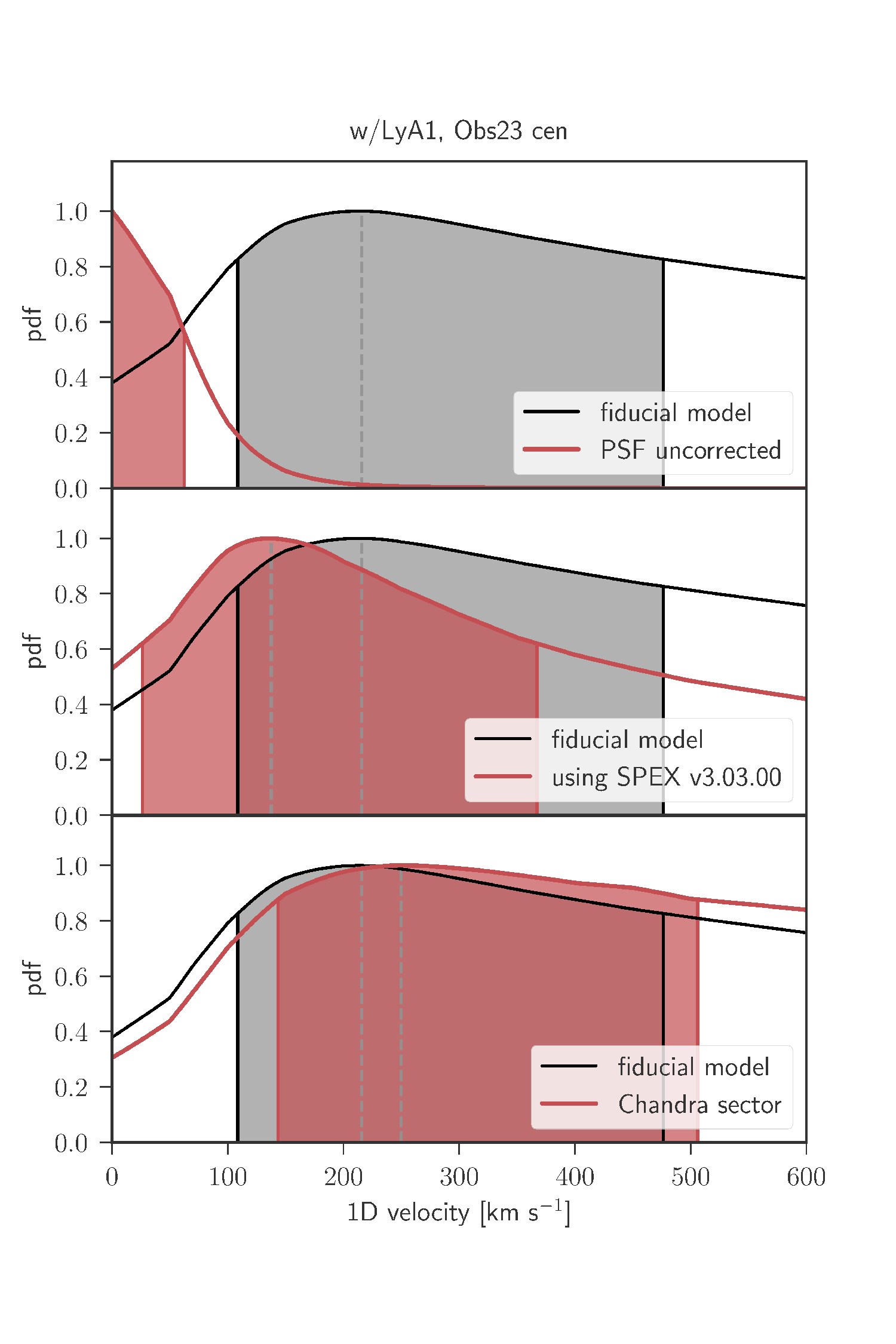}
\end{minipage}\hfill
\begin{minipage}{0.33\textwidth}
\includegraphics[trim=0 0 0 0,width=\textwidth]{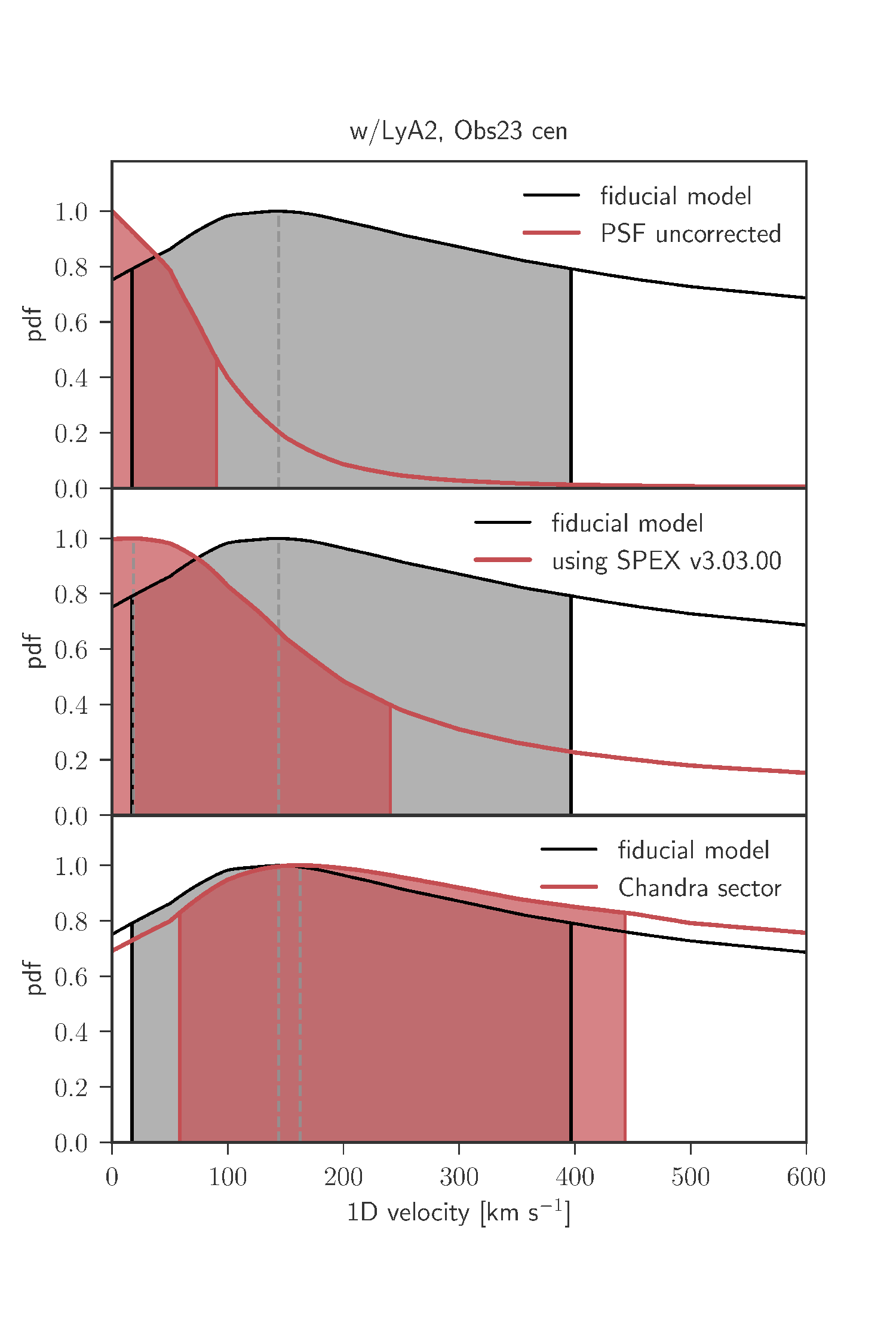}
\end{minipage}

\begin{minipage}{0.33\textwidth}
\includegraphics[trim=0 0 0 0,width=\textwidth]{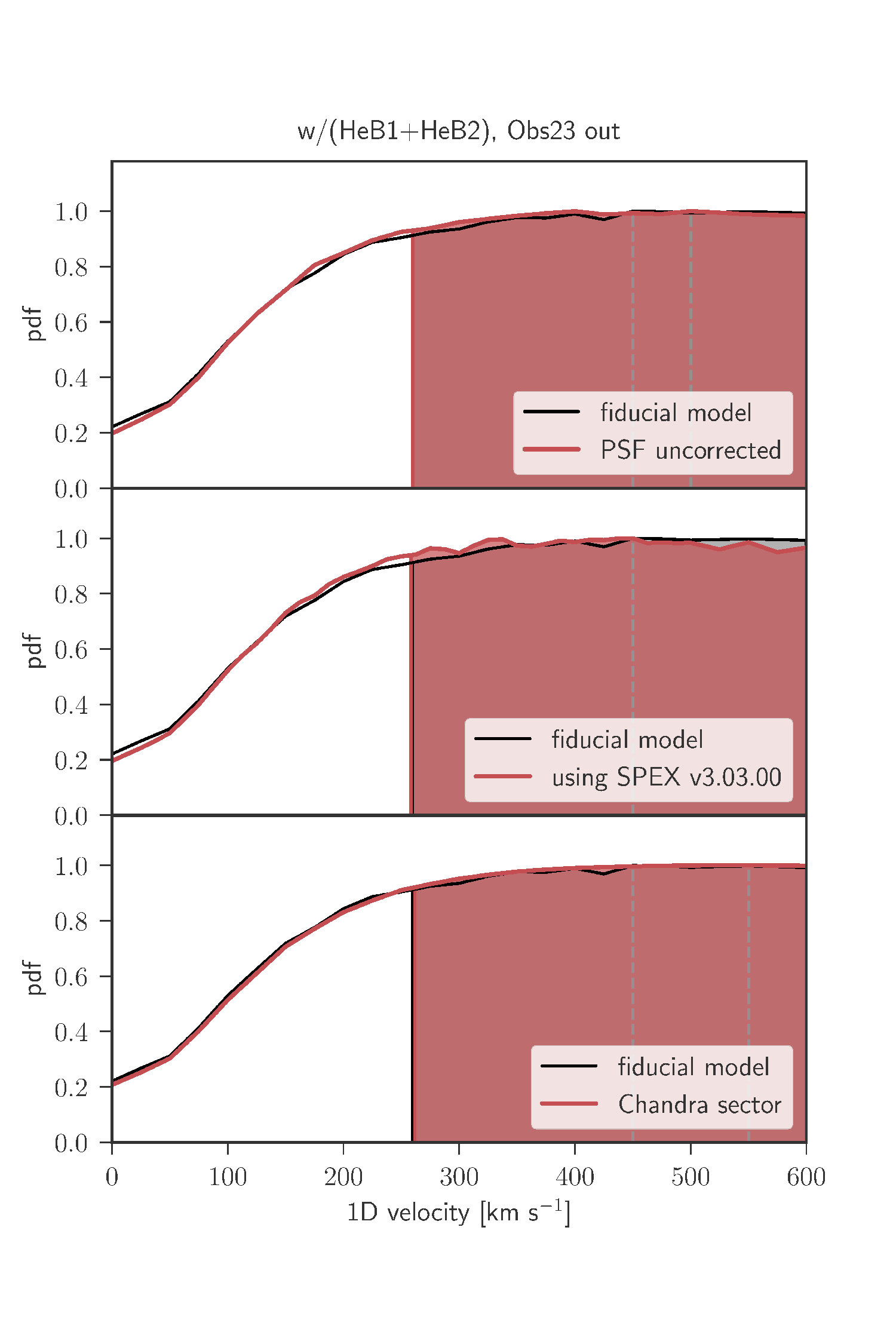}
\end{minipage}\hfill
\begin{minipage}{0.33\textwidth}
\includegraphics[trim=0 0 0 0,width=\textwidth]{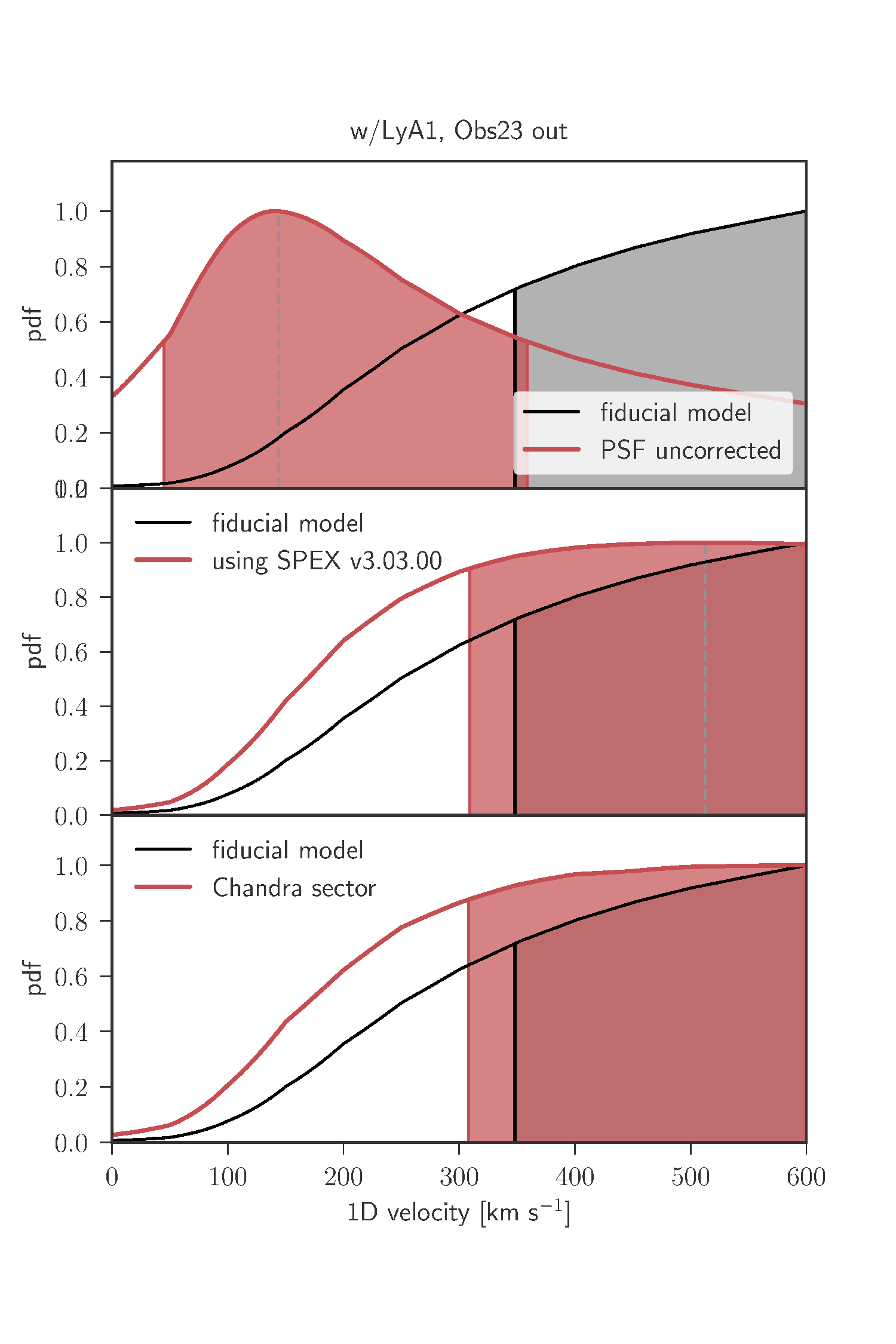}
\end{minipage}\hfill
\begin{minipage}{0.33\textwidth}
\includegraphics[trim=0 0 0 0,width=\textwidth]{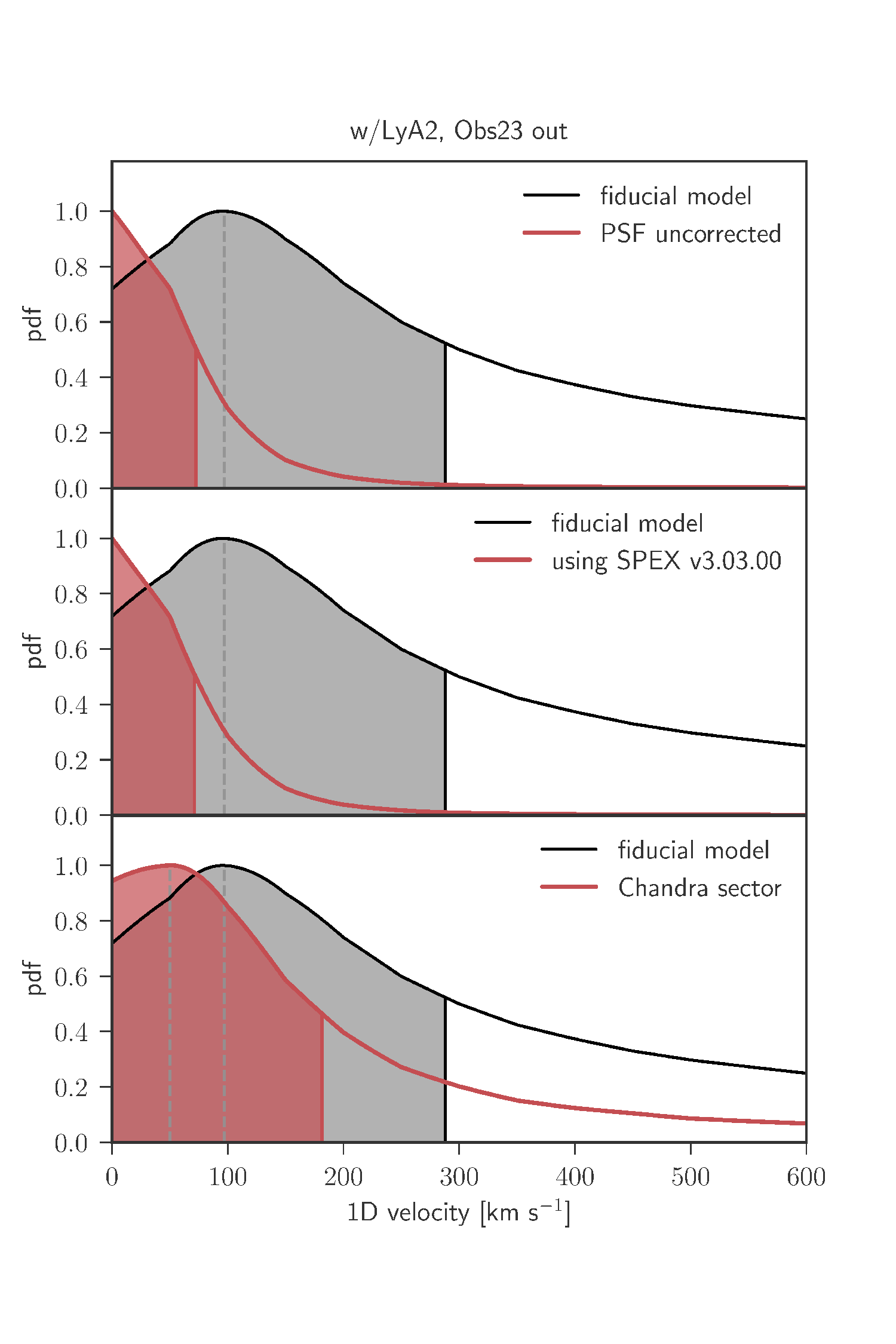}
\end{minipage}
\caption{Same as Figure \ref{fig:app_fig3} for the rest of considered line 
ratios. Results in the obs23\_cen (obs23\_out) are shown on the top (bottom) 
panels.  
\label{fig:app_fig4}
}
\end{figure*}

{\bf Hitomi PSF.} For correct interpretation of the observed line ratios, 
it is important to take the Hitomi PSF into account. 
The sensitivity of the results to the PSF is shown 
in the top panels in figures \ref{fig:app_fig3} and \ref{fig:app_fig4}, 
where the PSF-corrected (default) and PSF-uncorrected results are shown 
in gray and red, respectively. The peaks of the distributions 
for the $w/z$ ratio change by a factor of $\sim$ 2 and by $\sim$ 20\% 
in the central and outer regions, correspondingly. Results 
for the $w/He{\beta}$ ratio are almost unaffected by the PSF 
in contrast to $w/Ly{\alpha}$ ratios. The PSF correction always 
brings the peak velocity closer 
to the directly-measured value \citep{velocity}.

{\bf Model for the optically thin plasma.} Our simulations are based on the
APEC model for an optically thin plasma. However, the line emissivities are
slightly different in the SPEX plasma model \citep{Kaastra96}, 
as discussed in the ``A'' paper \citep{atomic}. 
Though such differences have little affect on the overall parameters 
of the best-fitting spectral models, they might be significant for
more subtle plasma diagnostics such as the RS.
We have therefore implemented the SPEX v3.03.00 model in our simulations 
and redid the analysis. Following the ``A'' paper, 
the ionization balance is set to \citet{Urd17} instead of 
the default one by \citet{Bry09}. The \citet{Urd17} 
calculations provide inner-subshells ionization contributions 
to the spectrum, which are compatible with the SPEX code.

Results based on both plasma models are consistent 
within the uncertainties as shown in figures \ref{fig:app_fig3} and 
\ref{fig:app_fig4} (middle panels). The measured 
$w/z$ line ratio velocity shifts from 
$150^{+80}_{-50}$ km/s ($162^{+78}_{-50}$ km/s) with APEC 
to $125^{+55}_{-48}$ km/s ($119^{+46}_{-36}$ km/s) with SPEX 
in obs23\_cen (obs23\_out), see figure \ref{fig:app_fig3}.

{\bf Spherical symmetry.} Our Perseus model is calculated assuming
spherical symmetry, which will not be correct in detail given the complex
structure of the cluster core. To test this assumption, we re-measured the
thermodynamic properties of the cluster from the Chandra data, limiting 
the analysis to a sector that contains Hitomi
pointings and repeated the analysis described above for this model.
The results for the $w/z$ ratio are essentially unaffected, as shown in 
figure \ref{fig:app_fig3}, bottom panels. The peak of the velocity 
distribution inferred from the $w/Ly{\alpha 2}$ line ratio 
decreases by a factor of $\sim$ 2 in the outer region, 
although the results 
remain consistent within uncertainties (figure \ref{fig:app_fig4}, 
bottom right panel). In all other cases the peak velocity 
changes by even smaller factor, typically by less than 10\%.

Improvements in the results from the $w/z$ line ratio will require 
refinements of the details of the plasma models for these lines. 
SPEX plasma model predicts a stronger RS effect in the cluster core 
than APEC. The $w/He{\beta}$ results, in contrast, are limited by 
statistical uncertainties and can easily be improved with longer, 
Hitomi-like observations.
Further improvements for the $w/Ly{\alpha 1}$ and $w/Ly{\alpha 2}$ 
line ratios will require a more detailed model of the Perseus Cluster, 
especially the contribution of the gas component with $T>5$ keV.

\section{Uncertainties in the atomic excitation rates}
\label{sec:uncertainties}

In this section, systematic uncertainties in the observed line ratios 
arising from uncertainties in calculated atomic parameters are discussed. 
The focus will be on parameters that most directly affect the manifestation 
of RS, i.e., the optical depth of the resonance line, $w$, and the intensity 
ratio $w/z$.

The optical depth at line center is proportional to a line's 
absorption oscillator strength \citep{Zhu13}, which depends on the upper 
level's radiative rate, and in turn, is directly related to the natural 
line width. Hence, as an estimate of the uncertainty of the oscillator 
strength, we compare the natural line width, $\Delta E^{\rm Nat.}$, of 
line $w$ in APEC (AtomDB 3.0.8) and SPEX 3.03 (which are based 
on different atomic 
structure calculations) to the measured line width from 
laboratory measurements by \citet{Rudolph13}. 
The agreement between the measured values and those found in SPEX and APEC 
is good: $\Delta E^{\rm SPEX}=$0.301 eV, 
APEC: $\Delta E^{\rm APEC}=$0.308 eV, 
$\Delta E^{\rm Meas.}=0.311 \pm 0.01$ eV \citep{Rudolph13}. 
Based on this comparison, the systematic error 
in the oscillator strength is estimated to be $<5$\%. 

The error associated with the optically thin intensity ratio $w/z$ is 
more complex. It includes errors in the total collisional excitation 
cross sections, errors associated with unresolved satellites, and 
contributions from charge exchange recombination. 
The dominant excitation mechanism for populating the upper state 
of line $w$ in a thermal plasma is electron impact excitation (EIE) 
from the ground state. The total effective EIE cross sections have been 
measured \citep{Wong95,Hell17} at a few single electron impact energies 
using an electron beam ion trap. These measurements do not include 
contributions from dielectronic satellites. While it is not possible to 
compare the results of the measurements directly to the output of SPEX 
and APEC (because neither model provides cross sections as a function of 
electron impact energy, but rather produces electron temperature dependent, 
unitless collision strengths) it is possible to compare the measurement 
results to the EIE cross sections calculated using the same theoretical 
method used to produce the collision strengths in APEC and SPEX, i.e., 
the methods of \citet{Aggarwal13} and of \citet{Zhang90}, respectively. 
This comparison shows good agreement, i.e., well within the $\sim10$\% 
error of the measurement \citep{Hell17}. Given this agreement, 
and the agreement among calculations, the error on 
the total electron impact excitation rate of line $w$ is estimated 
to be $<10$\%\@. 

The forbidden line $z$ has a significantly more complicated excitation 
structure. As detailed in the ``Atomic'' paper \citep{atomic}, the upper level 
of line $z$ is populated by a variety of mechanisms, including direct 
excitation from the ground state, excitation from cascades, and 
from innershell ionization of Li-like Fe\emissiontype{XXIV}. 
As a result, the uncertainty in the emissivity of line $z$ is coupled to 
detailed population kinetics, in addition to the plasma model. 
The excitation cross section has been measured \citep{Wong95} 
using an electron beam ion trap and the agreement 
with theory is good. 
However, this is insufficient to estimate the total error 
in the line strength for a 4 keV thermal plasma. 

\begin{figure}[tbh]
 \begin{center}
  \includegraphics[width=8cm]{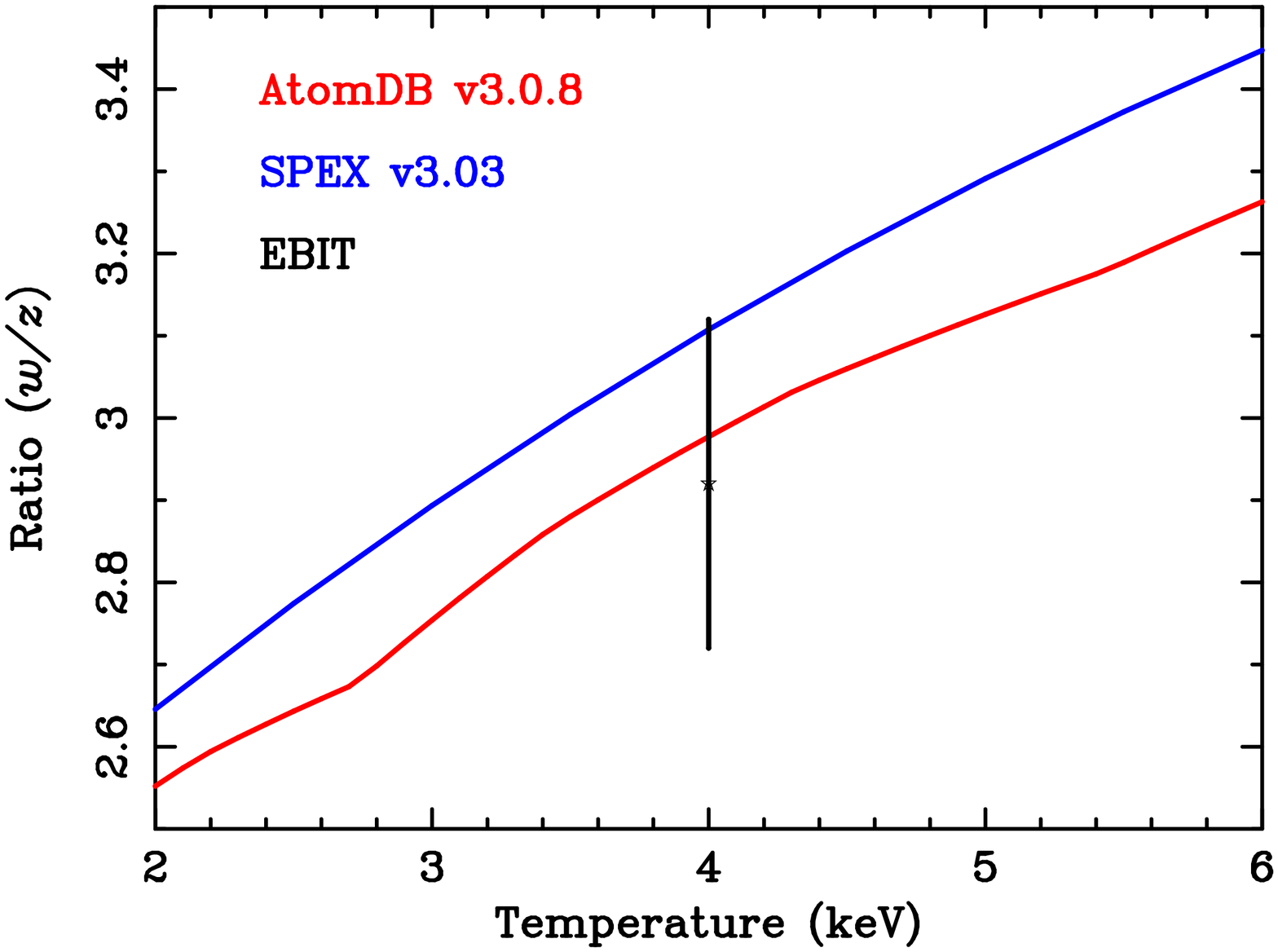}
 \end{center}
\caption{
Temperature dependence of the $w/z$ ratio for AtomDB version 3.0.8, 
SPEC v3.03, and EBIT measurement.
}
\label{fig:diff-w2z}
\end{figure}

To estimate the errors associated with $w/z$ for a 4 keV thermal, 
optically thin plasma directly, the values predicted by AtomDB and SPEX 
are compared to results of laboratory measurement from plasmas with 
Maxwellian electron temperatures at coronal densities. 
Here, the calculations are compared to measurement using the Maxwellian 
simulator mode \citep{Savin00,Savin08} employed at the LLNL EBIT-I 
electron beam ion trap facility. Using this mode, the spectrum of 
the He$\alpha$ complex of Fe\emissiontype{XXV} including satellites 
has been measured \citep{Gu12}. Caveats of the measurement include 
the fact that the average charge balance produced using the Maxwellian 
simulator is underionized (and thus the charge balance is not 
representative of a true Maxwellian) and the fact that line emission produced 
in an EBIT is, in general, polarized, hence, when comparing to line ratios 
measured form celestial sources, polarization effects must be taken into 
account. The amount of polarization depends on the electron impact energy. 
Hence, the line emission measured using EBIT-I's 
Maxwellian mode, 
where the electron beam energy is swept across a large range, 
may have a range of polarization values. The calculated polarization 
of line $w$ ranges from 0.6 to 0.4 between the 
threshold for excitation 
and an electron impact energy of 24 keV. The polarization of line $z$ 
is $\sim -0.08$ near threshold and $-0.22$ above threshold for population 
by cascades \citep{Hakel07}. 
The agreement
with theory is good although the uncertainty in the 
measurements remains relatively large, 
i.e., on the order of  20--30\% 
\citep{Beiersdorfer96,Hakel07}.

Here, polarization effects are taken into account by setting the 
polarization of line $w$ to $P=0.5$ and of line $z$ to $P=-0.22$. 
No energy dependence is included because the correction factor to 
the $w/z$ line ratio across the entire range of polarization only varies 
by $\sim 5$\%\@, and because the true polarization of lines $z$ and $w$ 
is not known due to depolarization effects \citep{Gu99}. 
The difference between the polarization corrected ratio and uncorrected 
ratio is 11\%, i.e., the polarization-corrected $w/z$ ratio is 
$2.92 \pm 0.2$ and the uncorrected ratio is $3.25 \pm 0.07$.  
The errors in these ratios include the uncertainty 
in the polarization, the spectrometer response, and statistics.  

The measured ratio is also systematically lower than the ratio for 
a true Maxwellian because the large amount of Li-like Fe\emissiontype{XXIV} 
present in the EBIT results in a larger contribution to $z$ from 
innershell ionization. The measured ratio has to be corrected
based on a comparison of the ionization balance in the EBIT in its Maxwellian 
simulator mode with the true thermal ionization balance in a 4 keV plasma, 
and the fraction of the emissivity in $z$ due to innershell ionization.
However, according to calculations, 
for a 4 keV plasma, the contribution from inner shell ionization 
is only a $\sim 10$\% enhancement of line $z$ in the EBIT case. 
While the measured value at EBIT is therefore $w/z = 2.92 \pm 0.2$, 
the corrected ratio may be as high as 
$\approx 1.1 \times 2.92 = 3.21 \pm 0.2$. 
This value is fully consistent with the ratio at 
$kT = 4$ keV in APEC of $w/z = 2.98$ and SPEX v3.03 of $w/z=3.11$ 
(see figure \ref{fig:diff-w2z}).
It should be noted that the Fe\emissiontype{XXV} He$\alpha$ complex, 
as well as other He-like systems, have been measured in tokamak 
plasmas where no polarization effects are present 
\citep{Bitter08}; however, those measurements are at lower temperatures. 
Comparison of the lower temperature data with the predictions 
from SPEX and APEC would be useful. 
 
In a 4 keV plasma, unresolved dielectronic recombination (DR) satellites 
contribute to the flux of line $w$ and line $z$, and uncertainties 
in their contributions to $w$ and $z$ should be estimated.  
In the case of line $w$, the emission from high-$n$ DR ($n \geq 3$)  
satellites blends with line $w$. These satellite intensities have been 
measured in detail by \citet{Beiersdorfer92a} and also by 
\citet{Watanabe01}. The agreement
between theory and experiment is good \citep{Watanabe01}.
We can get a rough estimate of the uncertainty in the intensity of $w$ 
due to uncertainties in the intensities of the unresolved 
satellites as follows. In the ``Atomic'' paper, 
calculated satellite 
intensities are listed in Table 11, where calculations from three 
separate plasma models are compared, assuming collisional equilibrium 
at $kT = 4$ keV. Four $n=3$ satellites have a summed intensity of 0.065 
of the intensity of $w$. The dispersion between the models has a standard 
deviation of about 10\% of this relative intensity. 
Likewise, three $n=4$ satellites have a summed intensity relative to $w$ 
of 0.018, with a standard deviation of about 10\% of this value. 
The dispersion between the models compares well to the intensity measurement 
error of about 10\% quoted by \citet{Watanabe01}. From these data 
we estimate that the uncertainty in the intensity of $w$ due to uncertainties 
in the unresolved satellites is of order 1--2\% (allowing for the presence 
of weaker $n > 4$ satellites).

In the case of line $z$, the most significant contribution is from 
the DR channel of 
the $1s2p^2\ ^2D_{5/2} \rightarrow 1s^2 2p\ ^2P_{3/2}$
transition, 
known as line $j$ \citep{Gabriel72}, which, although it is $\sim 8$ eV 
away from line $z$, is still only marginally separated 
in the Perseus spectrum. In addition, there is also emission from 
the $1s2p^2\ ^2D_{3/2} \rightarrow 1s^2 2p\ ^2P_{3/2}$,
known as line $l$ \citep{Gabriel72}, which is only about 1.5 eV above $z$, 
but its strength is only about 1/10 of the strength of line $j$.  
The strengths of both line $j$ and line $l$ have been measured 
\citep{Beiersdorfer92b}, with an estimated accuracy of about $\lesssim 20$\%. 
The agreement with theoretical calculations 
is better than 15\%. 
In collisional equilibrium at $kT = 4$ keV, the intensities of $j$ and $l$ 
relative to $z$ are approximately 0.33 and 0.025, respectively. 
The uncertainty in the intensities of $j$ and $l$, if completely unresolved 
from $z$, therefore translates into an uncertainty of about 6.5\% 
in the estimated intensity of $z$; in practice, 
the dominant contributor, $j$, is partially resolved from $z$ in our data, 
so the uncertainty due to the contribution of the satellites is 
significantly smaller than 6.5\%. If we assign an uncertainty of 3\% 
to $z$ due to the uncertainty in the satellite contribution, 
and 2\% to $w$, the ratio $w/z$ has an uncertainty of about 4\% due to 
possible errors in the satellite contribution. 

Finally, our model for the emissivity of Fe\emissiontype{XXV} does not 
contain excitation by charge exchange of neutral H with Fe\emissiontype{XXVI}.
The core region of the cluster does contain neutral H in close contact 
with the hot ICM, as is evident from the filamentary H$\alpha$ emission 
from the core \citep{Salome11}.
Charge exchange contributes to the intensity of both the $w$ and the $z$ 
lines, but it contributes more strongly to $z$ than to $w$, and 
the effect therefore would tend to reduce the ratio $w/z$ from its pure 
CIE value. In the ``Atomic'' paper, an explicit model 
for the charge exchange emission is fitted along with a CIE model. 
Based on the best-fit parameters for the charge exchange 
(which are constrained mainly by the charge-exchange predicted intensities 
of higher-order series members),
we estimate that the process could contribute 8\% of the observed flux 
in $z$, and 2\% of the observed flux in $w$. That would lower the ratio 
$w/z$ by 6\% from its value in CIE in the central region of the cluster, 
which is significantly less than the suppression we observe. Moreover, 
charge exchange is not likely to contribute much to the emission
we observe in the $1-2$ arcmin range, where $w/z$ is also significantly 
suppressed with respect to its value in CIE.

To summarize, when we compare the measured $w/z$ ratios to the ratios 
predicted by the best-fitting CIE model without radiative transfer, 
we see a significant suppression in the innermost
regions of the cluster. In the innermost region, the measured ratio 
is $w/z = 2.43$, while the predicted ratio is 2.98 (AtomDB 3.0.8) 
or 3.11 (SPEX 3.03), and the ratio measured
at EBIT, corrected for systematic errors, is at least $2.92 \pm 0.2$. 
Errors in the model fluxes of unresolved satellite lines to $w$ and $z$ 
cannot account for more than a few percent of this suppression. 
Charge exchange excitation could account for a suppression of 6\%, 
but only in the innermost ($r < 1$ arcmin) bin.
As is expected if the suppression is due to resonance scattering of 
$w$ photons, $w/z$ tends to the 'optically thin' CIE value 
in the outer regions of the field, and this would not be the case 
if the optically thin, CIE only, value we use were simply incorrect. 

Comparing the results of the measured $w/z$ ratio from obs23 given 
in table \ref{tab:ratios} with the values predicted for an optically 
thin plasma, and taking into account known uncertainties in the predicted 
values, there is very good evidence for RS in line $w$.

\section{Conclusions}
\label{sec:conclusions}

We have showed evidence for the resonance scattering  
in the core of the Perseus Cluster observed with Hitomi. 
Namely, we observe: i) the characteristic suppression of the flux of 
the resonance line in the Fe\emissiontype{XXV} He$\alpha$ complex seen 
towards the center of the cluster; ii) the expected decrease of this 
suppression with distance from the cluster center; and 
iii) an additional broadening of the resonance line compared to 
other lines from the same ion. Fitting the spectra with a combination 
of an emission model for optically thin plasma supplemented with 
individual Gaussian emission lines, we measure the ratios of the resonance 
line flux to the fluxes in the forbidden line in the Fe\emissiontype{XXV} 
He$\alpha$ complex, two Fe\emissiontype{XXV} He$\beta$ lines, and 
the Fe\emissiontype{XXVI} Ly$\alpha$ lines. 
To interpret the observed results, we perform radiative transfer 
Monte Carlo simulations, assuming a spherically symmetric model 
for the cluster and plausible velocity fields based on direct velocity 
measurements \citep{velocity} including an isotropic field. 
Comparing the observed line ratios and the simulated values, we infer 
velocities of gas motions that are consistent with direct velocity 
measurements from line broadening. 
We investigate systematic uncertainties 
in the analysis, including the assumption of spherical symmetry, the  
modeling of the ICM properties, the uncertainties in line emissivities, 
and the contribution of charge exchange excitation. 

Future, non-dispersive high resolution spectroscopy such as the Hitomi SXS 
observations will allow us to explore the effect of RS in even more detail, 
which, in combination with the direct velocity measurements, will provide 
us with a unique tool to probe the anisotropy and spatial scales of 
gas motions. 
It is important to take the RS effect into account when measuring 
plasma properties from high resolution X-ray spectra of galaxy clusters. 
The effect can be even stronger in lower-energy lines 
in cooler, gas-rich systems, such as galaxy groups and large 
elliptical galaxies.

Hitomi's lifetime was unfortunately short. However, 
the micro-calorimeter at the heart of the SXS has already provided 
new insights with its high energy resolution. Future X-ray missions 
with micro-calorimeters, XARM and Athena, will be indispensable in the 
investigation of cluster physics.

\begin{trueauthors}
K. Sato is the sub leader of the Perseus RS team, led the data analysis, 
numerical simulations and prepared the manuscript.
He also contributed to the SXS hardware development, integration tests, 
launch campaign, in-orbit operation, and calibration. 
I. Zhuravleva is the sub leader of the RS team and led the data analysis, 
numerical simulations and prepared the manuscript. 
F. Paerels is the leader of the RS team, led the study of uncertainties 
in the atomic excitation rate and prepared the manuscript.
M. Furukawa, M. Ohno, K. Matsushita, and Y. Fukazawa performed numerical 
simulations with {\tt Geant4} and participated in discussions. 
M. Ohno, Y. Fukazawa, and A. Furuzawa contributed the Hitomi hardware and 
in-orbit operation.
A. Ogorzalek performed numerical simulations with the ICMMC code and 
studied the uncertainties of the plasma codes.
G. Brown, A. Foster, L. Gu, and M. Leutenegger studied the uncertainties 
of the atomic excitation rates and codes.
M. Eckart, C. Kilbourne, and M. Leutenegger contributed to the SXS hardware 
design, development, integration tests, launch campaign, and calibration.
S. Nakashima, T. Sasaki and H. Yamaguchi helped with the modeling of the 
ICM structure in data analysis.
S. Allen, C. Kilbourne, C. Pinto, M. Tsujimoto, M. Ozaki, and A. Simionescu 
helped to improve the manuscript.

\end{trueauthors}

\begin{ack}
We thank the support from the JSPS Core-to-Core Program.
We acknowledge all the JAXA members who have contributed to the ASTRO-H (Hitomi)
project.
All U.S. members gratefully acknowledge support through the NASA Science Mission
Directorate. SLAC members acknowledge support via DoE contract to SLAC
National Accelerator Laboratory DE-AC3-76SF00515. Part of this work was performed under
the auspices of the U.S. DoE by LLNL under Contract DE-AC52-07NA27344.
Support from the European Space Agency is gratefully acknowledged.
French members acknowledge support from CNES, the Centre National d'\'{E}tudes Spatiales.
SRON is supported by NWO, the Netherlands Organization for Scientific Research.  Swiss
team acknowledges support of the Swiss Secretariat for Education, Research and
Innovation (SERI).
The Canadian Space Agency is acknowledged for the support of Canadian members.  
We acknowledge support from JSPS/MEXT KAKENHI grant numbers 15J02737,
15H00773, 15H00785, 15H02090, 15H03639, 15H05438, 15K05107, 15K17610,
15K17657, 16J00548, 16J02333, 16H00949, 16H06342, 16K05295, 16K05296,
16K05300, 16K13787, 16K17672, 16K17673, 21659292, 23340055, 23340071,
23540280, 24105007, 24244014, 24540232, 25105516, 25109004, 25247028,
25287042, 25400236, 25800119, 26109506, 26220703, 26400228, 26610047,
26800102, JP15H02070, JP15H03641, JP15H03642, JP15H06896,
JP16H03983, JP16K05296, JP16K05309, JP16K17667, and JP16K05296.
The following NASA grants are acknowledged: NNX15AC76G, NNX15AE16G, NNX15AK71G,
NNX15AU54G, NNX15AW94G, and NNG15PP48P to Eureka Scientific.
H. Akamatsu acknowledges support of NWO via Veni grant.  
C. Done acknowledges STFC funding under grant ST/L00075X/1.  
A. Fabian and C. Pinto acknowledge ERC Advanced Grant 340442.
P. Gandhi acknowledges JAXA International Top Young Fellowship and UK Science and
Technology Funding Council (STFC) grant ST/J003697/2. 
Y. Ichinohe, K. Nobukawa, and H. Seta are supported by the Research Fellow of JSPS for Young
Scientists.
N. Kawai is supported by the Grant-in-Aid for Scientific Research on Innovative Areas
``New Developments in Astrophysics Through Multi-Messenger Observations of Gravitational
Wave Sources''.
S. Kitamoto is partially supported by the MEXT Supported Program for the Strategic
Research Foundation at Private Universities, 2014-2018.
B. McNamara and S. Safi-Harb acknowledge support from NSERC.
T. Dotani, T. Takahashi, T. Tamagawa, M. Tsujimoto and Y. Uchiyama acknowledge support
from the Grant-in-Aid for Scientific Research on Innovative Areas ``Nuclear Matter in
Neutron Stars Investigated by Experiments and Astronomical Observations''.
N. Werner is supported by the Lend\"ulet LP2016-11 grant from the Hungarian Academy of
Sciences.
D. Wilkins is supported by NASA through Einstein Fellowship grant number PF6-170160,
awarded by the Chandra X-ray Center, operated by the Smithsonian Astrophysical
Observatory for NASA under contract NAS8-03060.

We thank contributions by many companies, including in particular, NEC, Mitsubishi Heavy
Industries, Sumitomo Heavy Industries, and Japan Aviation Electronics Industry. Finally,
we acknowledge strong support from the following engineers.  JAXA/ISAS: Chris Baluta,
Nobutaka Bando, Atsushi Harayama, Kazuyuki Hirose, Kosei Ishimura, Naoko Iwata, Taro
Kawano, Shigeo Kawasaki, Kenji Minesugi, Chikara Natsukari, Hiroyuki Ogawa, Mina Ogawa,
Masayuki Ohta, Tsuyoshi Okazaki, Shin-ichiro Sakai, Yasuko Shibano, Maki Shida, Takanobu
Shimada, Atsushi Wada, Takahiro Yamada; JAXA/TKSC: Atsushi Okamoto, Yoichi Sato, Keisuke
Shinozaki, Hiroyuki Sugita; Chubu U: Yoshiharu Namba; Ehime U: Keiji Ogi; Kochi U of
Technology: Tatsuro Kosaka; Miyazaki U: Yusuke Nishioka; Nagoya U: Housei Nagano;
NASA/GSFC: Thomas Bialas, Kevin Boyce, Edgar Canavan, Michael DiPirro, Mark Kimball,
Candace Masters, Daniel Mcguinness, Joseph Miko, Theodore Muench, James Pontius, Peter
Shirron, Cynthia Simmons, Gary Sneiderman, Tomomi Watanabe; ADNET Systems: Michael
Witthoeft, Kristin Rutkowski, Robert S. Hill, Joseph Eggen; Wyle Information Systems:
Andrew Sargent, Michael Dutka; Noqsi Aerospace Ltd: John Doty; Stanford U/KIPAC: Makoto
Asai, Kirk Gilmore; ESA (Netherlands): Chris Jewell; SRON: Daniel Haas, Martin Frericks,
Philippe Laubert, Paul Lowes; U of Geneva: Philipp Azzarello; CSA: Alex Koujelev, Franco
Moroso.

\end{ack}


\appendix

\section{Differences between AtomDB version 3.0.8 and 3.0.9}
\label{sec:v309}

\begin{figure*}[h]

\begin{minipage}{0.33\textwidth}
\FigureFile(\textwidth,\textwidth){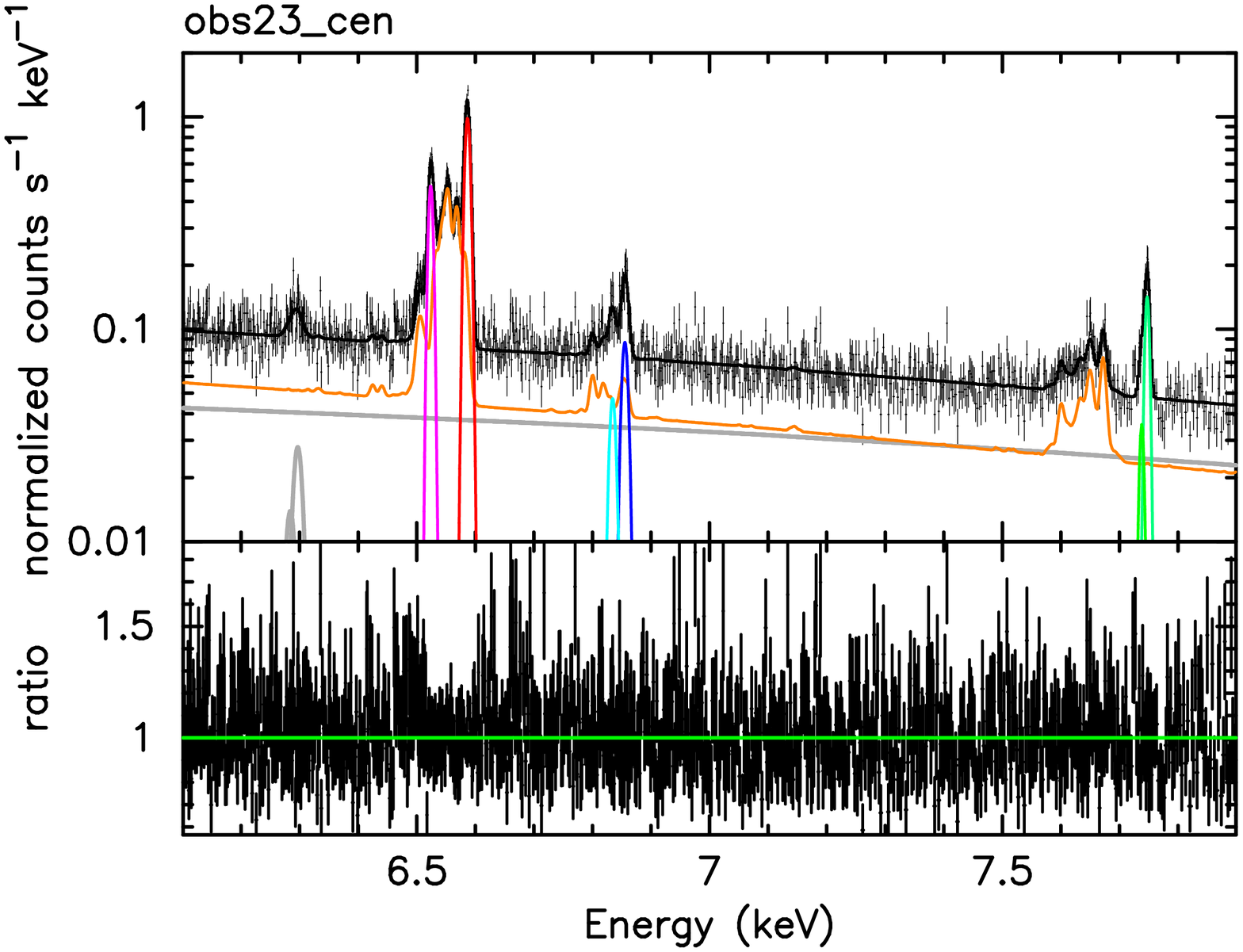}
\end{minipage}\hfill
\begin{minipage}{0.33\textwidth}
\FigureFile(\textwidth,\textwidth){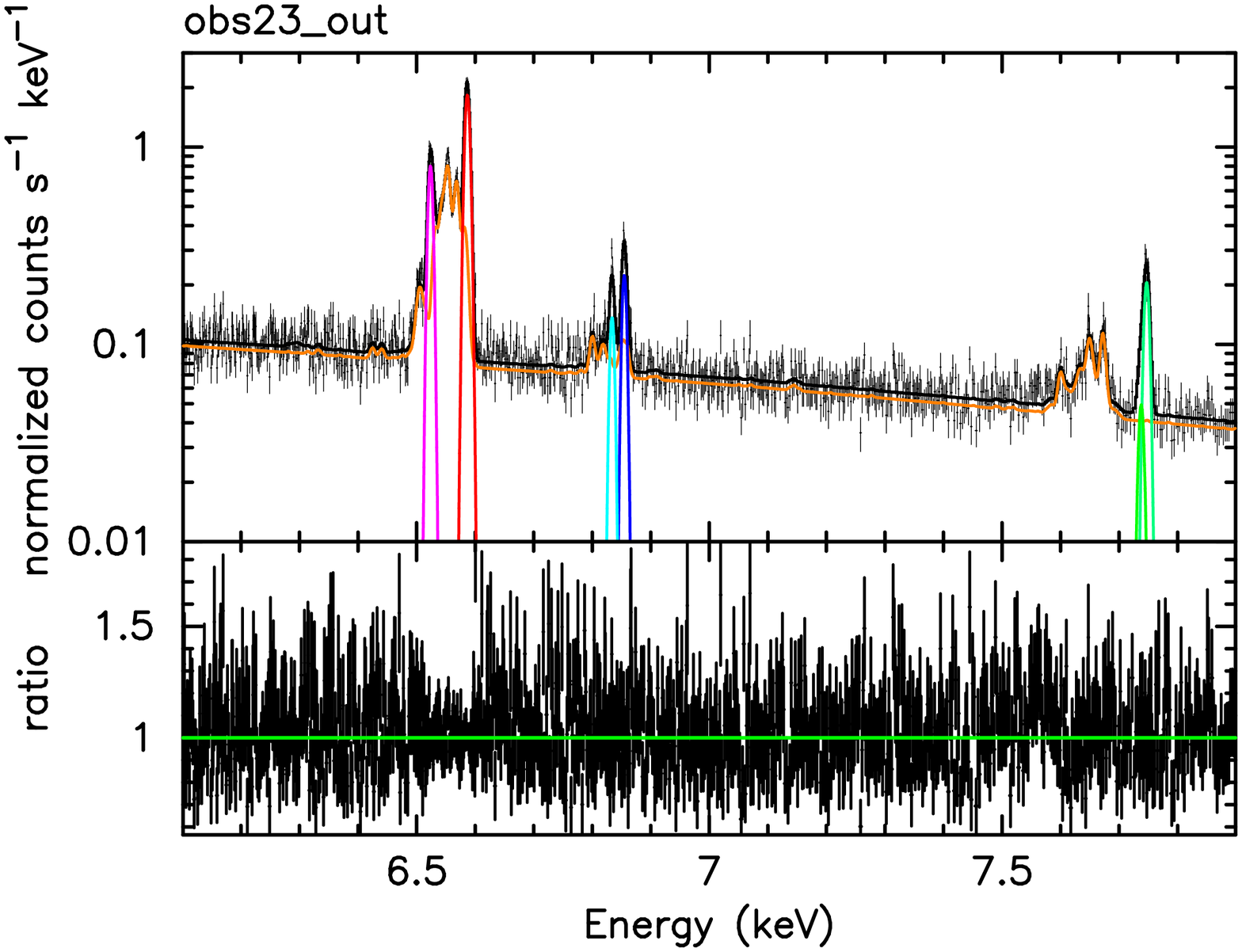}
\end{minipage}\hfill
\begin{minipage}{0.33\textwidth}
\FigureFile(\textwidth,\textwidth){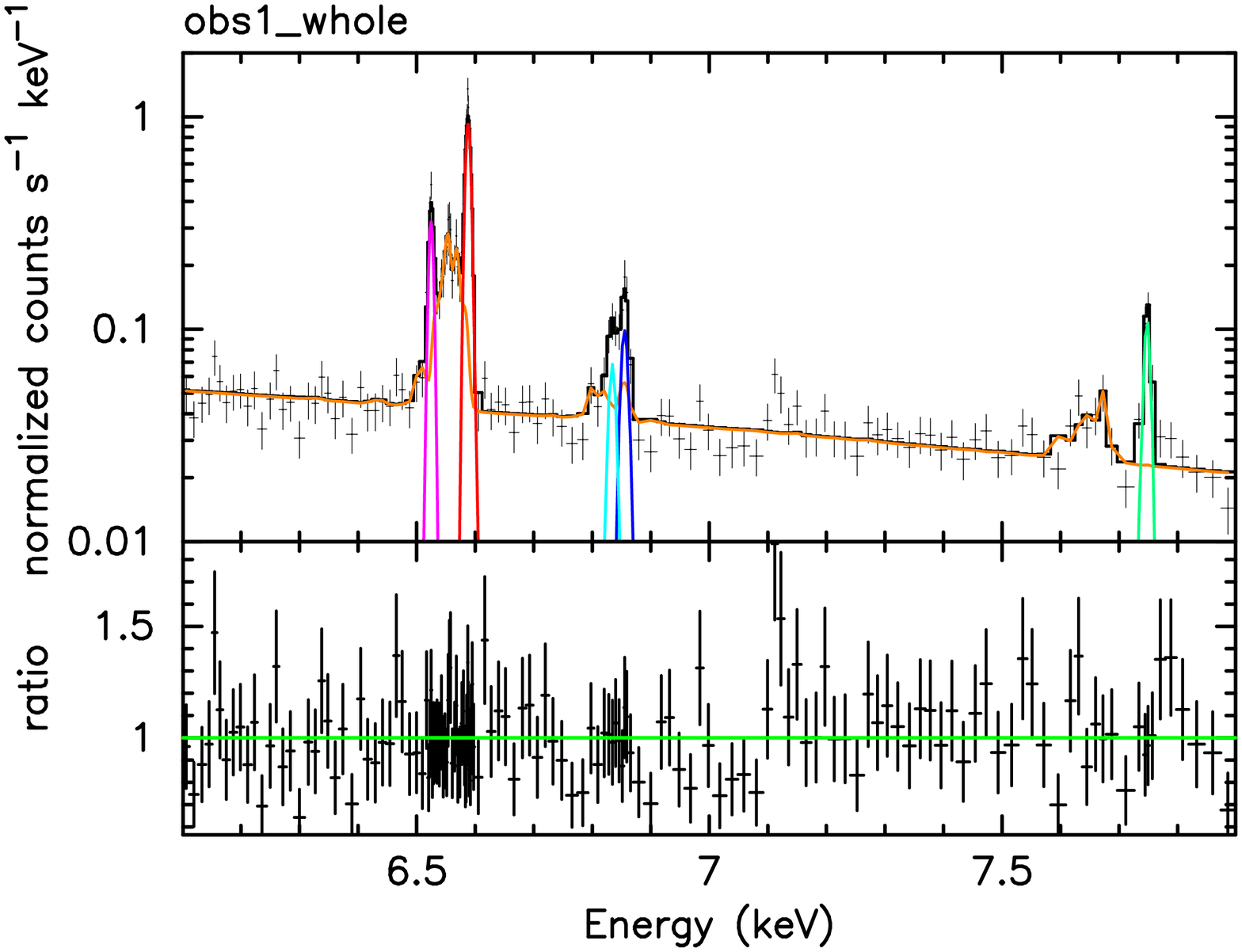}
\end{minipage}

\begin{minipage}{0.33\textwidth}
\FigureFile(\textwidth,\textwidth){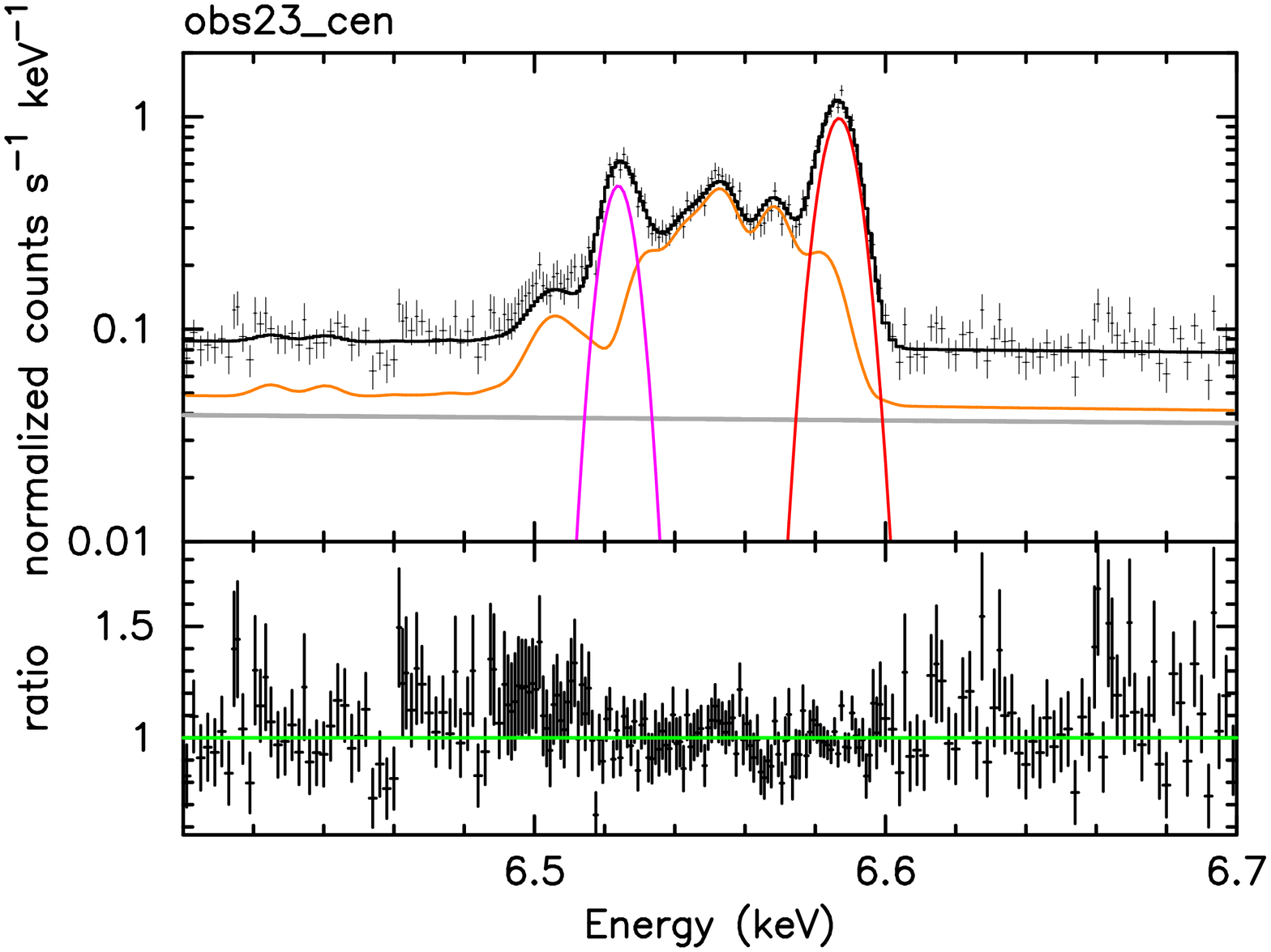}
\end{minipage}\hfill
\begin{minipage}{0.33\textwidth}
\FigureFile(\textwidth,\textwidth){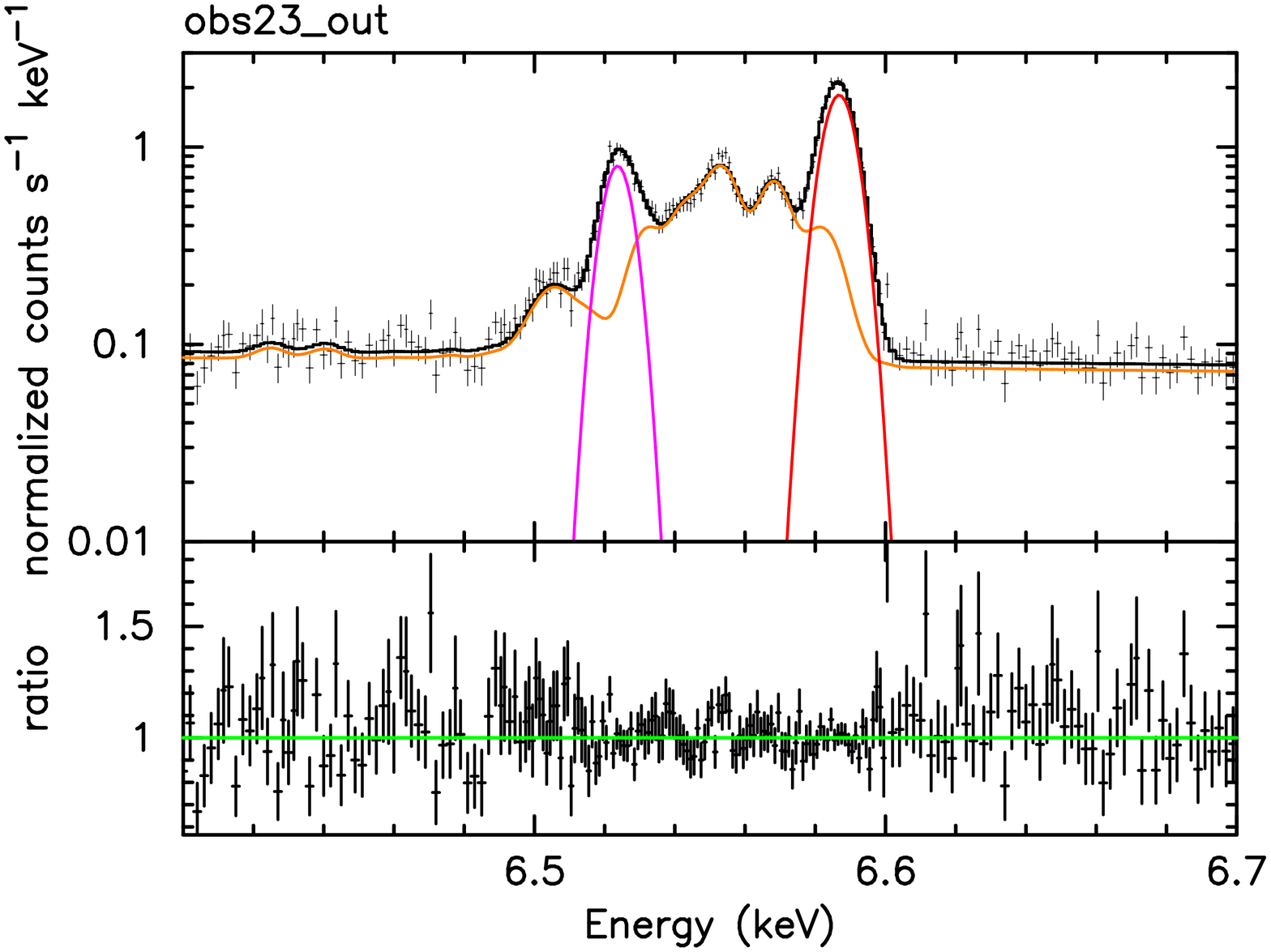}
\end{minipage}\hfill
\begin{minipage}{0.33\textwidth}
\FigureFile(\textwidth,\textwidth){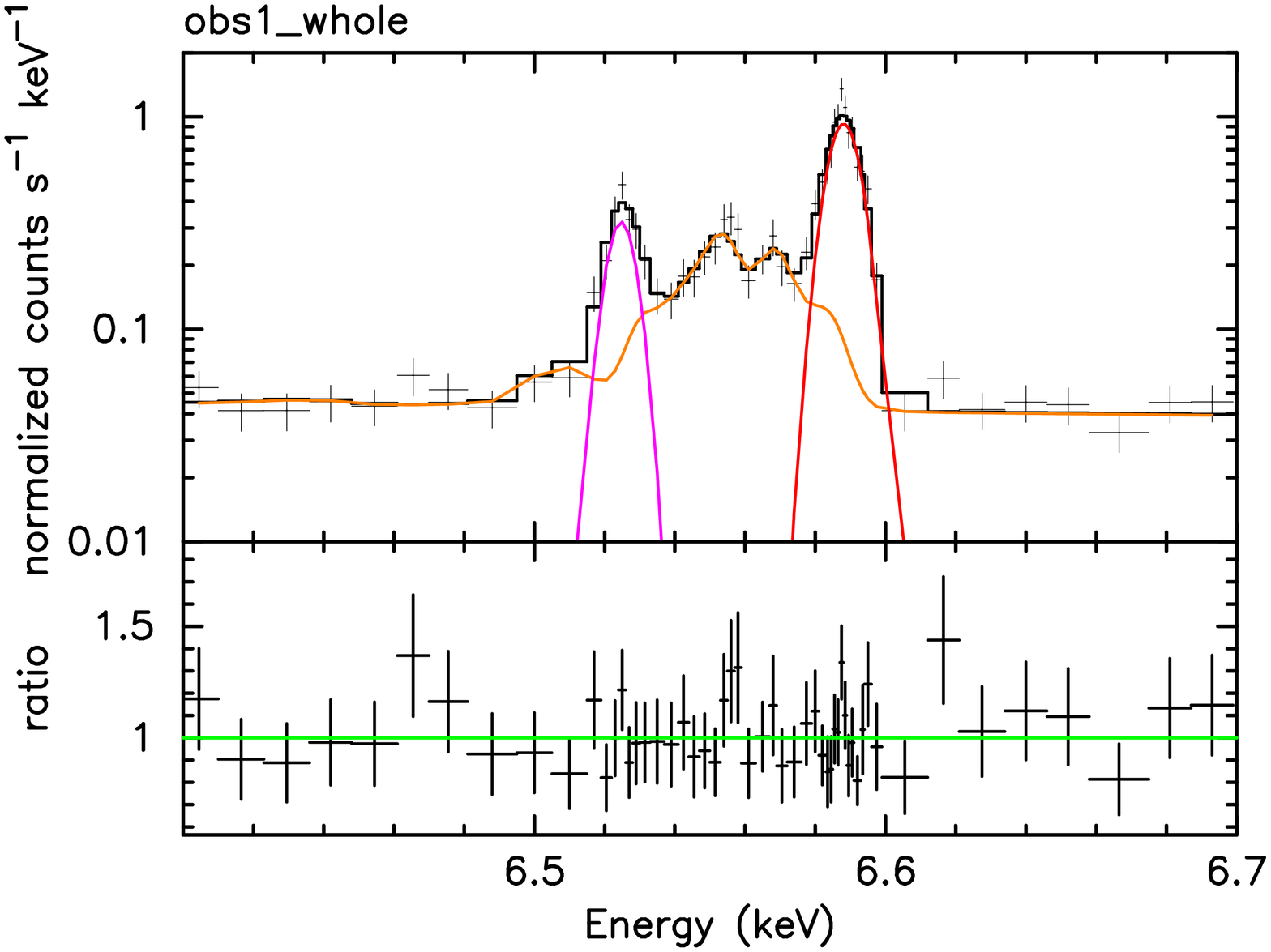}
\end{minipage}

\begin{minipage}{0.33\textwidth}
\FigureFile(\textwidth,\textwidth){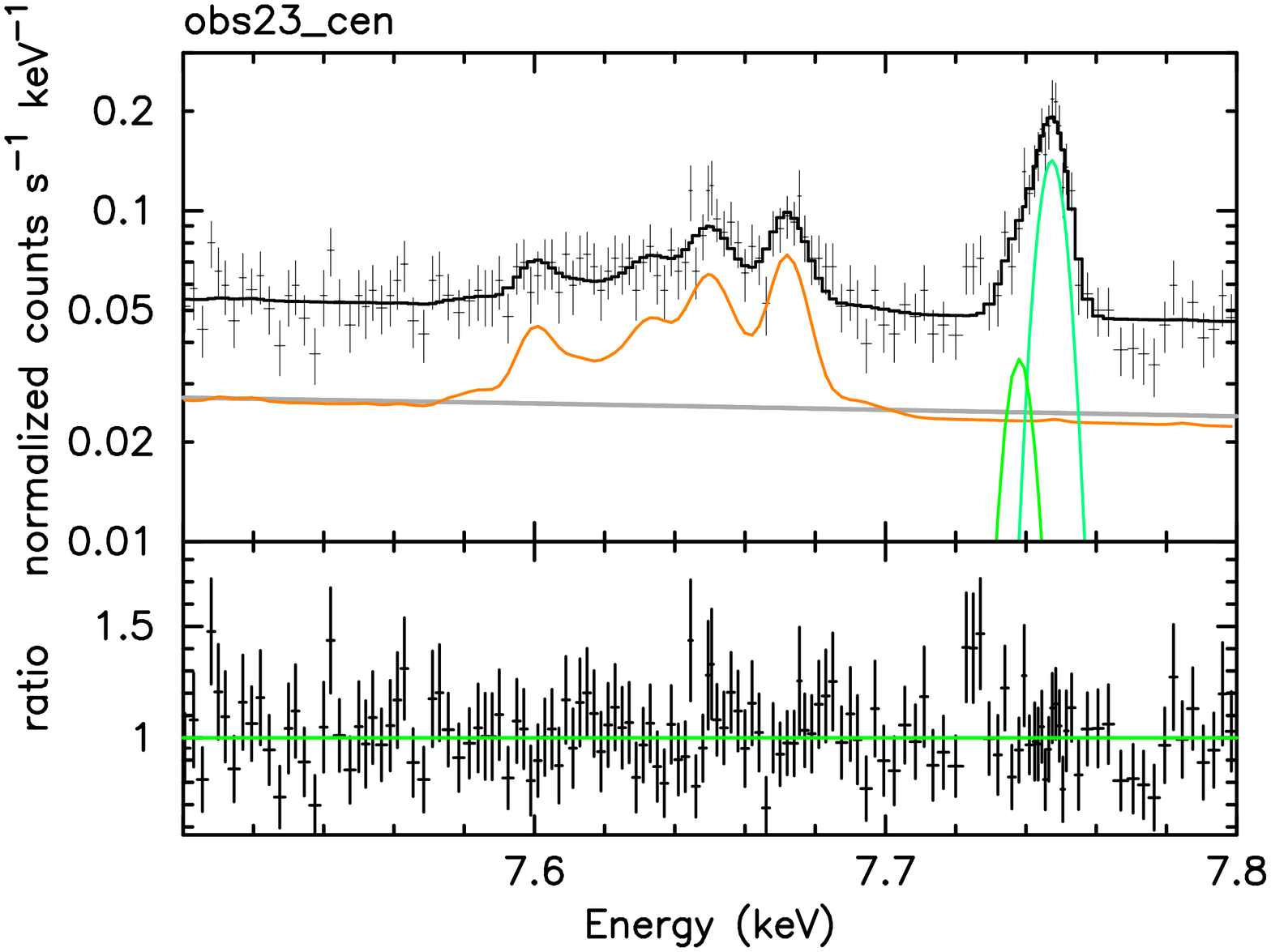}
\end{minipage}\hfill
\begin{minipage}{0.33\textwidth}
\FigureFile(\textwidth,\textwidth){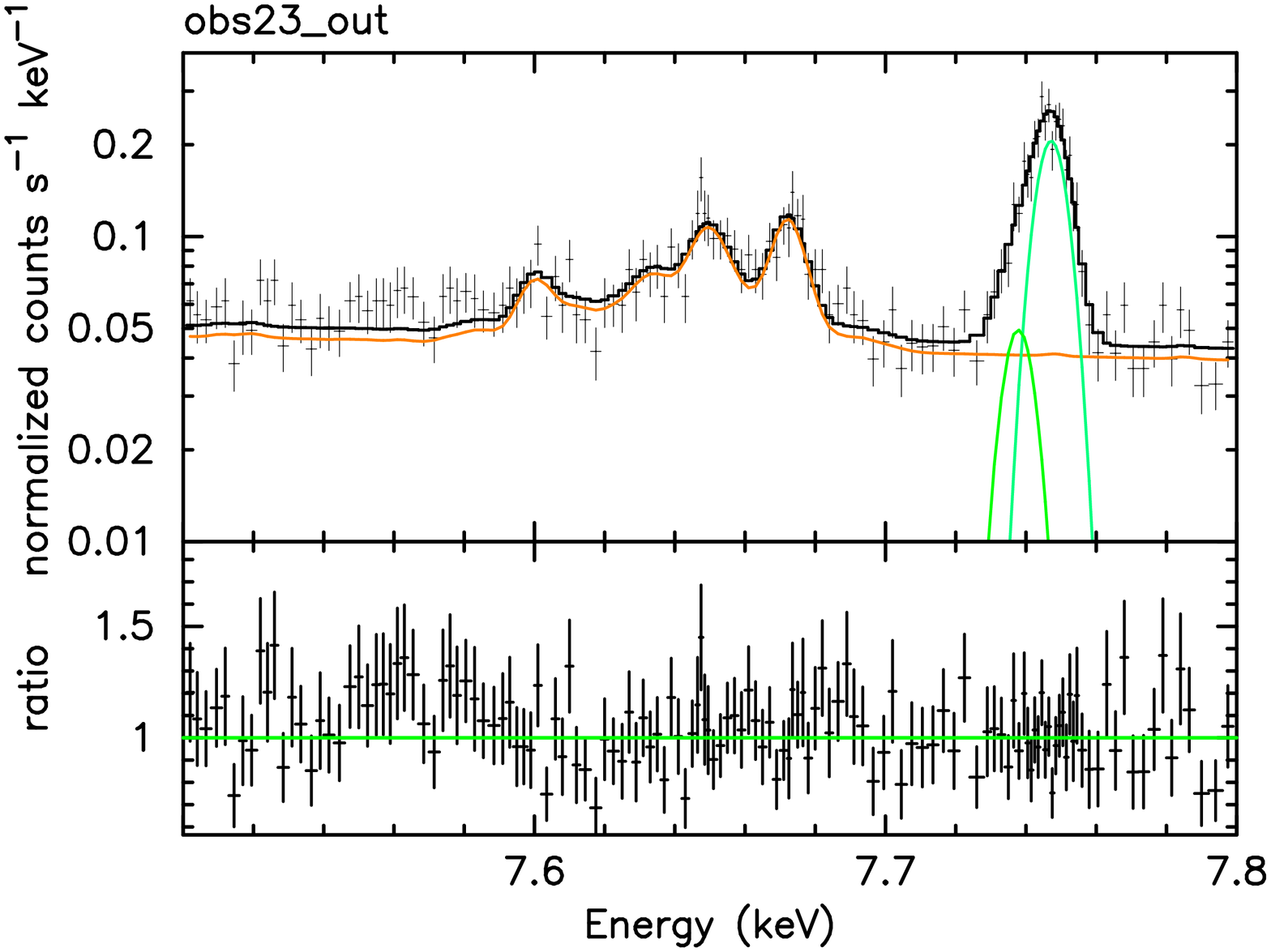}
\end{minipage}\hfill
\begin{minipage}{0.33\textwidth}
\FigureFile(\textwidth,\textwidth){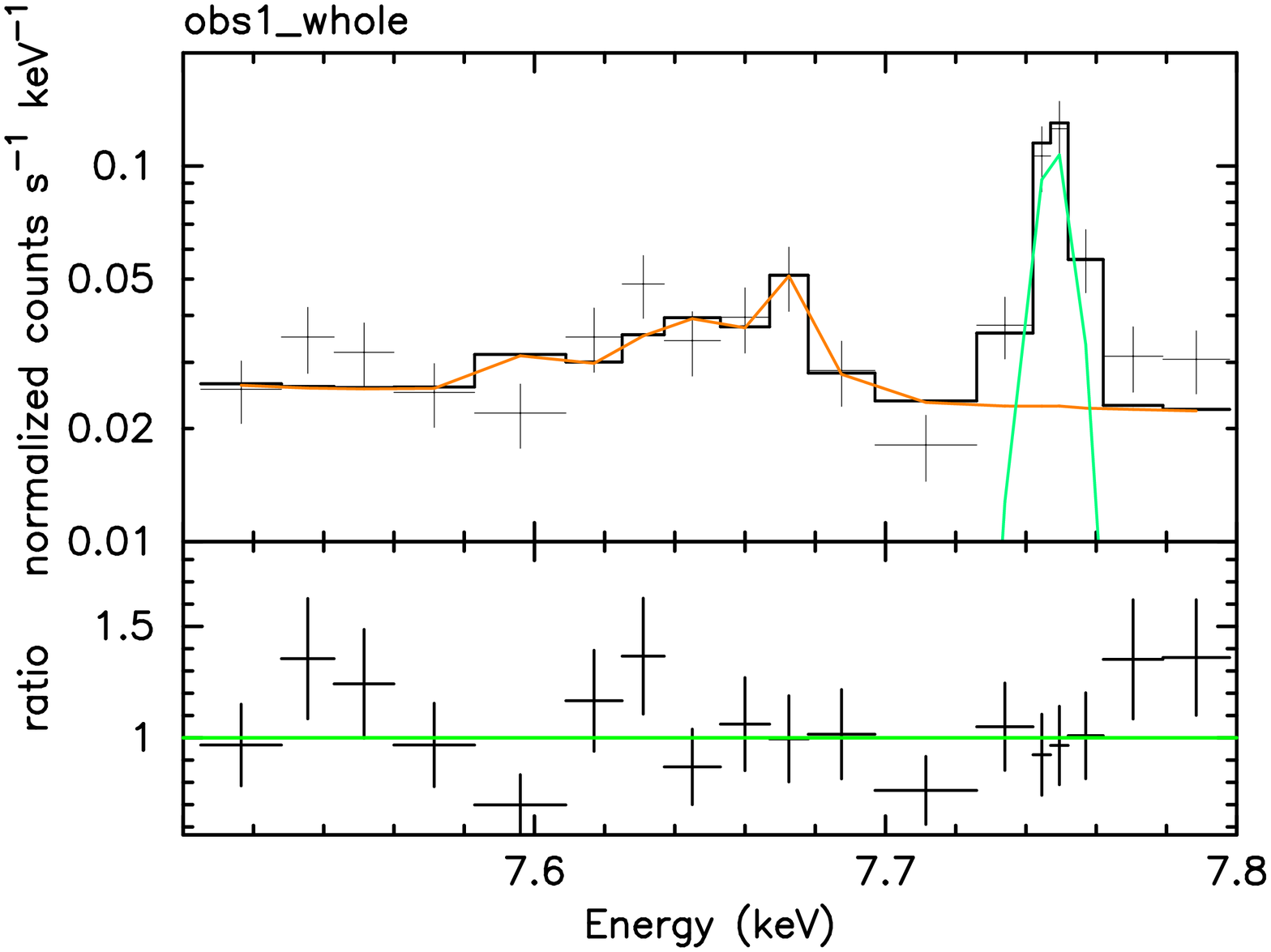}
\end{minipage}

\begin{minipage}{0.33\textwidth}
\FigureFile(\textwidth,\textwidth){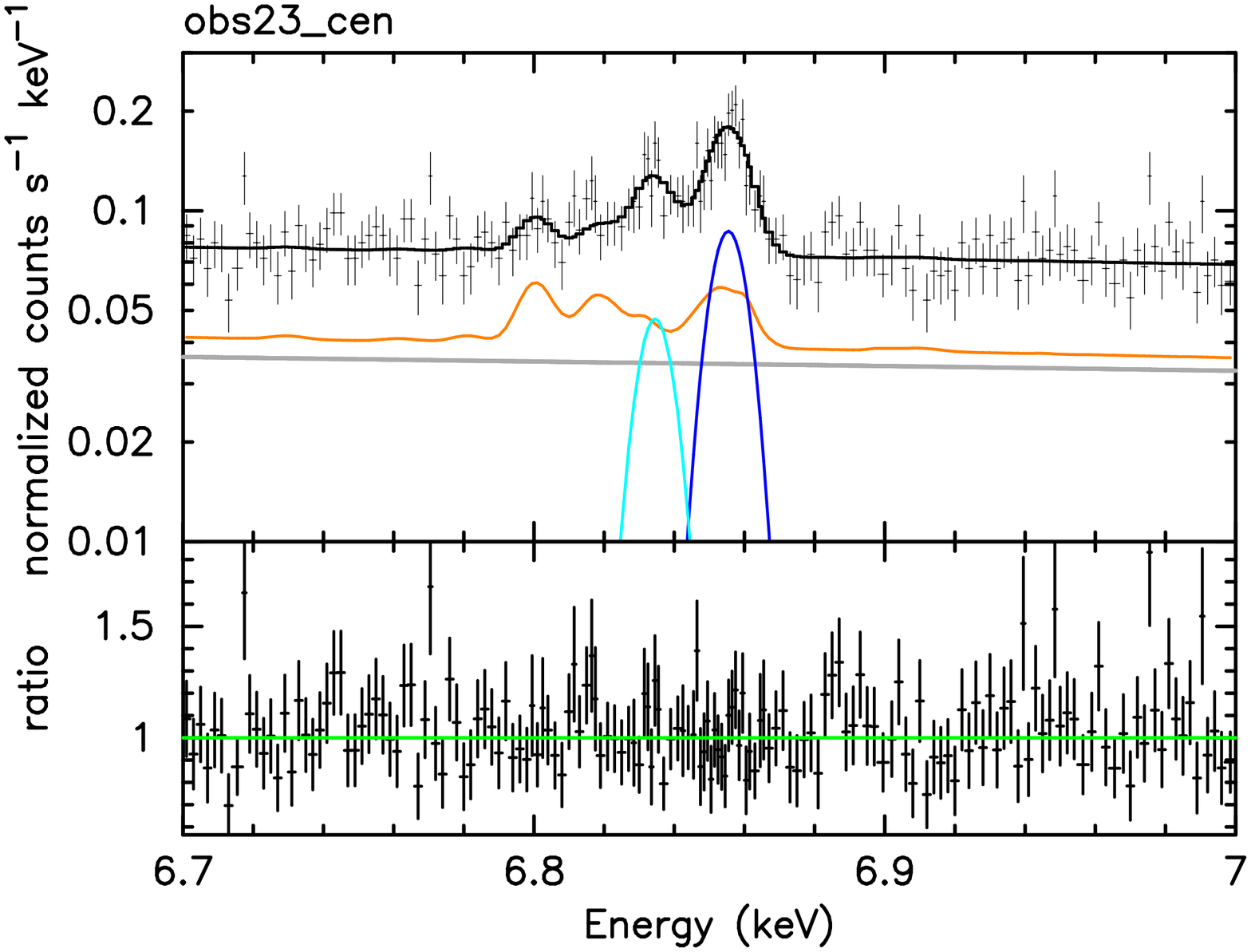}
\end{minipage}\hfill
\begin{minipage}{0.33\textwidth}
\FigureFile(\textwidth,\textwidth){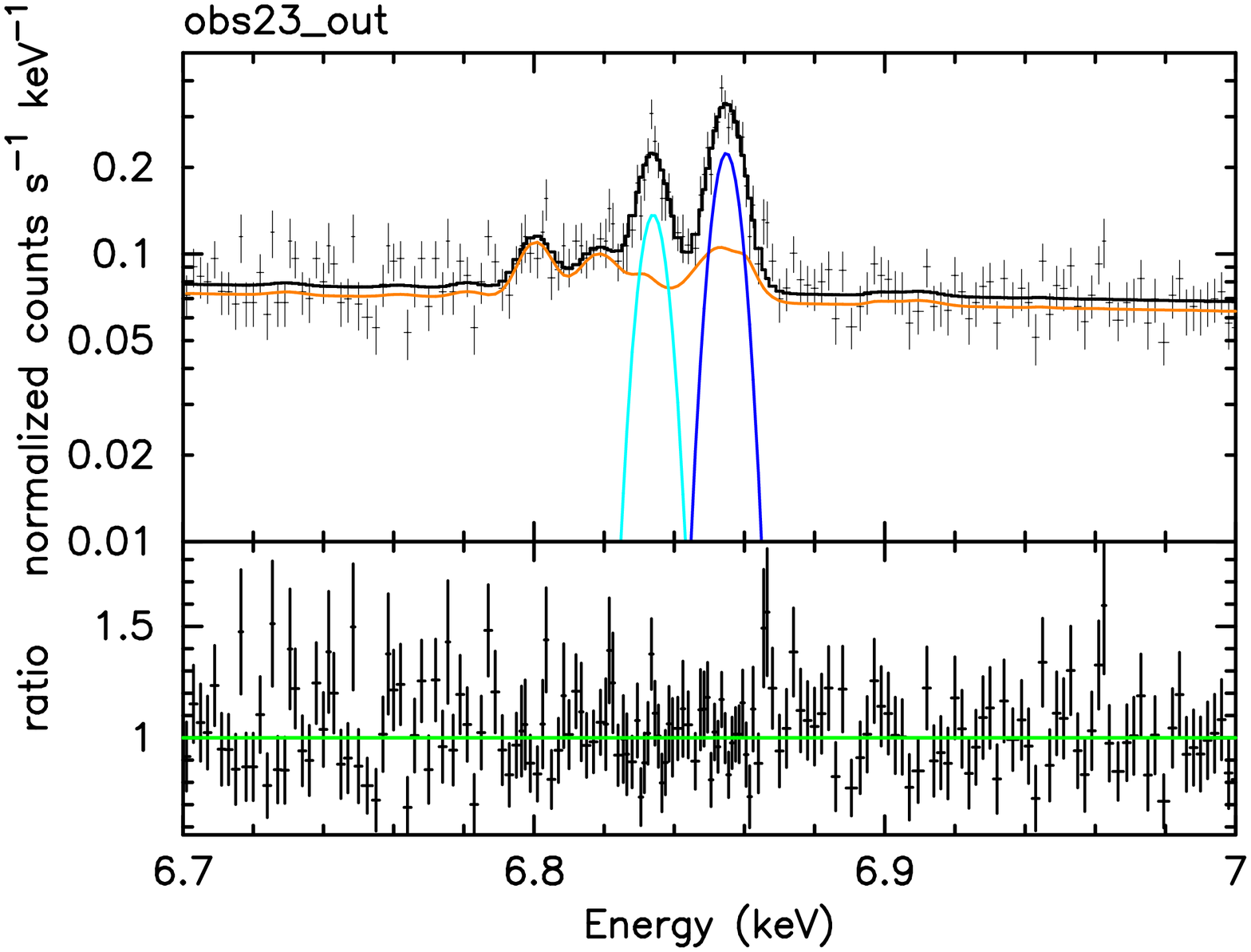}
\end{minipage}\hfill
\begin{minipage}{0.33\textwidth}
\FigureFile(\textwidth,\textwidth){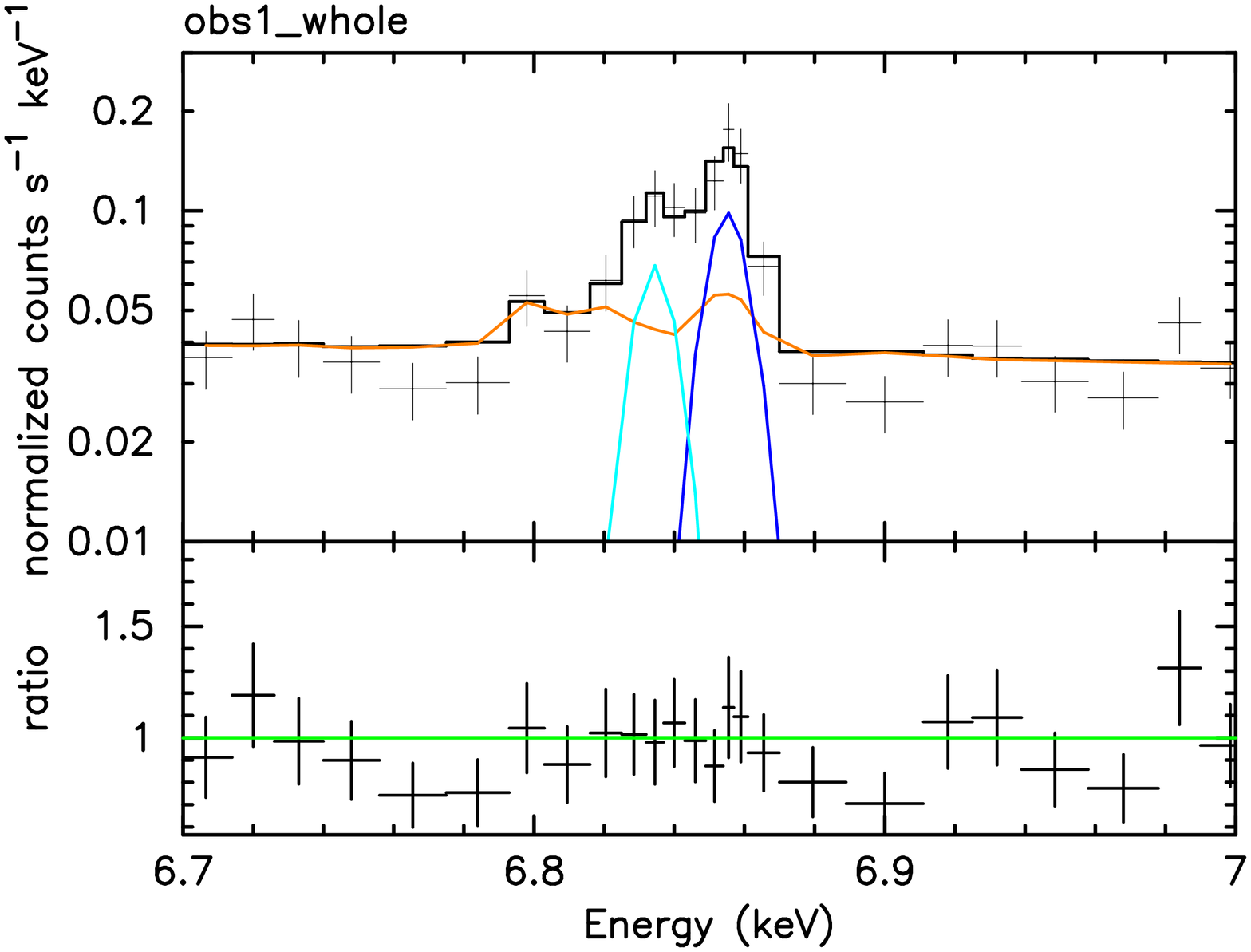}
\end{minipage}

\vspace*{2ex}
\caption{ Same as figure \ref{fig:spectra-all} but with AtomDB version 3.0.9.}
\label{fig:spectra-allv309}
\end{figure*}

As described in section \ref{sec:spectralfits}, we found 
the residuals around 6.55 keV in the spectral fit with 
AtomDB version 3.0.8\@. This comes from the overestimation 
of the Li-like Fe\emissiontype{XXIV} lines in AtomDB 3.0.8\@.
Since the corrected version 3.0.9 will be release in a few month, 
we examined reanalysis with AtomDB version 3.0.9. 
Figure \ref{fig:spectra-allv309} and table \ref{tab:ratiosv309} 
show the resultant spectral fits and parameters with AtomDB version 
3.0.9\@. The resultant parameters slightly changed comparing to 
the results in main text within the statistical errors, 
but the differences are negligible for our results.
Figure \ref{fig:opticaldepth_v309} and table \ref{tb:felines_v309} 
show the optical depth and line properties with AtomDB version 3.0.9.
The changes in Li-like Fe\emissiontype{XXIV} lines are separated 
from the He-like Fe\emissiontype{XXV} resonance and forbidden 
lines, these are also negligible for our simulations and conclusions.

\begin{table*}[t]
  \caption{Same as table \ref{tab:ratios} but with AtomDB version 3.0.9}
\label{tab:ratiosv309}
\begin{center}
\begin{tabular}{lrrrrrrr} \hline
Region ID & $kT^{\ast}$ & Fe$^{\ast}$ & $\sigma_{\rm v}^{\ast}$ & C-stat/d.o.f$^{\ast}$ & C-stat/d.o.f$^{\dagger}$ & & \\
  & keV & solar & km sec$^{-1}$ & 1.8--20 keV & 6.1--7.9 keV & & \\\hline
obs23\_cen  & $3.85 \pm 0.03$ & $0.73 \pm 0.01$ & $161 \pm 7$ & 10600/11151 & 1797/1784  &  & \\
obs23\_out  & $3.98 \pm 0.01$ & $0.73 \pm 0.01$ & $147 \pm 5$ & 14518/11744 & 1930/1784  &  & \\
obs1\_whole & $4.99 \pm 0.09$ & $0.57 \pm 0.03$ & $163 \pm 18$ & 6324/6930   & 1277/1494  & & \\
\hline \hline
Region ID & $w/z^{\dagger}$ & $w$/He$\beta^{\dagger}$ & $w$/Ly$\alpha_1^{\dagger}$ & $w$/Ly$\alpha_2^{\dagger}$ & $z$/He$\beta^{\dagger}$ & $z$/Ly$\alpha_1^{\dagger}$ & $z$/Ly$\alpha_2^{\dagger}$ \\ 
Line ratio & & & & & & & \\ 
\hline
obs23\_cen  & $2.34 \pm 0.10$ & $5.80 \pm 0.55$ & $9.62 \pm 0.95$ & $17.74 \pm 2.04$ & $2.48 \pm 0.25$ & $4.10 \pm 0.42$ & $7.57 \pm 0.30$ \\
obs23\_out  & $2.50 \pm 0.08$ & $6.06 \pm 0.56$ & $9.26 \pm 0.57$ & $15.16 \pm 1.22$ & $2.43 \pm 0.23$ & $3.70 \pm 0.24$ & $6.07 \pm 0.21$ \\
obs1\_whole & $3.18 \pm 0.33$ & $6.21 \pm 0.92$ & $6.86 \pm 1.12$ & $9.74 \pm 1.95$  & $1.95 \pm 0.33$ & $2.16 \pm 0.39$ & $3.07 \pm 0.36$ \\
\hline
Region ID & $w^{\dagger}$ & $z^{\dagger}$ & Ly$\alpha^{\dagger}$ & He$\beta^{\dagger}$ & & &  \\ 
Line width ($\sigma_{\rm v+th}$) &  eV & eV & eV & eV & & & \\ 
\hline
obs23\_cen  & $4.40 \pm 0.11$ & $3.78 \pm 0.22$ & $5.29 \pm 0.58$ & $3.43 \pm 0.51$ & & &  \\
obs23\_out  & $4.12 \pm 0.08$ & $3.68 \pm 0.15$ & $3.44 \pm 0.26$ & $4.26 \pm 0.45$ & & &  \\
obs1\_whole & $4.30 \pm 0.22$ & $3.78 \pm 0.51$ & $6.09 \pm 1.02$ & $4.90 \pm 0.90$ & & &  \\
\hline
\hline\\[-1ex]
\multicolumn{8}{l}{\parbox{0.9\textwidth}{\footnotesize
\footnotemark[$\ast$]
Fits in the broad band of 1.8--20.0 keV with the AGN and modified {\it bvvapec} models which exclude only the resonance line and add the line with the Gaussian model. $\sigma_{\rm v}$ is a turbulent velocity in {\it bvvapec} model without the resonance line. The numbers in this table are slightly smaller than those in V paper \citep{velocity} which are from the difference of the energy band in the spectral fits.
}}\\
\multicolumn{8}{l}{\parbox{0.9\textwidth}{\footnotesize
\footnotemark[$\dagger$]
Fits in the narrow band of 6.1--7.9 keV with the AGN and modified {\it bvvapec} models which exclude the He-$\alpha$ resonance and forbidden, He-$\beta$1\&2, and Ly-$\alpha$1\&2 lines.
}}\\
\end{tabular}
\end{center}
\end{table*}

\ifnum1=0
\begin{figure*}[h]
\begin{center}
\begin{minipage}{0.25\textwidth}
\FigureFile(\textwidth,\textwidth){./fig_v309_rev/scatt_w2z_20170802.eps}
\end{minipage}\hfill
\begin{minipage}{0.25\textwidth}
\FigureFile(\textwidth,\textwidth){./fig_v309_rev/scatt_z2hb_20170802.eps}
\end{minipage}\hfill
\begin{minipage}{0.25\textwidth}
\FigureFile(\textwidth,\textwidth){./fig_v309_rev/scatt_wwid_20170802.eps}
\end{minipage}\hfill
\begin{minipage}{0.25\textwidth}
\FigureFile(\textwidth,\textwidth){./fig_v309_rev/scatt_zwid_20170802.eps}
\end{minipage}
\vspace*{5ex}
\end{center}
\caption{
{\bf [\textcolor{red}{AtomDB v3.0.9}]}
Scattering plots between the gain corrected and uncorrected data for (a) 
the w/z line ratio, (b) the z/He$\beta$ line ratio, 
(c) w line width ($\sigma_{\rm v+th}$) and (d) z line width 
($\sigma_{\rm v+th}$).
Open circles, squares, and triangles correspond to the measurements in the obs23\_cen, obs23\_out, and obs1\_whole regions, respectively.
}
\label{fig:scatplot}
\end{figure*}
\fi

\begin{figure*}[tb]
\begin{minipage}{0.45\textwidth}
\includegraphics[width=\textwidth]{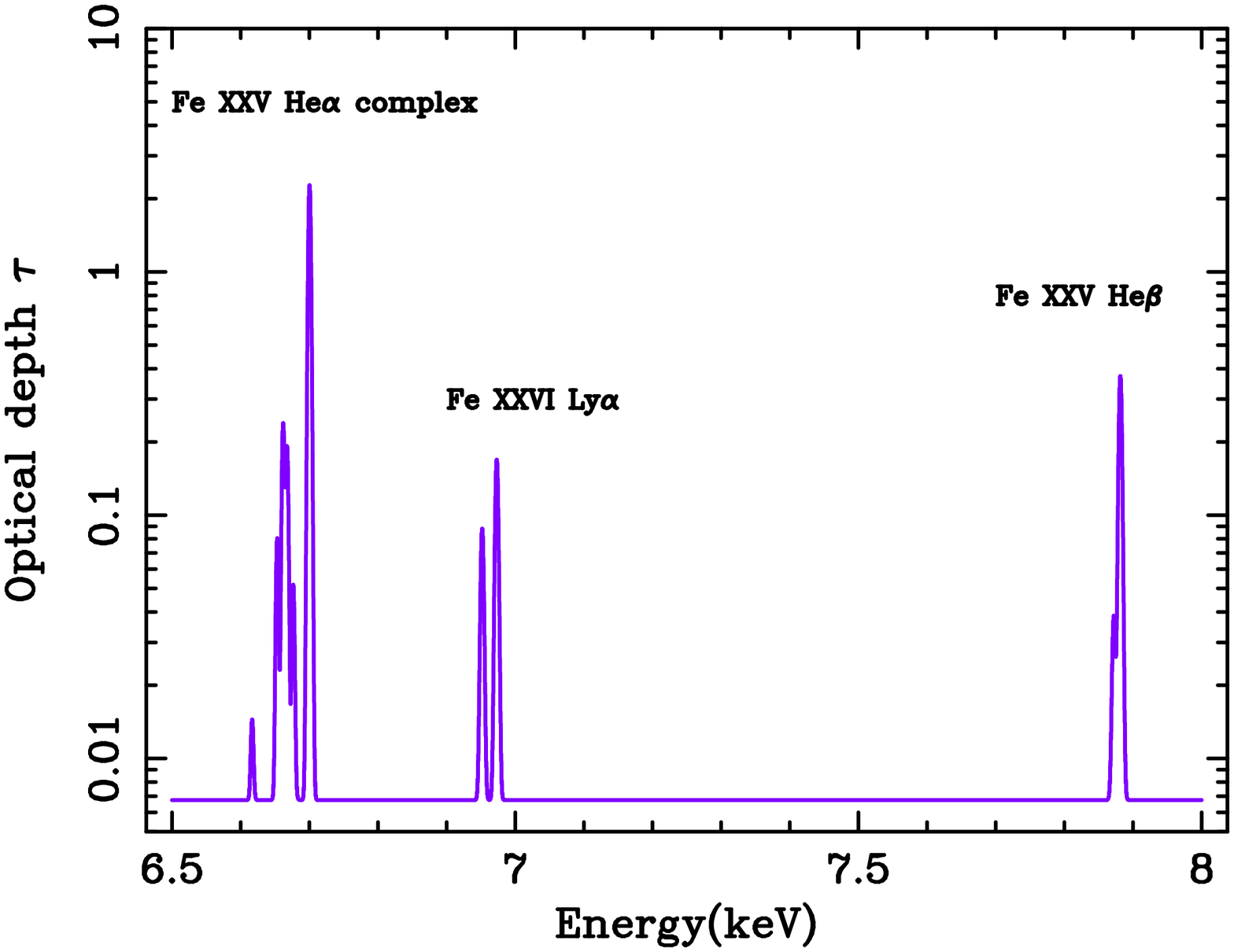}
\end{minipage}\hfill
\begin{minipage}{0.45\textwidth}
\includegraphics[width=\textwidth]{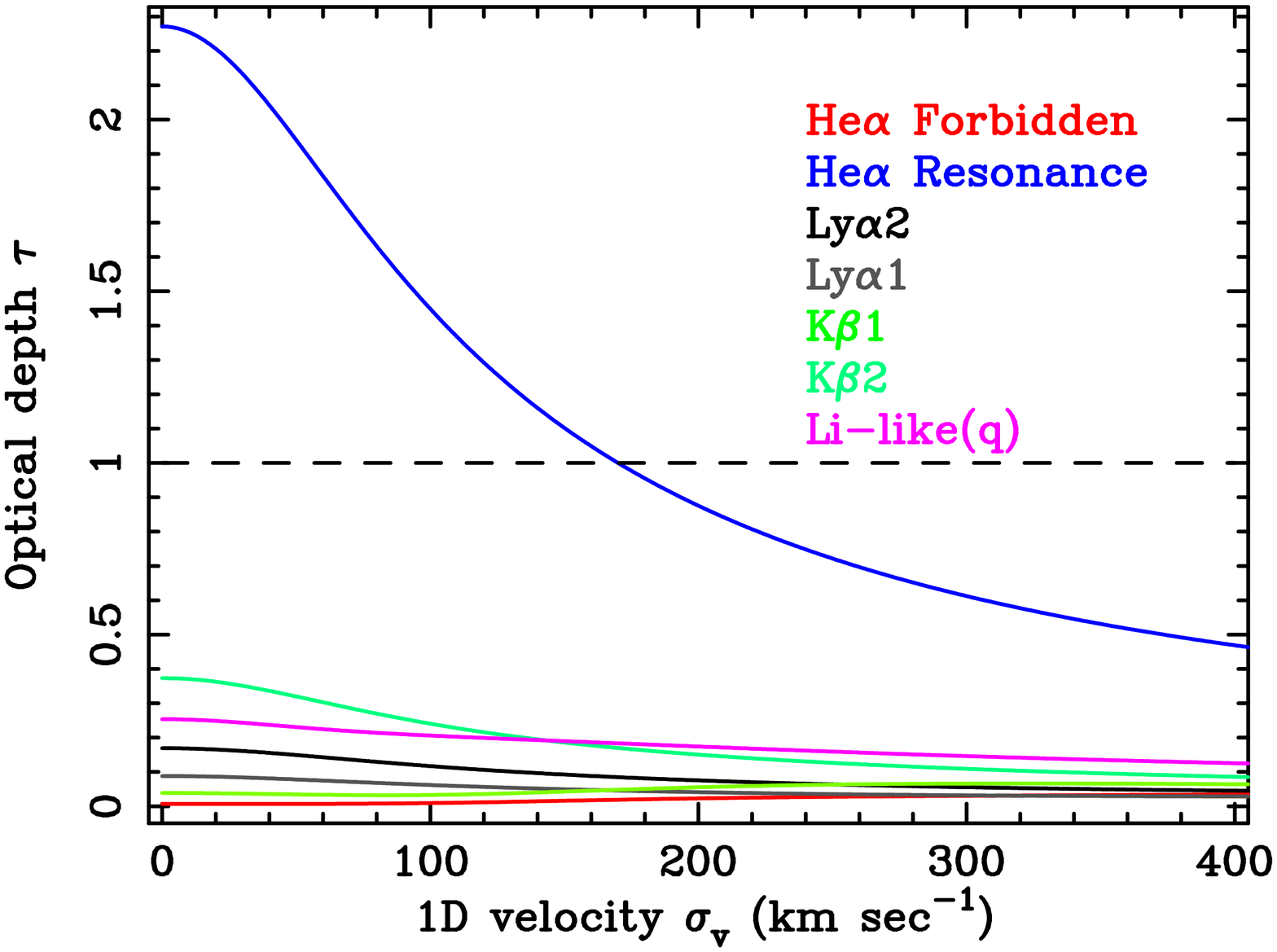}
\end{minipage}
\vspace*{1ex}
\caption{
Same as figure \ref{fig:opticaldepth} but with AtomDB version 3.0.9.
}
\label{fig:opticaldepth_v309}
\end{figure*}

\begin{table*}[tb]
\caption{Same as table \ref{tb:felines} with AtomDB 3.0.9.}
\label{tb:felines_v309}
\begin{center}
  \begin{tabular}{l c c c c c l} \hline \hline
    \makebox[40pt] [l]{Ion} & \makebox[40pt] {Energy} & \makebox[40pt] {Lower Level$^\ast$} & \makebox[40pt] {Upper Level$^\ast$} & \makebox[70pt] {Oscillator strength} & \makebox[70pt] {Optical depth $\tau$} & \makebox[70pt] [l]{Comments$^\ast$}\\
    & (eV)&  &  & & $\sigma_{\rm v}=0$ km sec$^{-1}$ & \\ \hline
    Fe\emissiontype{XXIV}& 6616.73 & $1s^2 2s_{1/2}\ ^2S_{1/2}$ & $1s_{1/2}2s_{1/2} 2p_{1/2}\ ^4P_{3/2}$  & 1.63$\times 10^{-2}$ & 1.45$\times10^{-2}$ & $u$ \\
    Fe\emissiontype{XXV} & 6636.58 & $1s^2\ ^1S_0$             & $1s2s\ ^3S_1$                        & 3.03$\times 10^{-7}$ & 6.75$\times 10^{-3}$ & He$\alpha$, $z$ \\
    Fe\emissiontype{XXIV}& 6653.30 & $1s^2 2s_{1/2}\ ^2S_{1/2}$ & $1s_{1/2}2s_{1/2} 2p_{1/2}\ ^2P_{1/2}$  & 1.57$\times 10^{-1}$ & 8.04$\times 10^{-2}$ & $r$ \\
    Fe\emissiontype{XXIV}& 6661.88 & $1s^2 2s_{1/2}\ ^2S_{1/2}$ & $1s_{1/2}2s_{1/2} 2p_{3/2}\ ^2P_{3/2}$  & 4.89$\times 10^{-1}$ & 2.39$\times 10^{-1}$ & $q$ \\
    Fe\emissiontype{XXV} & 6667.55 & $1s^2\ ^1S_0$             & $1s_{1/2}2p_{1/2}\ ^3P_{1}$            & 5.79$\times 10^{-2}$ & 1.92$\times 10^{-1}$ & He$\alpha$, $y$ \\
    Fe\emissiontype{XXIV}& 6676.59 & $1s^2 2s_{1/2}\ ^2S_{1/2}$ & $1s_{1/2}2s_{1/2} 2p_{3/2}\ ^2P_{1/2}$  & 9.62$\times 10^{-2}$ & 5.18$\times 10^{-2}$ & $t$ \\
    Fe\emissiontype{XXV} & 6682.30 &  $1s^2\ ^1S_0$            &  $1s_{1/2}2p_{3/2}\ ^3P_{2}$           & 1.70$\times 10^{-5}$ & 7.26$\times 10^{-3}$ & He$\alpha$, $x$ \\
    Fe\emissiontype{XXV} & 6700.40 &  $1s^2\ ^1S_0$            &  $1s_{1/2}2p_{3/2}\ ^1P_{1}$           & 7.19$\times 10^{-1}$ & 2.27 & He$\alpha$, $w$\\
    Fe\emissiontype{XXVI}& 6951.86 &  $1s$                     & $2p_{1/2}$                           & 1.36$\times 10^{-1}$ & 8.81$\times 10^{-2}$ & Ly$\alpha_2$\\
    Fe\emissiontype{XXVI}& 6973.07 &  $1s$                     & $2p_{3/2}$                           & 2.73$\times 10^{-1}$ & 1.69$\times 10^{-1}$ & Ly$\alpha_1$\\
    Fe\emissiontype{XXV} & 7872.01 & $1s^2\ ^1S_0$             & $1s 3p\ ^3P_1$                       & 1.18$\times 10^{-2}$ & 3.87$\times 10^{-2}$ & He$\beta_2$, intercomb.\\
    Fe\emissiontype{XXV} & 7881.52 & $1s^2\ ^1S_0$             & $1s 3p\ ^1P_1$                       & 1.37$\times 10^{-1}$ & 3.73$\times 10^{-1}$ & He$\beta_1$, resonance\\
\hline \hline\\[-1ex]
\multicolumn{7}{l}{\parbox{0.9\textwidth}{\footnotesize
\footnotemark[$\ast$]
Letter designations for the transitions as per \citet{Gabriel72}
}}\\
  \end{tabular}
\end{center}
\end{table*}

\end{document}